\documentclass[prb,showpacs,twocolumn,amsmath,amssymb,floatfix]{revtex4-1}
\usepackage{graphicx}
\usepackage{bm}
\usepackage{bbm}
\usepackage{color}
\usepackage{amsmath}
\usepackage{amssymb}
\usepackage{color}
\usepackage{footnote}
\usepackage{epstopdf}
\usepackage{hyperref}
\usepackage{appendix}
\usepackage[applemac]{inputenc}

\begin{document}
\title{Odd-frequency superconducting pairing in junctions with Rashba spin-orbit coupling}
\author{Jorge Cayao and Annica M. Black-Schaffer}
\affiliation{Department of Physics and Astronomy, Uppsala University, Box 516, S-751 20 Uppsala, Sweden}
\date{\today} 
\begin{abstract}
We consider normal-superconductor (NS) and superconductor-normal-superconductor (SNS) junctions based on one-dimensional nanowires with Rashba spin-orbit coupling and proximity-induced $s$-wave spin-singlet superconductivity and analytically demonstrate how both even- and odd-frequency and spin-singlet and -triplet superconducting pair correlations are always present.
In particular, by using a fully quantum mechanical scattering approach, we show that Andreev reflection induces mixing of spatial parities at interfaces, thus being the unique process which generates odd-frequency pairing; on the other hand, both  Andreev and normal reflections contribute to even-frequency pairing. 
We further  find that locally neither odd-frequency nor spin-triplet correlations are induced, but only even-frequency spin-singlet pairing. 
In the superconducting regions of NS junctions, the interface-induced amplitudes decay into the bulk, with the odd-frequency components being generally much larger than the even-frequency components at low frequencies.  
 The odd-frequency pairing also develops short and long-period oscillations due to the chemical potential and spin-orbit coupling, respectively, leading to a visible beating feature in their magnitudes.
Moreover, we find that in short SNS junctions at $\pi$-phase difference and strong spin-orbit coupling, the odd-frequency spin-singlet and -triplet correlations strongly dominate with an alternating spatial pattern for a large range of sub-gap frequencies.
\end{abstract}

\maketitle
\section{Introduction}
\label{sect0}
Conventional superconductors are characterized by an equal-time pairing amplitude, but in its most general form superconducting pairing  can also occur at different times, or equivalently at finite frequency. Such finite frequency pairing opens for an odd-frequency dependence of the pairing amplitudes, or Cooper pair wave-functions, which have the only formal requirement that they need to be fully antisymmetric under the interchange of all quantum numbers (frequency, spin, and spatial coordinates), due to Fermi-Dirac statistics.
For even-frequency dependence this leads to the usual classification of symmetries of Cooper pairs into spin-singlet and even parity in spatial coordinates, such as $s$-wave (in short ESE for its frequency, spin, and spatial symmetries) or spin-triplet and odd in spatial coordinates, e.g., $p$-wave (ETO). But the antisymmetry condition also allows for more exotic odd-frequency pairing, where the pairing function can instead have spin-triplet and even spatial parity ($s$-wave) symmetry (OTE) or spin-singlet and odd in spatial parity ($p$-wave) symmetry (OSO).

Originally,\citet{bere74} envisaged a possible odd-frequency dependence in the superconducting order parameter in the context of superfluid $^3$He with OTE symmetry.  Later, \citet{PhysRevB.45.13125} investigated an odd-frequency order parameter in spin-singlet superconductors with OSO symmetry and discussed interesting peculiarities of such a non-trivial state.
Although the initial excitement of odd-frequency superconductivity was related to the frequency dependence of the order parameter itself, no experimental evidence of such a state  has been reported so far. The search has instead mainly turned towards the study of odd-frequency pair correlations in systems possessing conventional even-frequency order parameters.
In this regard, it is important to mention that even such odd-frequency pair correlations can arise due to intrinsic properties of the system, as found for example in multiband superconductors,\cite{PhysRevB.88.104514,PhysRevB.92.094517,PhysRevB.92.224508,PhysRevLett.119.087001,PhysRevB.97.064505} or even due to more exotic mechanisms.\cite{PhysRevLett.66.1533,PhysRevB.46.8393,PhysRevB.60.3485,PhysRevB.45.13125,PhysRevB.52.1271,0953-8984-9-2-002}

Odd-frequency pairing has also been extensively studied in hybrid systems that include superconductor-ferromagnet junctions,\cite{longrangeExp,PhysRevLett.86.4096,PhysRevLett.90.117006,RevModPhys.77.1321,RevModPhys.77.935,PhysRevB.73.104412,PhysRevB.75.104509, PhysRevB.75.134510,
Eschrig2007,EschrigNat15,
PhysRevLett.104.137002,PhysRevB.82.100501,PhysRevB.83.144518, 7870d3ff91ed485fa3e55e901ff81c80,visani12,PhysRevB.85.184526,PhysRevB.87.104513,
PhysRevB.89.180506,0953-8984-26-45-453201,PhysRevB.92.060501,PhysRevB.92.014508,bernardo15,linder15A, 0034-4885-78-10-104501,LinderNat15,hwang17,jeon18} normal-superconducror (NS) junctions,\cite{PhysRevLett.98.037003,PhysRevLett.99.037005,Eschrig2007,PhysRevB.76.054522,PhysRevLett.98.037003,Nagaosa12,PhysRevLett.98.037003,PhysRevB.92.134512,PhysRevB.95.184518,Ebisu16,PhysRevB.95.224502,Triola18} topological insulators-superconductor junctions,\,\cite{PhysRevB.86.144506,PhysRevB.87.220506,bo2016,PhysRevB.92.205424,PhysRevB.92.100507,PhysRevB.96.155426,PhysRevB.96.174509,PhysRevB.97.075408,PhysRevLett.120.037701,PhysRevB.97.134523}
as well as  in inhomogeneous systems under time-dependent fields.\cite{PhysRevB.94.094518,triola17}
In NS junctions, odd-frequency pairing is generated due to the interface breaking the spatial parity, which allows the transformation from even $s$-wave to odd $p$-wave symmetry, while conserving the spin structure. Likewise, under the presence of spin active fields for example created by a ferromagnet, a spin-singlet to -triplet conversion can take place, which has raised strong interest due to its fundamental importance in unconventional superconductivity\cite{RevModPhys.63.239,Maeno94,PhysRevLett.80.3129,PhysRevLett.107.077003} and also due to applications in superconducting spintronics.\cite{EschrigNat15,7870d3ff91ed485fa3e55e901ff81c80,EschrigNat15,0034-4885-78-10-104501}

Although the existence of intrinsic odd-frequency pairing has been debated, the emergence of induced odd-frequency pairing in hybrid systems is nowadays well established. 
For instance, the induced long-range superconducting correlations in superconductor-ferromagnet junctions \cite{longrangeExp} are only explained by considering odd-frequency superconducting pairing.\cite{PhysRevLett.86.4096,PhysRevLett.90.117006,RevModPhys.77.1321} Similarly, the enhancement of local density of states at interfaces of hybrid junctions, \cite{PhysRevLett.99.037005,PhysRevB.76.054522,PhysRevB.92.205424,PhysRevB.96.155426} as well as Majorana bound states in topological superconductors have been demonstrated to be due to odd-frequency pairing.\cite{PhysRevLett.70.2960,Nagaosa12,PhysRevB.87.104513,PhysRevB.92.014513,PhysRevB.92.121404,lutchyn16,PhysRevB.92.205424,PhysRevB.95.174516,PhysRevB.96.155426} For a recent review on progress on odd-frequency superconductivity, see Ref.\,[\onlinecite{Balatsky2017}].

While interfaces generally mix spatial parities, spin active fields also allow mixing of spin states, which opens for spin-triplet pairing amplitudes using simple conventional spin-singlet superconductors.
Among the alternatives to achieve spin-triplet odd-frequency superconducting pairing, hybrid structures with magnetic fields have mainly been considered so far.  Another intriguing possibility is offered by spin-orbit (SO) coupling,\cite{PhysRev.100.580,Rashba1960,rashba84a} which exhibits large intrinsic values in many materials, thus avoiding the need of external magnetic fields and explicit breaking of time-reversal symmetry. The SO coupling can be of different nature depending on the crystal symmetries,\cite{SAMOKHIN20092385} and can, in some situations, even be controlled by voltage gates.\cite{0268-1242-11-8-009,PhysRevLett.78.1335,doi:10.1021/nl301325h,Takase17}
Only very few and recent works have so far investigated this possibility of SO coupling generating significant odd-frequency pairing and then with their major findings focused on two-dimensional situations.\cite{PhysRevB.92.134512,PhysRevLett.116.257001,PhysRevB.95.184518,Tamura18} Additional studies are thus desirable for a better understanding of the induced odd-frequency superconducting pairing in systems with SO coupling. In particular, one-dimensional systems are highly interesting in this regard as they both provide possibilities for analytical treatment within a fully quantum picture without any notable approximations and because of the huge recent experimental interest in SO coupled one-dimensional nanowires.\cite{chang15,Higginbotham,Krogstrup15,0957-4484-26-21-215202,Deng16,Albrecht16,zhang16,gulonder,Gazibegovic17,PhysRevMaterials.2.044202,vaitienkenas17,zhang18,deng18}

In this work we therefore investigate analytically and within a fully quantum mechanical framework the emergence of odd-frequency superconducting pairing in one-dimensional nanowire-based junctions with SO coupling. In particular, we consider NS and short superconductor-normal-superconductor (SNS) junctions with Rashba SO coupling,\cite{PhysRev.100.580,Rashba1960,rashba84a} which arises due to the lack of structural inversion symmetry. These are both highly relevant experimental situations as large values of intrinsic Rashba SO coupling have been reported in InAs,\cite{chang15,Higginbotham,Krogstrup15,Deng16,Albrecht16,vaitienkenas17,deng18} InSb,\cite{zhang16,0957-4484-26-21-215202,Gazibegovic17,zhang18} and InAsSb nanowires.\cite{PhysRevMaterials.2.044202} 
Additionally, a strong superconducting proximity effect has recently been observed in such nanowires, as clearly reflected in induced hard gaps,\cite{chang15,Higginbotham,Krogstrup15,0957-4484-26-21-215202,Deng16,Albrecht16,zhang16,Gazibegovic17,PhysRevMaterials.2.044202,vaitienkenas17,zhang18,deng18} and thus providing the necessary superconducting order parameter in the S regions of the wire.

Using retarded Green's functions extracted from scattering states, which allow us to fully analytically extract all pairing amplitudes, we demonstrate that both even- and odd-frequency spin-singlet (ESE and OSO) and spin-triplet (ETO and OTE) pairings are induced in SO coupled NS and SNS nanowires even without any external magnetic fields. The former two arise due to translational invariance breaking,\cite{PhysRevB.76.054522,PhysRevLett.98.037003,Eschrig2007} while the latter two are due to the SO field causing mixing between spin-singlet and -triplet spin states.\cite{PhysRevLett.87.037004,PhysRevLett.92.027003,PhysRevLett.92.097001,Reyren07,doi:10.1143/JPSJ.76.051008,PhysRevB.79.094504,PhysRevLett.113.227002,0034-4885-80-3-036501} 
We find that locally only ESE pairing survives due to the specific scattering processes present in junctions with Rashba SO coupling, while non-local pairing correlations, including both even and odd parities, are non-zero in all symmetry classes.

More specifically for the normal region of NS junctions, we find that all pairing amplitudes coexist and all are proportional solely to the Andreev processes at the interface. However, in the superconducting region, the pairing amplitudes acquire contributions from both the bulk and interface through Andreev as well as normal reflections processes. The interface terms exhibit an exponential and oscillatory decay from the interface, such that their magnitudes develop a beating profile  determined by  the SO coupling. 
Interestingly, we reveal that the odd-frequency components in both the normal and superconducting regions are only determined by Andreev processes. This we demonstrate as a direct consequence of Andreev scattering being responsible for mixing the spatial parities at interfaces. Although Andreev terms also contribute to the even-frequency components, these also contain additional contributions from normal reflections. 
At low-frequencies, both  odd-frequency spin-singlet and triplet pairings (OSO and OTE) are stronger than  even-frequency terms and thus explain the large values of the local density of states (LDOS) in the S region. 

In short SNS junctions we find that pairing amplitudes are phase ($\phi$) dependent 
and sense the emergence of Andreev bound states in the junction. Also in this case, we reveal that all odd-frequency components are solely generated by Andreev scattering.
We also show that at zero phase all the interface amplitudes vanish, provided full transparency and large chemical potentials in the junction, and they only acquire larger finite values as the phase approaches $\phi=\pi$. While at low SO coupling both the even- and odd-frequency components exhibit large values, at strong SO coupling odd-frequency pairing dominates, especially when frequencies are larger than the Andreev state energies and below the superconducting gap. Thus the odd-frequency pairing serves as an indicator for the reduction of the minigap at $\phi = \pi$ at large SO coupling.

The remainder of this paper is organized as follows. In Sec.\,\ref{sect1} we present the model and discuss the method based on retarded Green's functions calculated from scattering states. In Sec.\,\ref{sec2} we analytically derive all induced pairing amplitudes in NS and short SNS junctions at finite SO coupling. Finally, we present some concluding remarks in Sec.\,\ref{concl}. For completeness, in the Appendices we provide all the details on the derivation of the analytical calculations reported in this work and also show how odd-frequency spin-singlet pairing emerges in NS and short SNS junctions at zero SO coupling.

 \begin{figure*}[!ht]
\begin{minipage}[t]{\linewidth}
\centering
\includegraphics[width=.99\textwidth]{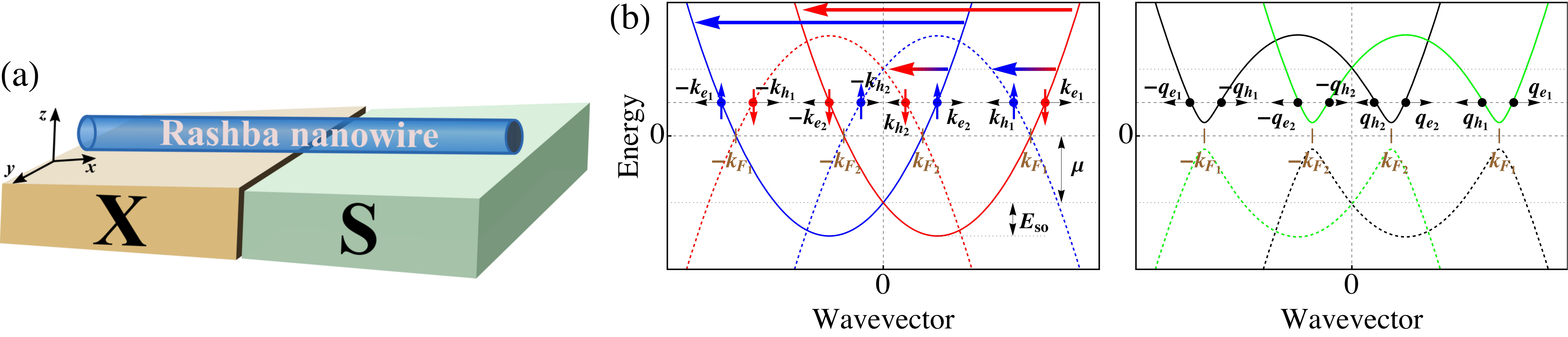} 
\caption{(Color online.) (a) Nanowire with Rashba SO coupling in contact with a conventional spin-singlet $s$-wave superconductor (S). Left region of the nanowire is in contact with either normal (X=N) or superconducting lead (X=S). 
(b) Energy dispersion with quasiparticles indicated for nanowire showing how SO coupling splits the normal bands around $k=0$ (left), while proximity to an $s$-wave superconductor opens gaps at the Fermi momenta $k_{F_{1,2}}$ (in brown), mixing electron and hole bands of different spins (right). Horizontal arrows indicate normal (solid color) and Andreev (changing colors) scattering processes. }
\label{fig1}
\end{minipage}
\end{figure*}

%%%%%%%%%%%%%%%%%%%%%%%%%%
%                   MODEL and METHOD                      %
%%%%%%%%%%%%%%%%%%%%%%%%%%

\section{Model and method}
\label{sect1}
We consider a one-dimensional single-mode nanowire with Rashba SO coupling\cite{PhysRev.100.580,Rashba1960,rashba84a} as shown in Fig.\,\ref{fig1},
whose right region is in proximity to a  spin-singlet $s$-wave superconductor (S), while the left region in contact either with a normal (X = N) or superconducting lead (X = S).  
Experimental advances have made very good contacts between nanowires and leads, which guarantees  induced superconducting correlations into the nanowire. The Bogoliubov-de Gennes Hamiltonian for the nanowire, in the basis $(\psi_{\uparrow},\psi_{\downarrow},\psi_{\uparrow}^{\dagger},\psi_{\downarrow}^{\dagger})$, can thus be written
\begin{equation}
\label{HBdG}
H_{\rm BdG}(x)=\Big(\frac{p_{x}^{2}}{2m}-\mu_{i}\Big)\sigma_{0}\tau_{z}+\frac{\alpha}{\hbar}\sigma_{z}\tau_{0}p_{x}+\Delta(x)\sigma_{y}\tau_{y}\,,
\end{equation}
where $p_{x}=-i\hbar\partial_{x}$, while $\sigma_{i}$ and $\tau_{i}$ are the $i$-Pauli matrices in spin and Nambu spaces, respectively. The first term is the kinetic energy of the nanowire with $\mu$ being the chemical potential. Due to different local environments of the wire controllable by gate voltages, we allow for different chemical potentials, $\mu_{\rm  N}$ and $\mu_{\rm S}$ in the normal and superconducting regions of the nanowires, respectively. 
The second term is the Rashba SO coupling with $\alpha$ being the SO strength. Note that the choice of SO direction along the $z$-axis is both most common and also allows us to directly compare with results recently obtained for the edge of a two-dimensional topological insulator.\cite{PhysRevB.96.155426} The third term represents the  induced conventional spin-singlet $s$-wave superconductivity with $\Delta(x)$ being the induced pairing potential which introduces a length scale $\xi=\hbar \sqrt{2\mu_{\rm S}/m}/\Delta$ known as the superconducting coherence length, where $\Delta$ is the value of the induced superconducting potential in the wire.
We consider for NS junctions  $\Delta(x)=\theta(x)\Delta$ and for SNS junctions $\Delta_{L/R}(x)=\Delta{\rm e}^{i\phi_{L/R}}$, with L/R denoting the left and right S regions,\footnote{Spatial variations of $\Delta(x)$ at the NS interfaces\cite{Fer18} do not alter the main conclusions of this work.} as shown in Fig.\,\ref{fig1}(a).  The interface between the X and S regions we model by adding the finite delta barrier $V\delta(x)$ to the above Hamiltonian, giving an interface transparency $Z=2mV/\hbar^{2}$. 

Our choice of system and Hamiltonian is strongly motivated by recent experiments where high quality junctions made of InAs\cite{chang15,Higginbotham,Krogstrup15,Deng16,Albrecht16,vaitienkenas17,deng18} and InSb\cite{zhang16,0957-4484-26-21-215202,Gazibegovic17,zhang18} nanowires with strong intrinsic SO coupling have been fabricated with extremely good contact to superconductors, resulting in hard induced gaps in the nanowires. Furthermore, a recent study reported that InAsSb\cite{PhysRevMaterials.2.044202} exhibits even larger values of SO coupling than previous two materials, making it important to study the impact of strong spin-orbit coupling.

The energy versus momentum for the Hamiltonian in Eq.~\eqref{HBdG} is plotted in Fig.\,\ref{fig1}(b) in the normal and superconducting regimes, respectively. The SO coupling splits the normal spin bands around $k=0$, while a finite superconducting potential $\Delta$ opens gaps at the Fermi momenta $k_{F_{1,2}}$, mixing electron and hole bands of different spins. Notice that at low energies, each energy corresponds to eight values of momentum, namely, $\pm k_{e_{1,2}}$ and $\pm k_{h_{1,2}}$ with
\begin{equation}
\label{kmomenta}
\begin{split}
k_{e_{1(2)}}(\omega)&=\pm k_{SO}+\sqrt{k_{SO}^{2}+k_{\mu_{i}}(1+\omega/\mu_{i})}\,
\end{split}
\end{equation}
where $k_{h_{1(2)}}(\omega)=k_{e_{1(2)}}(-\omega)$, $k_{SO}=m\alpha/\hbar^{2}$, and $k_{\mu_{i}}=\sqrt{2m\mu_{i}/\hbar^{2}}$ and $i$ denotes the normal (N) and superconducting (S) regions. In the S region at low energies but above the gap, the momenta acquire the same form but with $\omega\rightarrow \sqrt{\omega^{2}-\Delta^{2}}$. Notice how the SO coupling introduces a new energy scale associated with $k_{SO}$, which we refer to as the SO energy $E_{\rm so}=m\alpha^{2}/(2\hbar^{2})$.
Since there are no spin flip processes possible due to the absence of any magnetic fields, normal reflection  occurs between states of the same spin band, with different momenta e.g.~$k_{e_{1(2)}}\rightarrow -k_{e_{2(1)}}$ [see blue and red horizontal arrows in Fig.~\ref{fig1}(b)], while Andreev reflection takes place between states of different spin bands with momenta, e.g., $k_{e_{1(2)}}\rightarrow k_{h_{1(2)}}$ [see blue-red and red-blue gradient arrows]. This is different from the case in  topological insulators, where normal reflections are forbidden, despite  similarities in the SO term in the Hamiltonian (\ref{HBdG}).   

The aim of this work is to study the role of Rashba SO coupling  on the induced superconducting pairing in nanowire-based superconducting junctions. For this purpose we follow Ref.~[\onlinecite{PhysRevB.96.155426}] and first construct the retarded Green's function $G^{r}(x,x',\omega)$ with outgoing boundary conditions in each region derived from  the scattering processes at the interface.\cite{PhysRev.175.559} For a more detailed description, see Appendix \ref{AppG}. While the explicit forms of all Green's functions are given in  Appendix \ref{finiteSOC}, we focus in the main text primarily on their anomalous electron-hole components which directly determine the pairing amplitudes. In the chosen basis, the spin symmetry of the anomalous electron-hole component is obtained from
\begin{equation}
\label{decomposeSPIN}
G_{eh}^{r}(x,x',\omega)=(f^{r}_{0}\sigma_{0}+f^{r}_{j}\sigma_{j})i\sigma_{y}\,,
\end{equation}
where $\sigma_{i}$ is the $i$-Pauli matrix in spin space and repeated indices imply  summation. Here, $f^{r}_{0}$ corresponds to the spin-singlet ($\uparrow\downarrow-\downarrow\uparrow$), $f^{r}_{1,2}$ to the equal spin-triplet ($\downarrow\downarrow\pm\uparrow\uparrow$), and $f^{r}_{3}$ to the mixed spin-triplet ($\uparrow\downarrow+\downarrow\uparrow$) contributions. Due to the specific Fermi points of the Rashba bands [see Fig.\,\ref{fig1}(b)], only mixed spin-singlet $f_{0}^{r}$ and triplet $f_{3}^{r}$ states are expected,\cite{PhysRevLett.87.037004,PhysRevLett.92.027003,PhysRevLett.92.097001,Reyren07,doi:10.1143/JPSJ.76.051008,PhysRevB.79.094504,PhysRevLett.113.227002,0034-4885-80-3-036501} which is also what we explicitly find.

Since the equal spin-triplet amplitudes vanish, $f^{r}_{1,2}(x,x',\omega)=0$, for the remainder of this work we refer to the mixed spin-triplet amplitude $f^{r}_{3}$ simple as the spin-triplet component. Also, $f^{r}_{i}$ is referred to as pairing amplitude, while $|f^{r}_{i}|=\sqrt{f_{i}^{r}(f_{i}^{r})^{*}}$ denotes the pairing magnitude. Finally, we can extract the  even- and odd-frequency components by using
\begin{equation}
\label{EVENODD}
\begin{split}
f^{r,{\rm E}}_{i}(x,x',\omega)&=\frac{f^{r}_{i}(x,x',\omega)+f^{a}_{i}(x,x',-\omega)}{2}\,,\\
f^{r,{\rm O}}_{i}(x,x',\omega)&=\frac{f^{r}_{i}(x,x',\omega)-f^{a}_{i}(x,x',-\omega)}{2}\,,
\end{split}
\end{equation}
where $f^{a}_{i}$ necessarily correspond to the advanced Green's function $G^{a}(x,x',\omega)=[G^{r}(x',x,\omega)]^{\dagger}$.\cite{PhysRevB.96.155426}

In the Appendix \ref{finiteSOC} we provide the full expressions for the Green's functions, but in order to reach tractable analytical expressions in the main text we work with a few reasonable assumptions. We consider large chemical potentials in the nanowires such that $\mu_{i}+E_{\rm SO}\gg \omega,\Delta$. Then the wave vectors acquire the following form
\begin{equation}
\label{ksom}
\begin{split}
k_{e_{1(2)}}&=k_{F_{1(2)}}+\kappa^{\rm N}\,,
\end{split}
\end{equation} 
where $k_{F_{1(2)}}=\pm k_{\rm SO} + \bar{k}$, $\bar{k}=\sqrt{2m(\mu_{i}+E_{\rm SO})/\hbar^{2}}$, $\kappa^{N}=\bar{k}\omega/[2(\mu_{i}+E_{\rm SO})]$. In the superconducting region for $\omega<\Delta$ we have $k_{e_{1(2)}}^{S}=\pm k_{\rm SO} + \bar{k} + i\kappa$ with  $\kappa=\bar{k}\sqrt{\Delta^{2}-\omega^{2}}/[2(\mu_{S}+E_{\rm SO})]$.
Also to simplify some results we assume the same chemical potential $\mu_{\rm N} = \mu_{\rm S}$. Finally, in the main text we mainly consider fully transparent junctions, $Z = 0$, but whenever important we incorporate a finite $Z$ and the full expressions are always given in the appendix.

% -------------------------------------- %
%  PAIRING AMPLITUDE:
% -------------------------------------- %
\section{Pairing amplitude analysis}
\label{sec2}
Following the method outlined above we here analyze the pairing amplitudes and their symmetries in both NS and short SNS junctions based on nanowires with Rashba SO coupling. 

\subsection{NS junctions}
\label{sec3}
We first focus on a semi-infinite NS junction with the interface located at $x=0$.
The Green's function in each region of the hybrid junction is a $4 \times 4$ matrix in electron-hole and spin spaces due to the SO coupling and is calculated from the scattering processes at the interface. 
The pairing amplitudes are then directly obtained from the anomalous electron-hole  element of $G^{r}(x,x',\omega)$  using Eqs\,.(\ref{decomposeSPIN}) and \eqref{EVENODD}, assuming that $\mu_{i}+E_{\rm SO}\gg \omega,\Delta$ with $i=\text{N,S}$ in order to derive analytical expressions.  Full derivation and details on the calculations are given in Appendix \ref{finiteSOC}, such that we here can focus on the resulting even and odd-frequency pairing amplitudes.

In the normal N region we obtain for the even and odd-frequency pairing amplitudes 
\begin{equation}
\label{PairinNS_N}
\begin{split}
f_{0}^{r, {\rm E}}(x,x',\omega)&=\frac{\eta}{2i}{\rm cos}[\bar{k}(x-x')]\mathcal{A}^{-+}_{xx'}(\omega)\,,\\
f_{0}^{r, {\rm O}}(x,x',\omega)&=-\frac{\eta}{2}{\rm sin}[\bar{k}(x-x')]\mathcal{A}_{xx'}^{-+}(\omega)\,,\\
f_{3}^{r, {\rm E}}(x,x',\omega)&=\frac{\eta}{2i}{\rm cos}[\bar{k}(x-x')]{\rm sgn}(x-x')\mathcal{A}_{xx'}^{+-}(\omega)\,,\\
f_{3}^{r, {\rm O}}(x,x',\omega)&=-\frac{\eta}{2}{\rm sin}[\bar{k}|x-x'|]%{\rm sgn}(x-x')
\mathcal{A}_{xx'}^{+-}(\omega)\,,\\
\end{split}
\end{equation}
which correspond to ESE, OSO, ETO and OTE classes, respectively, as easily seen by their frequency, spin, and spatial parities. Here $\eta = 2m/\hbar^2$, while $\mathcal{A}_{xx'}^{mn}(\omega)={\rm e}^{-i\kappa_{\omega}^{N}(x+x')}\Big[A_{m}{\rm cos}[k_{\rm SO}|x-x'|]-A_{n}i{\rm sin}[k_{\rm SO}|x-x'|] \Big]$, 
and $A_{\pm}=a_{42}\pm a_{31}$, $a_{42}=-a_{31}$. The expressions for the Andreev coefficients $a_{ij}$  are given in the Appendix \ref{NSAppSOC}.

\begin{figure}[!ht]
%\begin{minipage}[t]{\linewidth}
\centering
\includegraphics[width=.48\textwidth]{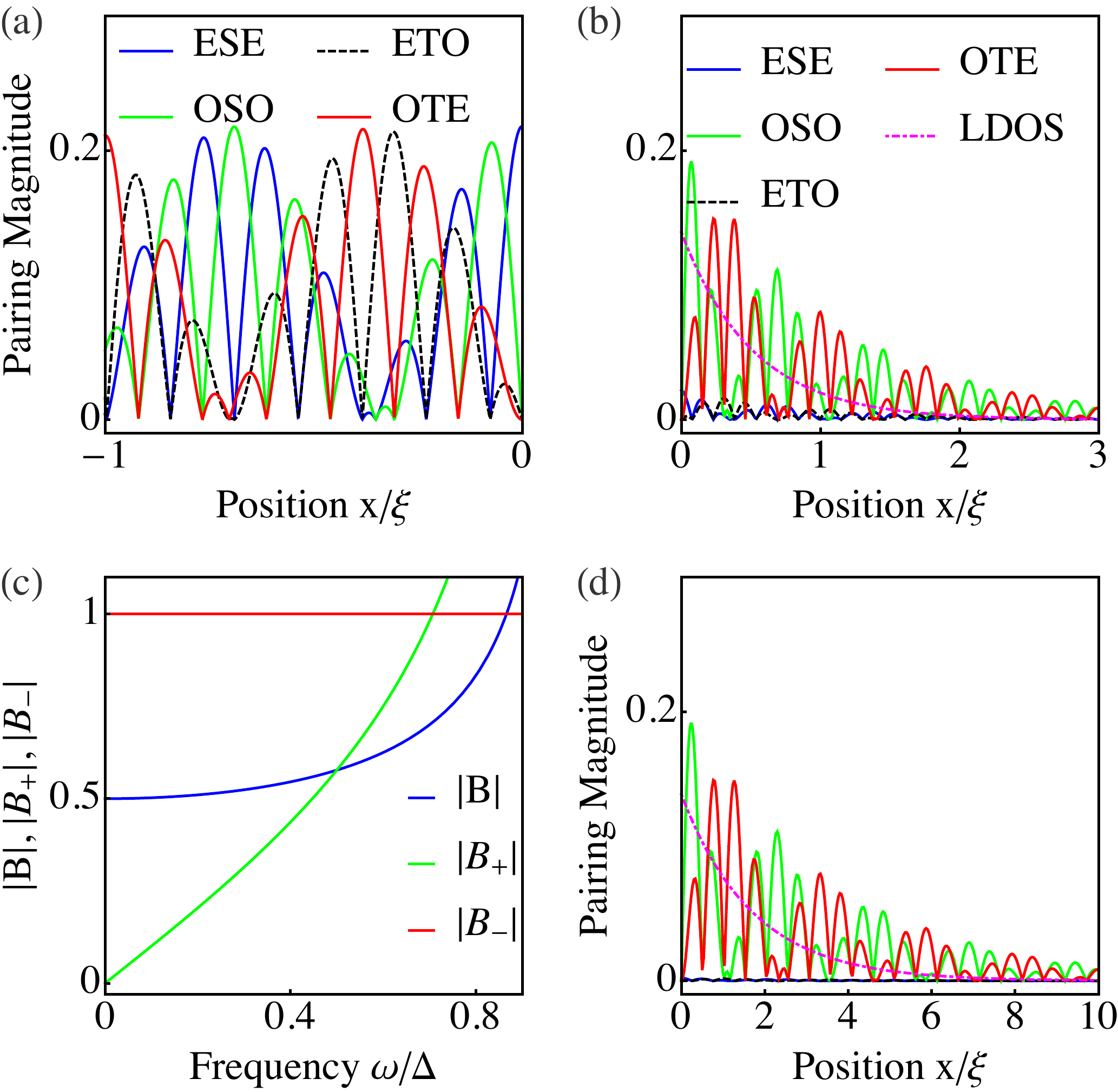} 
\caption{(Color online.) Spatial dependence of the pairing magnitudes in NS junctions in N region (a) and S region for $\omega=0.1\Delta$ (b) and $\omega=0.01\Delta$ (d), as well as frequency dependence for the factors $B$ and $B_{\pm}$ (c).  Magenta curve in (b,d) shows the spatial dependence of the LDOS. Parameters: $\omega=0.1\Delta$, $x'=0$, $E_{\rm so}=0.5\Delta$, $\mu_{N,S}=10\Delta$, $Z = 0$.}
\label{fig2}
%\end{minipage}
\end{figure}

The first observation is that all symmetry classes (ESE, OSO, ETO, and OTE) are present in N and proportional to the Andreev reflection coefficients $a_{ij}$, which forms the core of the proximity effect.\cite{Pannetier2000,Klapwijk2004} The finite values of these pairing amplitudes in N indeed represent the proximity-induced superconducting correlations, which interestingly, include both mixed spin-triplet pairing and odd-frequency components.
Notice that, at zero SO coupling, the spin symmetry is preserved and the last two expressions in Eqs.\,(\ref{PairinNS_N}) vanish, such that, as expected, only ESE and OSO amplitudes are induced into the N region due to the translation invariance breaking at the interface.\cite{PhysRevB.76.054522,PhysRevLett.98.037003,Eschrig2007} We have verified  this conclusion  by performing a full calculation at zero SO coupling, which is presented in Appendix \ref{zeroSOC}.
A finite Rashba SO induces spin-mixing,\cite{PhysRevLett.87.037004,PhysRevLett.92.027003,PhysRevLett.92.097001,Reyren07,doi:10.1143/JPSJ.76.051008,PhysRevB.79.094504,PhysRevLett.113.227002,0034-4885-80-3-036501} which gives rise to spin singlet $f_{0}^{r}$ and mixed spin-triplet correlations $f_{3}^{r}$ which are both non-local, i.e.,~they are finite only for $x\neq x'$, as also seen in Fig.\,\ref{fig2}(a). 

Secondly, all pairing amplitudes exhibit an oscillatory behavior determined by the SO coupling strength, through $\mathcal{A}_{xx'}^{mn}$, and by the chemical potential through $\bar{k}$, see Eqs.\,(\ref{PairinNS_N}) and Fig.\,\ref{fig2}(a). Note that our description is at zero temperature and therefore the induced pairing amplitudes in the N region do not decay but, instead, survive technically infinitely far away. Any finite temperature corrects this issue, one only needs to go into the Matsubara representation by $\omega\rightarrow i\omega$. This is a well-known behavior for proximity-induced superconductivity in the N region of a NS junction, since the decay length into N is $\xi^N \sim 1/T$, see e.g.~Refs.~[\onlinecite{PhysRevB.73.014503}] and [\onlinecite{PhysRevB.81.014517}].

Moreover, we find that, quite generally, at both zero and finite SO coupling, the Andreev reflection is solely responsible for mixing the spatial parities by mixing electron and hole wave vectors with spatial coordinates ($x$ and $x'$) and therefore all even and odd-frequency components are generated by Andreev reflection in the N region.
For instance, we find that the spin-triplet amplitude, see the derivation of Eqs.\,(\ref{PairinNS_N}) in Appendix \ref{NSAppSOC}, before decomposing into even and odd-frequency components is proportional to the Andreev coefficients (for right moving electron with spin up and down) multiplied by the term [$\theta(x-x'){\rm e}^{i(-k_{e_{1}}x+k_{h_{1}}x')}-\theta(x'-x){\rm e}^{i(-k_{e_{2}}x+k_{h_{2}}x')}$]. We directly see that this latter term mixes spatial coordinates ($x$ and $x'$) with electron and hole wave vectors of different spin bands, being therefore the generator of both even- and odd-parity amplitudes and subsequently also the generator of even- and odd-frequency components. 
We attribute this phenomenon to the Andreev process for the following reason. A right moving electron with spin down and momentum $k_{e_{1}}$ is Andreev reflected into a hole band with spin up and momentum $k_{h_{1}}$, leading to the first term in the exponent, $-k_{e_{1}}x+k_{h_{1}}x'$ [see red-blue gradient arrow in Fig.\,\ref{fig1}(b)]. Likewise, a right moving electron with spin up with momentum $k_{e_{2}}$ being Andreev reflected to a left moving hole with spin down with momentum $k_{h_{2}}$ and leading to a term $-k_{e_{2}}x+k_{h_{2}}x'$ [see blue-red gradient arrow in Fig.\,\ref{fig1}(b)].  
We can also understand the vanishing of the local ($x = x'$) spin-triplet amplitudes ETO and OTE as a direct consequence of the dependence on the Andreev reflection, by simply evaluating the exponential term given above at $x=x'$. Then we obtain for the two Andreev processes the exponents $-k_{e_{1}}+k_{h_{1}}$ and $-k_{e_{2}}+k_{h_{2}}$, which, interestingly, for a Rashba nanowire are the same and therefore the exponential term discussed above vanishes. This means vanishing local ETO and OTE amplitudes, in agreement with previous studies.\cite{PhysRevLett.113.227002,PhysRevB.92.134512} Similarly, we can explain the non-existence of a finite local OSO amplitude.

Although the expressions in Eqs.\,(\ref{PairinNS_N}) correspond to the large chemical potential limit, $\mu+E_{\rm SO}\gg \omega,\Delta$, it is straightforward to show that the existence of all four symmetry classes for the pairing amplitude remains under more general conditions. Also, the vanishing of odd-frequency and spin-triplet local $x=x'$ correlations is a general feature that does not rely on any approximation.

% IN S:
In the superconducting region of a NS junction the pairing amplitudes acquire a more complicated structure. In this case, the pairing amplitudes read as
\begin{widetext}
\begin{equation}
\label{PairinNS_S}
\begin{split}
f_{0,{\rm B}}^{r,{\rm E}}(x,x',\omega)&=B(\omega)2{\rm cos}[k_{SO}|x-x'|]{\rm e}^{-\kappa|x-x'|}
\bigg[\frac{{\rm e}^{i\bar{k}|x-x'|}}{k_{e_{1}}^{S}+k_{e_{2}}^{S}}
+\frac{{\rm e}^{-i\bar{k}|x-x'|}}{k_{h_{1}}^{S}+k_{h_{2}}^{S}}\bigg]\,,\\
f_{0,{\rm I}}^{r,{\rm E}}(x,x',\omega)&=B(\omega){\rm e}^{-\kappa(x+x')}
\bigg\{
\frac{{\rm e}^{i\bar{k}(x+x')}}{k_{e_{1}}^{S}+k_{e_{2}}^{S}}
\Big[ b_{51}{\rm e}^{ik_{SO}|x-x'|}+b_{62}{\rm e}^{-ik_{SO}|x-x'|}\Big]\\
&+
\frac{{\rm e}^{-i\bar{k}(x+x')}}{k_{h_{1}}^{S}+k_{h_{2}}^{S}}
\Big[ b_{82}{\rm e}^{ik_{SO}|x-x'|}+b_{71}{\rm e}^{-ik_{SO}|x-x'|}\Big]
\bigg\}
+B_{+}(\omega){\rm e}^{-\kappa(x+x')}{\rm cos}[\bar{k}(x-x')]\mathbb{A}_{xx'}^{+-}(\omega)\,,\\
f_{0,{\rm I}}^{r,{\rm O}}(x,x',\omega)&=B_{-}(\omega){\rm e}^{-\kappa(x+x')}i{\rm sin}[\bar{k}(x-x')]\mathbb{A}_{xx'}^{+-}(\omega)\,,\\
f_{3,{\rm B}}^{r,{\rm E}}(x,x',\omega)&=B(\omega)
(-2i){\rm sin}[k_{SO}|x-x'|]{\rm e}^{-\kappa|x-x'|}
\bigg[\frac{{\rm e}^{i\bar{k}|x-x'|}}{k_{e_{1}}^{S}+k_{e_{2}}^{S}}
+\frac{{\rm e}^{-i\bar{k}|x-x'|}}{k_{h_{1}}^{S}+k_{h_{2}}^{S}}\bigg]\,,\\
f_{3,{\rm I}}^{r,{\rm E}}(x,x',\omega)&=B(\omega){\rm e}^{-\kappa(x+x')}{\rm sgn}(x-x')
\bigg\{
\frac{{\rm e}^{i\bar{k}(x+x')}}{k_{e_{1}}^{S}+k_{e_{2}}^{S}}
\Big[ b_{62}{\rm e}^{-ik_{SO}|x-x'|}-b_{51}{\rm e}^{ik_{SO}|x-x'|}\Big]\\
&+
\frac{{\rm e}^{-i\bar{k}(x+x')}}{k_{h_{1}}^{S}+k_{h_{2}}^{S}}
\Big[ b_{71}{\rm e}^{-ik_{SO}|x-x'|}+b_{82}{\rm e}^{ik_{SO}|x-x'|}\Big]
\bigg\}+B_{+}(\omega){\rm e}^{-\kappa(x+x')}{\rm sgn}(x-x'){\rm cos}[\bar{k}(x-x')]\mathbb{A}_{xx'}^{-+}(\omega)\,,\\
f_{3,{\rm I}}^{r,{\rm O}}(x,x',\omega)&=B_{-}(\omega){\rm e}^{-\kappa(x+x')}{\rm sgn}(x-x')i{\rm sin}[\bar{k}(x-x')]\mathbb{A}_{xx'}^{-+}(\omega)\,,
\end{split}
\end{equation}
 \end{widetext}
 where $B(\omega)=\eta/[2i((u/v)-(v/u))]$, $B_{\pm}=(\eta/2i)(u^{2}\pm v^{2})/(u^{2}-v^{2})$, and $\mathbb{A}_{xx'}^{mn}(\omega)=A_{m}{\rm cos}[k_{SO}|x-x'|]-iA_{n}{\rm sin}[k_{SO}|x-x'|]$ with $A_{\pm}=\tilde{a}_{72}\pm a_{52}$. Also, $a_{52}=\tilde{a}_{72}$ and $b_{ij}$ are the Andreev and normal reflection coefficients, respectively, while $u$ and $v$ are the Bogoliubov coherence factors, all given by explicit expressions in Appendix \ref{finiteSOC}. The pairing amplitudes in Eq.~\eqref{PairinNS_S} are divided up into bulk (B) and interface (I) contributions, where bulk contributions do not depend  on any Andreev or normal reflections processes, as they are associated with the interface. 
 
Identifying symmetries we see that Eq.~\eqref{PairinNS_S} corresponds to ESE, OSO, ETO, and OTE symmetries, respectively, where the OSO and OTE pairing amplitudes vanish in the bulk, i.e.~ $f_{0(3){\rm B}}^{r,{\rm O}}(x,x',\omega)=0$. Furthermore, we notice that the even-frequency amplitudes (ESE and ETO) both have a bulk term and their interface contributions include normal (i.e.~proportional to $b_{ij}$) and Andreev (i.e.~proportional to $a_{ij}$) reflections, while the  odd-frequency amplitudes (OSO and OTE) contain solely interface contributions, which are proportional  only to Andreev reflection coefficients. The last conclusion also holds at zero SO coupling as reported in Appendix \ref{zeroSOC}. This result further supports the direct coupling between induced odd-frequency pairing and Andreev reflection, where the latter acts as the necessary mixer of even and odd-parities at the interface. Also, we see again how both spin-triplet ETO and OTE as well as the OSO amplitude only survive  non-locally, in full accordance with our results in the normal region.

Interestingly, all the amplitudes develop an oscillatory behavior, as seen in Figs.\,\ref{fig2}(b) and \ref{fig2}(d), which depends on the chemical potential $\mu_{S}$ for small SO contribution through $\bar{k}$ and on SO strength through $\mathbb{A}$.
The ESE and ETO classes also have an additional SO oscillatory dependence reflected in the terms in curly brackets stemming from normal reflections. However, for full transparent junctions ($Z=0$) only Andreev processes matter and thus the oscillations then only include short- and long-period oscillations that depend on the chemical potential ($\mu_{S}$) and SO coupling, respectively. 
The small period oscillations are given by the terms that include $\bar{k}$ which contain the chemical potential and SO coupling, for instance for OSO  the ${\rm sin}[\bar{k}(x-x')]$ term. On the other hand, the large period is given by the SO coupling itself, for instance by the ${\rm cos}[k_{SO}|x-x'|]$ term in the OSO amplitude. This multiple oscillations periods give rise to a considerable beating features for the OSO and OTE magnitudes, which therefore reveal a clear presence of SO coupling in the system. 
Moreover, in Figs.\,\ref{fig2}(b) and \ref{fig2}(d) we observe that the interface contributions exhibit an exponential  decay into the bulk of S with a decay length given by $1/\kappa^{\rm S}$. A strong SO coupling slows the decay of the pairing magnitudes.

Moreover, for the interface amplitudes, there is a competition between normal (terms with $b_{ij}$) and Andreev (terms with $a_{ij}$) processes, where the former terms also include $B(\omega)$ and the latter $B_{\pm}(\omega)$. The Andreev contributions to the even-frequency amplitudes (ESE and ETO) are proportional to $B_{+}$, while the Andreev contributions to the odd-frequency (OSO and OTE)  are proportional to $B_{-}$.
At low energies ($\omega<\Delta$), the factors $B$ and $B_{\pm}$ play an important role mainly because $B_{-}=1$ is larger than both $B_{+}$, which grows almost linearly, and $B(\omega)$, as seen in Fig.\,\ref{fig2}(c). Furthermore, for large chemical potentials, the case discussed here, the normal reflection magnitudes are very small in comparison to the Andreev magnitudes, provided there is no mismatch of chemical potentials and a fully transparent interface is guaranteed. 
The combined effect of these behaviors is that the OSO and OTE magnitudes are much larger than the ESE and ETO magnitudes, as clearly seen in Figs.\,\ref{fig2}(b) and \ref{fig2}(d). We have verified that varying the SO coupling does not alter the dominant behavior of odd-frequency pairing amplitudes.

Another interesting quantity here is the LDOS, which in the S region has contributions from both the bulk and interface as derived in Appendix \ref{NSAppSOC}. In particular, the interface term in the large chemical potential limit is given by $\rho_{\rm I}(x,\omega)=(-1/\pi){\rm Im}[\bar{\rho}_{\rm I}(x,\omega)]$ with
\begin{equation}
\label{LDOS}
\begin{split}
\bar{\rho}_{\rm I}(x,\omega)&=\frac{2\eta {\rm e}^{-2\kappa x}}{i(u^{2}-v^{2})}\bigg[\frac{u^{2}{\rm e}^{2i\bar{k} x}b_{51}}{k_{e_{1}}^{S}+k_{e_{2}}^{S}}+\frac{v^{2}{\rm e}^{-2i\bar{k} x}b_{71}}{k_{h_{1}}^{S}+k_{h_{2}}^{S}} \bigg]\\
&+\frac{4\eta uv a_{52}{\rm e}^{-2\kappa x}}{i(u^{2}-v^{2})}\,,
\end{split}
\end{equation}
where the first and second lines correspond to contributions from normal and Andreev reflections, respectively. In Fig.\,\ref{fig2}(b) and \ref{fig2}(d) the magenta curve shows the spatial dependence of the LDOS in the transparent limit $Z=0$. The major contribution to the LDOS for $Z=0$ comes from the Andreev reflection, $a_{52}$, as in this regime normal reflections are small. 
We observe that the LDOS decays exponentially in a very similar way to the pairing amplitudes but does not exhibit the oscillatory behavior due to the SO coupling. 
However, the local ESE at low energies is rather small due to the factor $B_{+}$, as also seen in Fig.\,\ref{fig2}(b,c,d), and thus does not explain the large value of LDOS. We are therefore forced to attribute non-local pairing and the finite odd-frequency magnitudes to the large values of LDOS in the S region.

We conclude this part by pointing out that SO coupling  mixes spin states giving rise to a coexistence of spin-singlet and mixed spin-triplet pairing amplitudes. Locally, only even-frequency spin-singlet (ESE) survives due to the special Andreev reflections in a Rashba SO coupled nanowire. However, large odd-frequency OSO and OTE dominate all non-local pairing correlations, including at short distances even compared to the superconducting coherence length $\xi$.
We also stress that, within our scattering approach, the Andreev reflection is the sole process that mixes spatial parities and therefore gives rise to a simultaneous coexistence of even and odd-frequency pairs which decay from the interface in the S and N regions.

\subsection{Short SNS junctions}
%%%%%%%%%%%%%%%%%%%%%%%%%%%%
%       Short SNS: FINITE SPIN-ORBIT COUPLING     %
%%%%%%%%%%%%%%%%%%%%%%%%%%%%
Next, we turn to short nanowire SNS junctions at finite SO coupling with the interface located at $x=0$, where $L_{\rm N}\rightarrow0$ the length of the N region. We allow for a finite phase difference, where the phase in the right region is fixed to $\phi$ while zero in the left region. The Green's functions from the left and right S regions are calculated in a similar way as for NS junctions and they give the same information but now with phase-dependent properties. Thus, it is enough to focus on the right region only we we obtain (for details see Appendix \ref{SNSAppSOC}):
\begin{widetext}
\begin{equation}
\label{SNSpairing}
\begin{split}
f_{0,{\rm B}}^{r,{\rm E}}(x,x',\omega)&=2B(\omega)\bigg[\frac{{\rm e}^{i\bar{k}|x-x'|}}{k_{e_{1}}^{S}+k_{e_{2}}^{S}}
+\frac{{\rm e}^{-i\bar{k}|x-x'|}}{k_{h_{1}}^{S}+k_{h_{2}}^{S}}\bigg]{\rm cos}[k_{\rm so}|x-x'|]{\rm e}^{-\kappa|x-x'|}{\rm e}^{i\phi}\,,\\
f_{0,{\rm I}}^{r,{\rm E}}(x,x',\omega)&=B(\omega)
\bigg\{
{\rm e}^{i\bar{k}(x+x')}
C_{1,xx'}^{{\rm NR},+}(\omega,\phi)
+
{\rm e}^{-i\bar{k}(x+x')}
C_{2,xx'}^{{\rm NR},+}(\omega,\phi)
\,,\\
&+{\rm cos}[\bar{k}(x-x')] \bigg[
C_{1,xx'}^{{\rm AR},+}(\omega,\phi){\rm e}^{-ik_{\rm so}|x-x'|}
+
C_{2,xx'}^{{\rm AR},+}(\omega,\phi){\rm e}^{ik_{\rm so}|x-x'|}
{\rm e}^{ik_{\rm so}|x-x'|}
\bigg]\bigg\}{\rm e}^{-\kappa(x+x')}{\rm e}^{i\phi}\,,\\
%f_{0,{\rm B}}^{r,{\rm O}}(x,x',\omega)&=0\,,\\
f_{0,{\rm I}}^{r,{\rm O}}(x,x',\omega)&=iB(\omega)\bigg[
C_{1,xx'}^{{\rm AR},-}(\omega,\phi){\rm e}^{-ik_{\rm so}|x-x'|}
+
C_{2,xx'}^{{\rm AR},-}(\omega,\phi){\rm e}^{ik_{\rm so}|x-x'|}
\bigg]{\rm sin}[\bar{k}(x-x')]{\rm e}^{-\kappa(x+x')}{\rm e}^{i\phi}\,,\\
f_{3,{\rm B}}^{r,{\rm E}}(x,x',\omega)&=-2iB(\omega)\bigg[\frac{{\rm e}^{i\bar{k}|x-x'|}}{k_{e_{1}}^{S}+k_{e_{2}}^{S}}
+\frac{{\rm e}^{-i\bar{k}|x-x'|}}{k_{h_{1}}^{S}+k_{h_{2}}^{S}}\bigg]{\rm sgn}(x-x'){\rm sin}[k_{\rm so}|x-x'|]{\rm e}^{-\kappa|x-x'|}{\rm e}^{i\phi}\,,\\
f_{3,{\rm I}}^{r,{\rm E}}(x,x',\omega)&=B(\omega)
\bigg\{
C_{1,xx'}^{{\rm NR},-}(\omega,\phi){\rm e}^{i\bar{k}(x+x')}
+
C_{2,xx'}^{{\rm NR},-}(\omega,\phi){\rm e}^{-i\bar{k}(x+x')}
\,,\\
&+{\rm cos}[\bar{k}(x-x')] \bigg[
C_{1,xx'}^{{\rm AR},+}(\omega,\phi){\rm e}^{-ik_{\rm so}|x-x'|}
-
C_{2,xx'}^{{\rm AR},+}(\omega,\phi){\rm e}^{ik_{\rm so}|x-x'|}
\bigg]\bigg\}{\rm sgn}(x-x'){\rm e}^{-\kappa(x+x')}{\rm e}^{i\phi}\,,\\
%f_{3,{\rm B}}^{r,{\rm O}}(x,x',\omega)&=0\,,\\
f_{3,{\rm I}}^{r,{\rm O}}(x,x',\omega)&=
iB(\omega)
 \bigg[
C_{1,xx'}^{{\rm AR},-}(\omega,\phi){\rm e}^{-ik_{\rm so}|x-x'|}
-
C_{2,xx'}^{{\rm AR},-}(\omega,\phi){\rm e}^{ik_{\rm so}|x-x'|}
\bigg]{\rm sin}[\bar{k}|x-x'|]{\rm e}^{-\kappa(x+x')}{\rm e}^{i\phi}
\,,\\
\end{split}
\end{equation}
\end{widetext}
where 
$C_{1,xx'}^{{\rm NR},\pm}=[b_{62}{\rm e}^{-ik_{\rm so}|x-x'|}\pm b_{51}{\rm e}^{ik_{\rm so}|x-x'|}]/(k_{e_{1}}^{S}+k_{e_{2}}^{S})$, 
$C_{2,xx'}^{{\rm NR},\pm}=[b_{71}{\rm e}^{-ik_{\rm so}|x-x'|}\pm b_{82}{\rm e}^{ik_{\rm so}|x-x'|}]/(k_{h_{1}}^{S}+k_{h_{2}}^{S})$, 
$C^{{\rm AR},\pm}_{1,xx'}=(\tilde{a}_{61}\frac{u}{v}\pm a_{61}\frac{v}{u})$, and $C^{{\rm AR},\pm}_{2,xx'}=(\tilde{a}_{52}\frac{u}{v}\pm a_{52}\frac{v}{u})$. Here the label NR (AR) in the coefficients $C$ stands for normal (Andreev) reflection with accompanied coefficients $b_{ij}$ $(a_{ij})$ and where $\tilde{a}_{ij}$ signifies the corresponding reflection for the conjugated process (see Appendix \ref{SNSAppSOC} for more details).

The pairing amplitudes in Eq.~\eqref{SNSpairing} correspond to ESE, OSO, ETO, and OTE amplitudes, respectively, where there are zero bulk (B) terms for the odd-frequency components $f_{0,3,{\rm B}}^{r,{\rm O}}(x,x',\omega)=0$. On the other hand, the interface (I) terms are present for all symmetries and involve both normal and Andreev reflections.
Normal reflections contribute to interface even-frequency amplitudes (ESE and ETO), while Andreev reflections contribute to all symmetry classes. 
Interestingly, Eqs.\,(\ref{SNSpairing}) also indicate that  the odd-frequency amplitudes (OSO and OTE) are proportional only to the Andreev coefficients, with neither bulk nor normal reflection terms, in agreement with the results for NS junctions. In addition, we observe in Eqs.\,(\ref{SNSpairing}) that local ($x=x'$) odd-frequency and spin-triplet pairings (OSO, ETO, and OTE) all vanish, also in agreement with our findings for NS junctions.
Yet another similarly to NS junctions is how the interface amplitudes exponentially decay into the bulk of the S region with a decay length $1/\kappa$ and an oscillatory behavior determined by the chemical potential (through $\bar{k}$) and SO coupling (terms in square brackets). 

\begin{figure}[!ht]
%\begin{minipage}[t]{\linewidth}
\centering
\includegraphics[width=.48\textwidth]{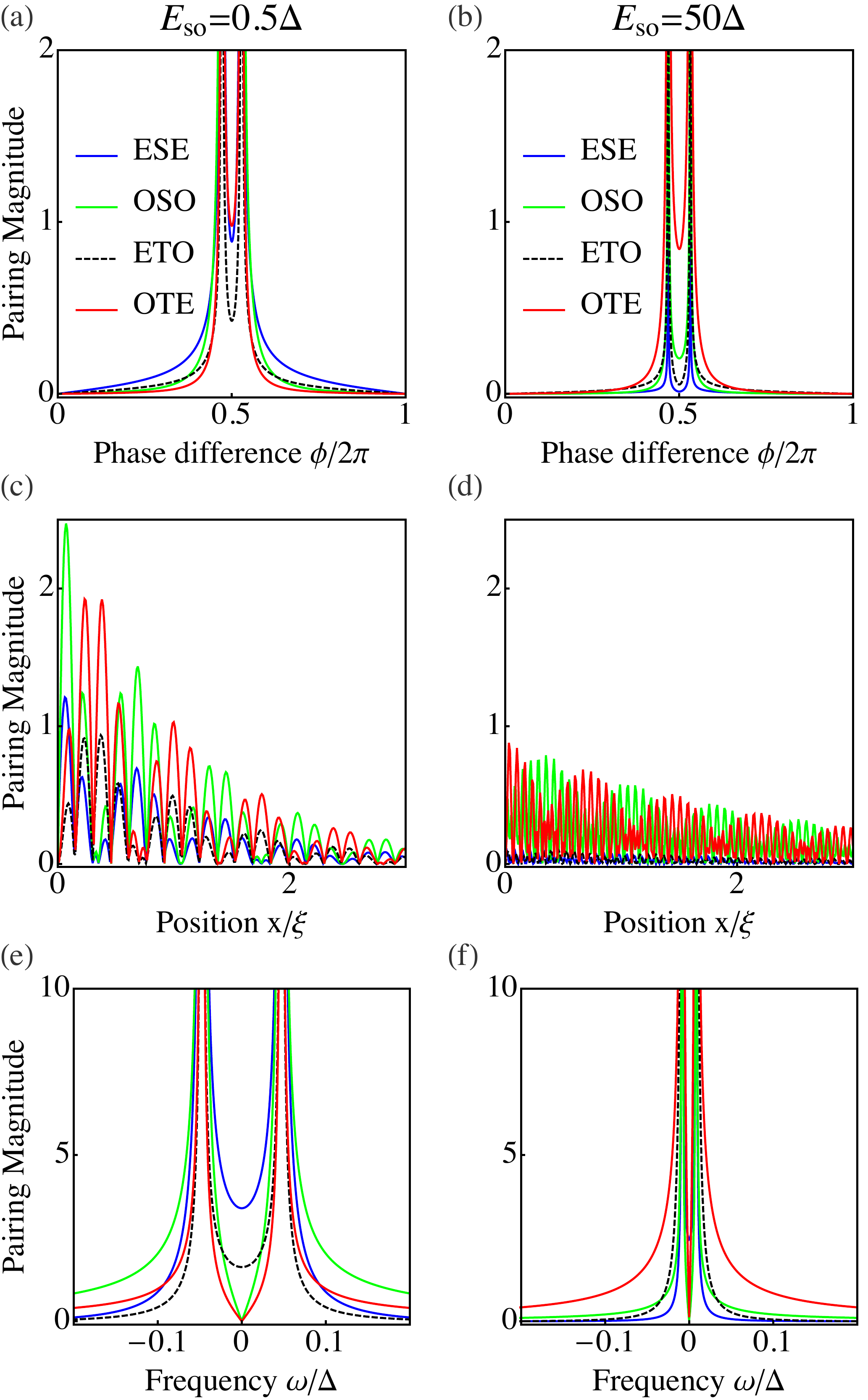} 
\caption{(Color online.) Phase dependent at $x=0.1\xi$ [(a), (b)], spatial dependent at $\phi=\pi$ [(c), (d)], and frequency-dependent at $\phi=\pi$ and at $x=0.1\xi$ [(e), (f)] pairing magnitudes in short SNS junctions for $E_{\rm SO}=0.5\Delta$ [(a), (c), (e)] and $E_{\rm SO}=50\Delta$ [(b), (d), (f)].  
Parameters: $\omega=0.1\Delta$, $x'=0$, $\mu_{\rm N, S}=10\Delta$, $Z = 0$.}
\label{fig3}
%\end{minipage}
\end{figure}

While many overall properties are similar for short SNS junctions and NS junctions, short SNS junctions acquire a unique dependence of their pairing functions on the superconducting  phase difference $\phi$. The normal and Andreev reflection coefficients $b_{ij}$ and $a_{ij}$ also acquire a phase dependence with important consequences. The most striking result of this is  that in SNS junctions a pair of Andreev bound states ($\omega_{ABS,\pm}$)  emerge within $\Delta$. 
The emergence of these states is reflected in the coefficients $a_{ij}$ and $b_{ij}$ and therefore captured in the pairing amplitudes as resonances in the phase-dependent pairing magnitudes, as shown in Figs.\,\ref{fig3}(a,b) at or close to $\phi = \pi$.

At zero phase-difference, $\phi=0$, and at large chemical potentials and fully transparent interface ($Z=0$), i.e.~the regime discussed here, the interface pairing magnitudes vanish, as observed in Fig.\,\ref{fig3}(a) and \ref{fig3}(b). This stems from vanishing normal and Andreev reflection coefficients in this regime.
We have checked that a finite value of $Z$ gives rise to finite 
normal coefficients at $\phi=0$ which induce finite even-frequency interface pairing, while the Andreev coefficients remain zero and lead to zero odd-frequency terms. 
A finite phase-difference results in finite even- and odd- frequency pairing amplitudes, which all exhibit resonances signaling the emergence of Andreev bound states [see Fig.\,\ref{fig3}(a), \ref{fig3}(b), \ref{fig3}(e), and \ref{fig3}(f)].

At $\phi=\pi$, Andreev and normal coefficients are all finite even for fully transparent junctions ($Z=0$), with their behavior strongly dependent on the strength of the SO coupling and chemical potential as any of these two parameters drives the system into the so-called Andreev approximation, where the SO energy or effective chemical potential are the largest energy scales. 
In the case of small SO coupling [see Fig.\,\ref{fig3}(c)], both normal and Andreev coefficients are large which translates to a situation where even- and odd-frequency amplitudes are all large. When studied as a function of position, the odd-frequency components are somewhat larger in magnitude at $\phi =\pi$ [see Fig.\,\ref{fig3}(c)]. The coexistence of even and odd-frequency amplitudes remains at finite transparency ($Z\neq0$), where the even-frequency can acquire larger values by increasing $Z$.

The situation is different for strong SO coupling at $\phi=\pi$, where we find that the odd-frequency completely dominates over the even-frequency amplitudes [see Figs.\,\ref{fig3}(b) and \ref{fig3}(d)]. In this case, the normal reflection coefficients are heavily reduced, inducing a suppression of even-frequency amplitudes, while the Andreev terms remain at around the same values, leading to dominating odd-frequency pairing at $\phi=\pi$. It is important to mention here that the dominant behavior we find is not restricted to low frequencies $\omega$. In fact, we find that the dominant behavior of odd-frequency remains for energies larger than the energy of the Andreev states but below the superconducting gap, namely, for all frequencies $\omega_{{\rm ABS},\pm}<\omega<\Delta$, as can be seen in Fig.\,\ref{fig3}(f). At very low frequency $\omega\approx0$ the odd-frequency terms practically vanish (as necessary for an odd function) and the even-frequency amplitudes dominate. When finite transparency is allowed ($Z\neq0$) the normal reflections become larger and induce larger even-frequency amplitudes, which can be comparable or even larger than the odd-frequency terms.
Although, Figs.\,\ref{fig3}(b), \ref{fig3}(d), and \ref{fig3}(f) show pairing magnitudes for quite large and possibly unrealistic SO coupling strengths, as they were chosen to highlight the role of SO, we have verified that similar results are obtained by choosing experimentally relevant SO coupling strength in the very large chemical potential limit. 

Furthermore, it has been demonstrated that the SO coupling in short junctions does not split the ABSs\cite{PhysRevB.77.045311,Dimitrova2006,PhysRevB.77.045311,SanJoseNJP:13,PhysRevB.96.014519,PhysRevB.96.075404,Cayao17b,cayao18a} 
but that it has a clear effect on the minigap at $\phi=\pi$,\cite{SanJoseNJP:13,PhysRevB.96.014519,cayao18a} significantly reducing its size when the SO gets stronger. This is also what we observe by comparing Fig.\,\ref{fig3}(e) and (f), where the frequency dependent pairing magnitudes capture this effect. This thus shows that the closing of this minigap by SO coupling acts as an indicator of dominant odd-frequency pairing for $\omega_{ABS,\pm}<\omega<\Delta$. This phenomenon can be seen as a crossing in the phase-dependent spectrum or  as a sawtooth profile at $\phi=\pi$ in the phase-dependent supercurrent of short SNS junctions.\cite{PhysRevB.96.014519,cayao18a}

% -------------------------------------- %
% CONCLUSIONS:
% -------------------------------------- %
\section{Concluding remarks}
\label{concl}
In this work we have studied the emergence of odd-frequency superconducting pairing in NS and SNS junctions with Rashba spin-orbit (SO) coupling. We have analytically found that, as expected, translational symmetry breaking at interfaces induces even- and odd-frequency spin-singlet components (ESE and OSO)\cite{PhysRevB.76.054522,PhysRevLett.98.037003,Eschrig2007} and, interestingly, singlet to triplet conversion due to SO coupling  induces also even- and odd-frequency mixed spin-triplet amplitudes (ETO and OTE). 
Importantly, we have demonstrated that, both at zero and finite SO coupling, Andreev reflection is solely responsible for mixing of spatial parities at interfaces and therefore acts as the generator of all odd-frequency components in both NS and SNS junctions. We have also obtained that locally, i.e.,~at $x = x'$, only the even-frequency spin singlet pairing (ESE) is finite due to the specific features of  Rashba SO coupling. However, non-local pairing correlations, including both even and odd parity, are non-zero in all symmetry classes.

In terms of NS junctions, we have shown that all pairing amplitudes coexist in the normal region and are solely proportional to the Andreev reflection, with an oscillatory behavior. 
In the superconducting region, the amplitudes also acquire contributions from the bulk and interface, the latter due to both normal and Andreev reflections. The interface terms exponential decay with both short and long-period oscillations due to the chemical potential and SO coupling, respectively. The large-period oscillations thus cause a prominent beating feature in the pairing magnitudes which is sensitive to the SO coupling. 
Interestingly, the odd-frequency terms emerge purely proportional to the Andreev processes, while even-frequency terms contain also contributions from normal reflections and the bulk.
Also, at very low frequencies ($\omega\ll\Delta$),  the odd-frequency spin-singlet and spin-triplet amplitudes (OSO and OTE) are much larger than the the even-frequency terms (ESE and ETO). This we have used to directly relate the high values of the LDOS we find at low frequencies in the superconducting region to odd-frequency pairing. 
In fact, we have found that the LDOS in the large chemical potential and high transparency limits is heavily dominated by one single Andreev process for a NS junction. The same Andreev process also determines the odd-frequency amplitudes, both OSO and OTE pairing. Thus, by measuring the LDOS and from there extract the associated Andreev coefficient, we can exactly resolve and determine all odd-frequency pairing amplitudes in NS junctions.  As a consequence, large LDOS indicates the presence of large odd-frequency pairing, a signature that could be observed experimentally.

In short SNS junctions, we have demonstrated that the pairing amplitudes become phase dependent and also capture the emergence of Andreev bound states in the junction. At zero phase, all amplitudes vanish at full transparency, while they acquire a finite value as the phase approaches $\phi=\pi$, where the odd-frequency components are also strongly dominating, especially for strong SO coupling and full transparency junctions. This behavior is preserved for frequencies larger than bound-state energies but below the superconducting gap. 
 Furthermore, we have showed that the odd- and even-frequency pairing amplitudes capture the reduction of the minigap in the low-energy spectrum at $\phi=\pi$ caused by large SO coupling, which thus serves as an indicator of odd-frequency dominant behavior for frequencies larger than bound-state energies but below the superconducting gap. The closing of the minigap, which leads to a sawtooth profile in current-phase curves, therefore corresponds to a strong experimental signature of large odd-frequency pairing. On the other hand, at very low frequencies $\omega\approx0$ and strong SO coupling,  the even-frequency amplitudes (ESE and ETO) are larger than the odd-frequency components (OSO and OTE).

To conclude,  Andreev reflection mixes spatial parities at interfaces and thus generates both even- and odd-frequency components. Adding SO coupling allows for a mixing of spin symmetries without breaking time-reversal symmetry and thus all possible symmetry classes of superconducting pairing, ESE, OSO, ETO, and OTE, generally appear in NS and SNS junctions in Rashba SO coupled nanowires. Importantly, all odd-frequency components are solely generated by Andreev reflection. This is both a significant conceptual result and can also be used to experimentally measure the odd-frequency components, as this quantity  can be obtained from LDOS or conductance measurements\cite{chang15,doi:10.1063/1.4971394,gulonder,Gazibegovic17,zhang18} which therefore allow to fully determine the odd-frequency amplitudes. 

%Thus,  both even- and odd-frequency superconducting pairing amplitudes induced at interfaces are equally important.

\section{Acknowledgements}
We thank A.~V.~Balatsky, A.~Bouhon, and C.~Reeg for interesting discussions and  C.~Triola for helpful comments on the manuscript.
This work was made possible by support from the Swedish Research Council (Vetenskapsr\aa det, 621-2014-3721), the G\"{o}ran Gustafsson Foundation, the Knut and Alice Wallenberg Foundation through the Wallenberg Academy Fellows program, and the European Research Council (ERC) under the European Union's Horizon 2020 research and innovation programme (ERC-2017-StG-757553).

% BIBLIOGRAPHY:
\bibliography{biblio}

\begin{thebibliography}{118}%
\makeatletter
\providecommand \@ifxundefined [1]{%
 \@ifx{#1\undefined}
}%
\providecommand \@ifnum [1]{%
 \ifnum #1\expandafter \@firstoftwo
 \else \expandafter \@secondoftwo
 \fi
}%
\providecommand \@ifx [1]{%
 \ifx #1\expandafter \@firstoftwo
 \else \expandafter \@secondoftwo
 \fi
}%
\providecommand \natexlab [1]{#1}%
\providecommand \enquote  [1]{``#1''}%
\providecommand \bibnamefont  [1]{#1}%
\providecommand \bibfnamefont [1]{#1}%
\providecommand \citenamefont [1]{#1}%
\providecommand \href@noop [0]{\@secondoftwo}%
\providecommand \href [0]{\begingroup \@sanitize@url \@href}%
\providecommand \@href[1]{\@@startlink{#1}\@@href}%
\providecommand \@@href[1]{\endgroup#1\@@endlink}%
\providecommand \@sanitize@url [0]{\catcode `\\12\catcode `\$12\catcode
  `\&12\catcode `\#12\catcode `\^12\catcode `\_12\catcode `\%12\relax}%
\providecommand \@@startlink[1]{}%
\providecommand \@@endlink[0]{}%
\providecommand \url  [0]{\begingroup\@sanitize@url \@url }%
\providecommand \@url [1]{\endgroup\@href {#1}{\urlprefix }}%
\providecommand \urlprefix  [0]{URL }%
\providecommand \Eprint [0]{\href }%
\providecommand \doibase [0]{http://dx.doi.org/}%
\providecommand \selectlanguage [0]{\@gobble}%
\providecommand \bibinfo  [0]{\@secondoftwo}%
\providecommand \bibfield  [0]{\@secondoftwo}%
\providecommand \translation [1]{[#1]}%
\providecommand \BibitemOpen [0]{}%
\providecommand \bibitemStop [0]{}%
\providecommand \bibitemNoStop [0]{.\EOS\space}%
\providecommand \EOS [0]{\spacefactor3000\relax}%
\providecommand \BibitemShut  [1]{\csname bibitem#1\endcsname}%
\let\auto@bib@innerbib\@empty
%</preamble>
\bibitem [{\citenamefont {Berezinskii}(1974)}]{bere74}%
  \BibitemOpen
  \bibfield  {author} {\bibinfo {author} {\bibfnamefont {V.~L.}\ \bibnamefont
  {Berezinskii}},\ }\href
  {http://www.jetpletters.ac.ru/ps/1792/article_27363.shtml} {\bibfield
  {journal} {\bibinfo  {journal} {JETP Lett.}\ }\textbf {\bibinfo {volume}
  {20}},\ \bibinfo {pages} {287} (\bibinfo {year} {1974})}\BibitemShut
  {NoStop}%
\bibitem [{\citenamefont {Balatsky}\ and\ \citenamefont
  {Abrahams}(1992)}]{PhysRevB.45.13125}%
  \BibitemOpen
  \bibfield  {author} {\bibinfo {author} {\bibfnamefont {A.}~\bibnamefont
  {Balatsky}}\ and\ \bibinfo {author} {\bibfnamefont {E.}~\bibnamefont
  {Abrahams}},\ }\href {\doibase 10.1103/PhysRevB.45.13125} {\bibfield
  {journal} {\bibinfo  {journal} {Phys. Rev. B}\ }\textbf {\bibinfo {volume}
  {45}},\ \bibinfo {pages} {13125} (\bibinfo {year} {1992})}\BibitemShut
  {NoStop}%
\bibitem [{\citenamefont {Black-Schaffer}\ and\ \citenamefont
  {Balatsky}(2013{\natexlab{a}})}]{PhysRevB.88.104514}%
  \BibitemOpen
  \bibfield  {author} {\bibinfo {author} {\bibfnamefont {A.~M.}\ \bibnamefont
  {Black-Schaffer}}\ and\ \bibinfo {author} {\bibfnamefont {A.~V.}\
  \bibnamefont {Balatsky}},\ }\href {\doibase 10.1103/PhysRevB.88.104514}
  {\bibfield  {journal} {\bibinfo  {journal} {Phys. Rev. B}\ }\textbf {\bibinfo
  {volume} {88}},\ \bibinfo {pages} {104514} (\bibinfo {year}
  {2013}{\natexlab{a}})}\BibitemShut {NoStop}%
\bibitem [{\citenamefont {Komendov\'a}\ \emph {et~al.}(2015)\citenamefont
  {Komendov\'a}, \citenamefont {Balatsky},\ and\ \citenamefont
  {Black-Schaffer}}]{PhysRevB.92.094517}%
  \BibitemOpen
  \bibfield  {author} {\bibinfo {author} {\bibfnamefont {L.}~\bibnamefont
  {Komendov\'a}}, \bibinfo {author} {\bibfnamefont {A.~V.}\ \bibnamefont
  {Balatsky}}, \ and\ \bibinfo {author} {\bibfnamefont {A.~M.}\ \bibnamefont
  {Black-Schaffer}},\ }\href {\doibase 10.1103/PhysRevB.92.094517} {\bibfield
  {journal} {\bibinfo  {journal} {Phys. Rev. B}\ }\textbf {\bibinfo {volume}
  {92}},\ \bibinfo {pages} {094517} (\bibinfo {year} {2015})}\BibitemShut
  {NoStop}%
\bibitem [{\citenamefont {Asano}\ and\ \citenamefont
  {Sasaki}(2015)}]{PhysRevB.92.224508}%
  \BibitemOpen
  \bibfield  {author} {\bibinfo {author} {\bibfnamefont {Y.}~\bibnamefont
  {Asano}}\ and\ \bibinfo {author} {\bibfnamefont {A.}~\bibnamefont {Sasaki}},\
  }\href {\doibase 10.1103/PhysRevB.92.224508} {\bibfield  {journal} {\bibinfo
  {journal} {Phys. Rev. B}\ }\textbf {\bibinfo {volume} {92}},\ \bibinfo
  {pages} {224508} (\bibinfo {year} {2015})}\BibitemShut {NoStop}%
\bibitem [{\citenamefont {Komendov\'a}\ and\ \citenamefont
  {Black-Schaffer}(2017)}]{PhysRevLett.119.087001}%
  \BibitemOpen
  \bibfield  {author} {\bibinfo {author} {\bibfnamefont {L.}~\bibnamefont
  {Komendov\'a}}\ and\ \bibinfo {author} {\bibfnamefont {A.~M.}\ \bibnamefont
  {Black-Schaffer}},\ }\href {\doibase 10.1103/PhysRevLett.119.087001}
  {\bibfield  {journal} {\bibinfo  {journal} {Phys. Rev. Lett.}\ }\textbf
  {\bibinfo {volume} {119}},\ \bibinfo {pages} {087001} (\bibinfo {year}
  {2017})}\BibitemShut {NoStop}%
\bibitem [{\citenamefont {Triola}\ and\ \citenamefont
  {Black-Schaffer}(2018)}]{PhysRevB.97.064505}%
  \BibitemOpen
  \bibfield  {author} {\bibinfo {author} {\bibfnamefont {C.}~\bibnamefont
  {Triola}}\ and\ \bibinfo {author} {\bibfnamefont {A.~M.}\ \bibnamefont
  {Black-Schaffer}},\ }\href {\doibase 10.1103/PhysRevB.97.064505} {\bibfield
  {journal} {\bibinfo  {journal} {Phys. Rev. B}\ }\textbf {\bibinfo {volume}
  {97}},\ \bibinfo {pages} {064505} (\bibinfo {year} {2018})}\BibitemShut
  {NoStop}%
\bibitem [{\citenamefont {Kirkpatrick}\ and\ \citenamefont
  {Belitz}(1991)}]{PhysRevLett.66.1533}%
  \BibitemOpen
  \bibfield  {author} {\bibinfo {author} {\bibfnamefont {T.~R.}\ \bibnamefont
  {Kirkpatrick}}\ and\ \bibinfo {author} {\bibfnamefont {D.}~\bibnamefont
  {Belitz}},\ }\href {\doibase 10.1103/PhysRevLett.66.1533} {\bibfield
  {journal} {\bibinfo  {journal} {Phys. Rev. Lett.}\ }\textbf {\bibinfo
  {volume} {66}},\ \bibinfo {pages} {1533} (\bibinfo {year}
  {1991})}\BibitemShut {NoStop}%
\bibitem [{\citenamefont {Belitz}\ and\ \citenamefont
  {Kirkpatrick}(1992)}]{PhysRevB.46.8393}%
  \BibitemOpen
  \bibfield  {author} {\bibinfo {author} {\bibfnamefont {D.}~\bibnamefont
  {Belitz}}\ and\ \bibinfo {author} {\bibfnamefont {T.~R.}\ \bibnamefont
  {Kirkpatrick}},\ }\href {\doibase 10.1103/PhysRevB.46.8393} {\bibfield
  {journal} {\bibinfo  {journal} {Phys. Rev. B}\ }\textbf {\bibinfo {volume}
  {46}},\ \bibinfo {pages} {8393} (\bibinfo {year} {1992})}\BibitemShut
  {NoStop}%
\bibitem [{\citenamefont {Belitz}\ and\ \citenamefont
  {Kirkpatrick}(1999)}]{PhysRevB.60.3485}%
  \BibitemOpen
  \bibfield  {author} {\bibinfo {author} {\bibfnamefont {D.}~\bibnamefont
  {Belitz}}\ and\ \bibinfo {author} {\bibfnamefont {T.~R.}\ \bibnamefont
  {Kirkpatrick}},\ }\href {\doibase 10.1103/PhysRevB.60.3485} {\bibfield
  {journal} {\bibinfo  {journal} {Phys. Rev. B}\ }\textbf {\bibinfo {volume}
  {60}},\ \bibinfo {pages} {3485} (\bibinfo {year} {1999})}\BibitemShut
  {NoStop}%
\bibitem [{\citenamefont {Abrahams}\ \emph {et~al.}(1995)\citenamefont
  {Abrahams}, \citenamefont {Balatsky}, \citenamefont {Scalapino},\ and\
  \citenamefont {Schrieffer}}]{PhysRevB.52.1271}%
  \BibitemOpen
  \bibfield  {author} {\bibinfo {author} {\bibfnamefont {E.}~\bibnamefont
  {Abrahams}}, \bibinfo {author} {\bibfnamefont {A.}~\bibnamefont {Balatsky}},
  \bibinfo {author} {\bibfnamefont {D.~J.}\ \bibnamefont {Scalapino}}, \ and\
  \bibinfo {author} {\bibfnamefont {J.~R.}\ \bibnamefont {Schrieffer}},\ }\href
  {\doibase 10.1103/PhysRevB.52.1271} {\bibfield  {journal} {\bibinfo
  {journal} {Phys. Rev. B}\ }\textbf {\bibinfo {volume} {52}},\ \bibinfo
  {pages} {1271} (\bibinfo {year} {1995})}\BibitemShut {NoStop}%
\bibitem [{\citenamefont {Coleman}\ \emph {et~al.}(1997)\citenamefont
  {Coleman}, \citenamefont {Georges},\ and\ \citenamefont
  {Tsvelik}}]{0953-8984-9-2-002}%
  \BibitemOpen
  \bibfield  {author} {\bibinfo {author} {\bibfnamefont {P.}~\bibnamefont
  {Coleman}}, \bibinfo {author} {\bibfnamefont {A.}~\bibnamefont {Georges}}, \
  and\ \bibinfo {author} {\bibfnamefont {A.~M.}\ \bibnamefont {Tsvelik}},\
  }\href {http://stacks.iop.org/0953-8984/9/i=2/a=002} {\bibfield  {journal}
  {\bibinfo  {journal} {J. Phys.: Condens. Matter}\ }\textbf {\bibinfo {volume}
  {9}},\ \bibinfo {pages} {345} (\bibinfo {year} {1997})}\BibitemShut {NoStop}%
\bibitem [{\citenamefont {Petrashov}\ \emph {et~al.}(1994)\citenamefont
  {Petrashov}, \citenamefont {Antonov}, \citenamefont {Maksimov},\ and\
  \citenamefont {Sha\v{l}kha\v{l}darov}}]{longrangeExp}%
  \BibitemOpen
  \bibfield  {author} {\bibinfo {author} {\bibfnamefont {V.~T.}\ \bibnamefont
  {Petrashov}}, \bibinfo {author} {\bibfnamefont {V.~N.}\ \bibnamefont
  {Antonov}}, \bibinfo {author} {\bibfnamefont {S.~V.}\ \bibnamefont
  {Maksimov}}, \ and\ \bibinfo {author} {\bibfnamefont {R.~S.}\ \bibnamefont
  {Sha\v{l}kha\v{l}darov}},\ }\href
  {http://www.jetpletters.ac.ru/ps/1309/article_19787.shtml} {\bibfield
  {journal} {\bibinfo  {journal} {Pis'ma Zh. Eksp. Teor. Fiz.}\ }\textbf
  {\bibinfo {volume} {59}},\ \bibinfo {pages} {523} (\bibinfo {year}
  {1994})}\BibitemShut {NoStop}%
\bibitem [{\citenamefont {Bergeret}\ \emph {et~al.}(2001)\citenamefont
  {Bergeret}, \citenamefont {Volkov},\ and\ \citenamefont
  {Efetov}}]{PhysRevLett.86.4096}%
  \BibitemOpen
  \bibfield  {author} {\bibinfo {author} {\bibfnamefont {F.~S.}\ \bibnamefont
  {Bergeret}}, \bibinfo {author} {\bibfnamefont {A.~F.}\ \bibnamefont
  {Volkov}}, \ and\ \bibinfo {author} {\bibfnamefont {K.~B.}\ \bibnamefont
  {Efetov}},\ }\href {\doibase 10.1103/PhysRevLett.86.4096} {\bibfield
  {journal} {\bibinfo  {journal} {Phys. Rev. Lett.}\ }\textbf {\bibinfo
  {volume} {86}},\ \bibinfo {pages} {4096} (\bibinfo {year}
  {2001})}\BibitemShut {NoStop}%
\bibitem [{\citenamefont {Volkov}\ \emph {et~al.}(2003)\citenamefont {Volkov},
  \citenamefont {Bergeret},\ and\ \citenamefont
  {Efetov}}]{PhysRevLett.90.117006}%
  \BibitemOpen
  \bibfield  {author} {\bibinfo {author} {\bibfnamefont {A.~F.}\ \bibnamefont
  {Volkov}}, \bibinfo {author} {\bibfnamefont {F.~S.}\ \bibnamefont
  {Bergeret}}, \ and\ \bibinfo {author} {\bibfnamefont {K.~B.}\ \bibnamefont
  {Efetov}},\ }\href {\doibase 10.1103/PhysRevLett.90.117006} {\bibfield
  {journal} {\bibinfo  {journal} {Phys. Rev. Lett.}\ }\textbf {\bibinfo
  {volume} {90}},\ \bibinfo {pages} {117006} (\bibinfo {year}
  {2003})}\BibitemShut {NoStop}%
\bibitem [{\citenamefont {Bergeret}\ \emph {et~al.}(2005)\citenamefont
  {Bergeret}, \citenamefont {Volkov},\ and\ \citenamefont
  {Efetov}}]{RevModPhys.77.1321}%
  \BibitemOpen
  \bibfield  {author} {\bibinfo {author} {\bibfnamefont {F.~S.}\ \bibnamefont
  {Bergeret}}, \bibinfo {author} {\bibfnamefont {A.~F.}\ \bibnamefont
  {Volkov}}, \ and\ \bibinfo {author} {\bibfnamefont {K.~B.}\ \bibnamefont
  {Efetov}},\ }\href {\doibase 10.1103/RevModPhys.77.1321} {\bibfield
  {journal} {\bibinfo  {journal} {Rev. Mod. Phys.}\ }\textbf {\bibinfo {volume}
  {77}},\ \bibinfo {pages} {1321} (\bibinfo {year} {2005})}\BibitemShut
  {NoStop}%
\bibitem [{\citenamefont {Buzdin}(2005)}]{RevModPhys.77.935}%
  \BibitemOpen
  \bibfield  {author} {\bibinfo {author} {\bibfnamefont {A.~I.}\ \bibnamefont
  {Buzdin}},\ }\href {\doibase 10.1103/RevModPhys.77.935} {\bibfield  {journal}
  {\bibinfo  {journal} {Rev. Mod. Phys.}\ }\textbf {\bibinfo {volume} {77}},\
  \bibinfo {pages} {935} (\bibinfo {year} {2005})}\BibitemShut {NoStop}%
\bibitem [{\citenamefont {Volkov}\ \emph {et~al.}(2006)\citenamefont {Volkov},
  \citenamefont {Anishchanka},\ and\ \citenamefont
  {Efetov}}]{PhysRevB.73.104412}%
  \BibitemOpen
  \bibfield  {author} {\bibinfo {author} {\bibfnamefont {A.~F.}\ \bibnamefont
  {Volkov}}, \bibinfo {author} {\bibfnamefont {A.}~\bibnamefont {Anishchanka}},
  \ and\ \bibinfo {author} {\bibfnamefont {K.~B.}\ \bibnamefont {Efetov}},\
  }\href {\doibase 10.1103/PhysRevB.73.104412} {\bibfield  {journal} {\bibinfo
  {journal} {Phys. Rev. B}\ }\textbf {\bibinfo {volume} {73}},\ \bibinfo
  {pages} {104412} (\bibinfo {year} {2006})}\BibitemShut {NoStop}%
\bibitem [{\citenamefont {Fominov}\ \emph {et~al.}(2007)\citenamefont
  {Fominov}, \citenamefont {Volkov},\ and\ \citenamefont
  {Efetov}}]{PhysRevB.75.104509}%
  \BibitemOpen
  \bibfield  {author} {\bibinfo {author} {\bibfnamefont {Y.~V.}\ \bibnamefont
  {Fominov}}, \bibinfo {author} {\bibfnamefont {A.~F.}\ \bibnamefont {Volkov}},
  \ and\ \bibinfo {author} {\bibfnamefont {K.~B.}\ \bibnamefont {Efetov}},\
  }\href {\doibase 10.1103/PhysRevB.75.104509} {\bibfield  {journal} {\bibinfo
  {journal} {Phys. Rev. B}\ }\textbf {\bibinfo {volume} {75}},\ \bibinfo
  {pages} {104509} (\bibinfo {year} {2007})}\BibitemShut {NoStop}%
\bibitem [{\citenamefont {Yokoyama}\ \emph {et~al.}(2007)\citenamefont
  {Yokoyama}, \citenamefont {Tanaka},\ and\ \citenamefont
  {Golubov}}]{PhysRevB.75.134510}%
  \BibitemOpen
  \bibfield  {author} {\bibinfo {author} {\bibfnamefont {T.}~\bibnamefont
  {Yokoyama}}, \bibinfo {author} {\bibfnamefont {Y.}~\bibnamefont {Tanaka}}, \
  and\ \bibinfo {author} {\bibfnamefont {A.~A.}\ \bibnamefont {Golubov}},\
  }\href {\doibase 10.1103/PhysRevB.75.134510} {\bibfield  {journal} {\bibinfo
  {journal} {Phys. Rev. B}\ }\textbf {\bibinfo {volume} {75}},\ \bibinfo
  {pages} {134510} (\bibinfo {year} {2007})}\BibitemShut {NoStop}%
\bibitem [{\citenamefont {Eschrig}\ \emph {et~al.}(2007)\citenamefont
  {Eschrig}, \citenamefont {L{\"o}fwander}, \citenamefont {Champel},
  \citenamefont {Cuevas}, \citenamefont {Kopu},\ and\ \citenamefont
  {Sch{\"o}n}}]{Eschrig2007}%
  \BibitemOpen
  \bibfield  {author} {\bibinfo {author} {\bibfnamefont {M.}~\bibnamefont
  {Eschrig}}, \bibinfo {author} {\bibfnamefont {T.}~\bibnamefont
  {L{\"o}fwander}}, \bibinfo {author} {\bibfnamefont {T.}~\bibnamefont
  {Champel}}, \bibinfo {author} {\bibfnamefont {J.~C.}\ \bibnamefont {Cuevas}},
  \bibinfo {author} {\bibfnamefont {J.}~\bibnamefont {Kopu}}, \ and\ \bibinfo
  {author} {\bibfnamefont {G.}~\bibnamefont {Sch{\"o}n}},\ }\href {\doibase
  10.1007/s10909-007-9329-6} {\bibfield  {journal} {\bibinfo  {journal} {J. Low
  Temp. Phys.}\ }\textbf {\bibinfo {volume} {147}},\ \bibinfo {pages} {457}
  (\bibinfo {year} {2007})}\BibitemShut {NoStop}%
\bibitem [{\citenamefont {Eschrig}\ and\ \citenamefont
  {L\"{o}fwander}(2008)}]{EschrigNat15}%
  \BibitemOpen
  \bibfield  {author} {\bibinfo {author} {\bibfnamefont {M.}~\bibnamefont
  {Eschrig}}\ and\ \bibinfo {author} {\bibfnamefont {T.}~\bibnamefont
  {L\"{o}fwander}},\ }\href {https://www.nature.com/articles/nphys831}
  {\bibfield  {journal} {\bibinfo  {journal} {Nat. Phys.}\ }\textbf {\bibinfo
  {volume} {4}},\ \bibinfo {pages} {138} (\bibinfo {year} {2008})}\BibitemShut
  {NoStop}%
\bibitem [{\citenamefont {Khaire}\ \emph {et~al.}(2010)\citenamefont {Khaire},
  \citenamefont {Khasawneh}, \citenamefont {Pratt},\ and\ \citenamefont
  {Birge}}]{PhysRevLett.104.137002}%
  \BibitemOpen
  \bibfield  {author} {\bibinfo {author} {\bibfnamefont {T.~S.}\ \bibnamefont
  {Khaire}}, \bibinfo {author} {\bibfnamefont {M.~A.}\ \bibnamefont
  {Khasawneh}}, \bibinfo {author} {\bibfnamefont {W.~P.}\ \bibnamefont
  {Pratt}}, \ and\ \bibinfo {author} {\bibfnamefont {N.~O.}\ \bibnamefont
  {Birge}},\ }\href {\doibase 10.1103/PhysRevLett.104.137002} {\bibfield
  {journal} {\bibinfo  {journal} {Phys. Rev. Lett.}\ }\textbf {\bibinfo
  {volume} {104}},\ \bibinfo {pages} {137002} (\bibinfo {year}
  {2010})}\BibitemShut {NoStop}%
\bibitem [{\citenamefont {Anwar}\ \emph {et~al.}(2010)\citenamefont {Anwar},
  \citenamefont {Czeschka}, \citenamefont {Hesselberth}, \citenamefont
  {Porcu},\ and\ \citenamefont {Aarts}}]{PhysRevB.82.100501}%
  \BibitemOpen
  \bibfield  {author} {\bibinfo {author} {\bibfnamefont {M.~S.}\ \bibnamefont
  {Anwar}}, \bibinfo {author} {\bibfnamefont {F.}~\bibnamefont {Czeschka}},
  \bibinfo {author} {\bibfnamefont {M.}~\bibnamefont {Hesselberth}}, \bibinfo
  {author} {\bibfnamefont {M.}~\bibnamefont {Porcu}}, \ and\ \bibinfo {author}
  {\bibfnamefont {J.}~\bibnamefont {Aarts}},\ }\href {\doibase
  10.1103/PhysRevB.82.100501} {\bibfield  {journal} {\bibinfo  {journal} {Phys.
  Rev. B}\ }\textbf {\bibinfo {volume} {82}},\ \bibinfo {pages} {100501}
  (\bibinfo {year} {2010})}\BibitemShut {NoStop}%
\bibitem [{\citenamefont {Usman}\ \emph {et~al.}(2011)\citenamefont {Usman},
  \citenamefont {Yates}, \citenamefont {Moore}, \citenamefont {Morrison},
  \citenamefont {Pecharsky}, \citenamefont {Gschneidner}, \citenamefont
  {Verhagen}, \citenamefont {Aarts}, \citenamefont {Zverev}, \citenamefont
  {Robinson}, \citenamefont {Witt}, \citenamefont {Blamire},\ and\
  \citenamefont {Cohen}}]{PhysRevB.83.144518}%
  \BibitemOpen
  \bibfield  {author} {\bibinfo {author} {\bibfnamefont {I.~T.~M.}\
  \bibnamefont {Usman}}, \bibinfo {author} {\bibfnamefont {K.~A.}\ \bibnamefont
  {Yates}}, \bibinfo {author} {\bibfnamefont {J.~D.}\ \bibnamefont {Moore}},
  \bibinfo {author} {\bibfnamefont {K.}~\bibnamefont {Morrison}}, \bibinfo
  {author} {\bibfnamefont {V.~K.}\ \bibnamefont {Pecharsky}}, \bibinfo {author}
  {\bibfnamefont {K.~A.}\ \bibnamefont {Gschneidner}}, \bibinfo {author}
  {\bibfnamefont {T.}~\bibnamefont {Verhagen}}, \bibinfo {author}
  {\bibfnamefont {J.}~\bibnamefont {Aarts}}, \bibinfo {author} {\bibfnamefont
  {V.~I.}\ \bibnamefont {Zverev}}, \bibinfo {author} {\bibfnamefont {J.~W.~A.}\
  \bibnamefont {Robinson}}, \bibinfo {author} {\bibfnamefont {J.~D.~S.}\
  \bibnamefont {Witt}}, \bibinfo {author} {\bibfnamefont {M.~G.}\ \bibnamefont
  {Blamire}}, \ and\ \bibinfo {author} {\bibfnamefont {L.~F.}\ \bibnamefont
  {Cohen}},\ }\href {\doibase 10.1103/PhysRevB.83.144518} {\bibfield  {journal}
  {\bibinfo  {journal} {Phys. Rev. B}\ }\textbf {\bibinfo {volume} {83}},\
  \bibinfo {pages} {144518} (\bibinfo {year} {2011})}\BibitemShut {NoStop}%
\bibitem [{\citenamefont {Eschrig}(2011)}]{7870d3ff91ed485fa3e55e901ff81c80}%
  \BibitemOpen
  \bibfield  {author} {\bibinfo {author} {\bibfnamefont {M.}~\bibnamefont
  {Eschrig}},\ }\href {\doibase 10.1063/1.3541944} {\bibfield  {journal}
  {\bibinfo  {journal} {Phys. Today}\ }\textbf {\bibinfo {volume} {64}},\
  \bibinfo {pages} {43} (\bibinfo {year} {2011})}\BibitemShut {NoStop}%
\bibitem [{\citenamefont {Visani}\ \emph {et~al.}(2012)\citenamefont {Visani},
  \citenamefont {Sefrioui}, \citenamefont {Tornos}, \citenamefont {Leon},
  \citenamefont {J.~Briatico}, \citenamefont {Barth\'{e}l\'{e}my},
  \citenamefont {Santamar\'{i}a},\ and\ \citenamefont {Villegas}}]{visani12}%
  \BibitemOpen
  \bibfield  {author} {\bibinfo {author} {\bibfnamefont {C.}~\bibnamefont
  {Visani}}, \bibinfo {author} {\bibfnamefont {Z.}~\bibnamefont {Sefrioui}},
  \bibinfo {author} {\bibfnamefont {J.}~\bibnamefont {Tornos}}, \bibinfo
  {author} {\bibfnamefont {C.}~\bibnamefont {Leon}}, \bibinfo {author}
  {\bibfnamefont {M.~B.}\ \bibnamefont {J.~Briatico}}, \bibinfo {author}
  {\bibfnamefont {A.}~\bibnamefont {Barth\'{e}l\'{e}my}}, \bibinfo {author}
  {\bibfnamefont {J.}~\bibnamefont {Santamar\'{i}a}}, \ and\ \bibinfo {author}
  {\bibfnamefont {J.~E.}\ \bibnamefont {Villegas}},\ }\href
  {https://www.nature.com/articles/nphys2318} {\bibfield  {journal} {\bibinfo
  {journal} {Nat. Phys.}\ }\textbf {\bibinfo {volume} {8}},\ \bibinfo {pages}
  {539} (\bibinfo {year} {2012})}\BibitemShut {NoStop}%
\bibitem [{\citenamefont {Witt}\ \emph {et~al.}(2012)\citenamefont {Witt},
  \citenamefont {Robinson},\ and\ \citenamefont
  {Blamire}}]{PhysRevB.85.184526}%
  \BibitemOpen
  \bibfield  {author} {\bibinfo {author} {\bibfnamefont {J.~D.~S.}\
  \bibnamefont {Witt}}, \bibinfo {author} {\bibfnamefont {J.~W.~A.}\
  \bibnamefont {Robinson}}, \ and\ \bibinfo {author} {\bibfnamefont {M.~G.}\
  \bibnamefont {Blamire}},\ }\href {\doibase 10.1103/PhysRevB.85.184526}
  {\bibfield  {journal} {\bibinfo  {journal} {Phys. Rev. B}\ }\textbf {\bibinfo
  {volume} {85}},\ \bibinfo {pages} {184526} (\bibinfo {year}
  {2012})}\BibitemShut {NoStop}%
\bibitem [{\citenamefont {Asano}\ and\ \citenamefont
  {Tanaka}(2013)}]{PhysRevB.87.104513}%
  \BibitemOpen
  \bibfield  {author} {\bibinfo {author} {\bibfnamefont {Y.}~\bibnamefont
  {Asano}}\ and\ \bibinfo {author} {\bibfnamefont {Y.}~\bibnamefont {Tanaka}},\
  }\href {\doibase 10.1103/PhysRevB.87.104513} {\bibfield  {journal} {\bibinfo
  {journal} {Phys. Rev. B}\ }\textbf {\bibinfo {volume} {87}},\ \bibinfo
  {pages} {104513} (\bibinfo {year} {2013})}\BibitemShut {NoStop}%
\bibitem [{\citenamefont {Kalcheim}\ \emph {et~al.}(2014)\citenamefont
  {Kalcheim}, \citenamefont {Felner}, \citenamefont {Millo}, \citenamefont
  {Kirzhner}, \citenamefont {Koren}, \citenamefont {Di~Bernardo}, \citenamefont
  {Egilmez}, \citenamefont {Blamire},\ and\ \citenamefont
  {Robinson}}]{PhysRevB.89.180506}%
  \BibitemOpen
  \bibfield  {author} {\bibinfo {author} {\bibfnamefont {Y.}~\bibnamefont
  {Kalcheim}}, \bibinfo {author} {\bibfnamefont {I.}~\bibnamefont {Felner}},
  \bibinfo {author} {\bibfnamefont {O.}~\bibnamefont {Millo}}, \bibinfo
  {author} {\bibfnamefont {T.}~\bibnamefont {Kirzhner}}, \bibinfo {author}
  {\bibfnamefont {G.}~\bibnamefont {Koren}}, \bibinfo {author} {\bibfnamefont
  {A.}~\bibnamefont {Di~Bernardo}}, \bibinfo {author} {\bibfnamefont
  {M.}~\bibnamefont {Egilmez}}, \bibinfo {author} {\bibfnamefont {M.~G.}\
  \bibnamefont {Blamire}}, \ and\ \bibinfo {author} {\bibfnamefont {J.~W.~A.}\
  \bibnamefont {Robinson}},\ }\href {\doibase 10.1103/PhysRevB.89.180506}
  {\bibfield  {journal} {\bibinfo  {journal} {Phys. Rev. B}\ }\textbf {\bibinfo
  {volume} {89}},\ \bibinfo {pages} {180506} (\bibinfo {year}
  {2014})}\BibitemShut {NoStop}%
\bibitem [{\citenamefont {Blamire}\ and\ \citenamefont
  {Robinson}(2014)}]{0953-8984-26-45-453201}%
  \BibitemOpen
  \bibfield  {author} {\bibinfo {author} {\bibfnamefont {M.~G.}\ \bibnamefont
  {Blamire}}\ and\ \bibinfo {author} {\bibfnamefont {J.~W.~A.}\ \bibnamefont
  {Robinson}},\ }\href {http://stacks.iop.org/0953-8984/26/i=45/a=453201}
  {\bibfield  {journal} {\bibinfo  {journal} {J. Phys.: Condens. Matter}\
  }\textbf {\bibinfo {volume} {26}},\ \bibinfo {pages} {453201} (\bibinfo
  {year} {2014})}\BibitemShut {NoStop}%
\bibitem [{\citenamefont {Kalcheim}\ \emph {et~al.}(2015)\citenamefont
  {Kalcheim}, \citenamefont {Millo}, \citenamefont {Di~Bernardo}, \citenamefont
  {Pal},\ and\ \citenamefont {Robinson}}]{PhysRevB.92.060501}%
  \BibitemOpen
  \bibfield  {author} {\bibinfo {author} {\bibfnamefont {Y.}~\bibnamefont
  {Kalcheim}}, \bibinfo {author} {\bibfnamefont {O.}~\bibnamefont {Millo}},
  \bibinfo {author} {\bibfnamefont {A.}~\bibnamefont {Di~Bernardo}}, \bibinfo
  {author} {\bibfnamefont {A.}~\bibnamefont {Pal}}, \ and\ \bibinfo {author}
  {\bibfnamefont {J.~W.~A.}\ \bibnamefont {Robinson}},\ }\href {\doibase
  10.1103/PhysRevB.92.060501} {\bibfield  {journal} {\bibinfo  {journal} {Phys.
  Rev. B}\ }\textbf {\bibinfo {volume} {92}},\ \bibinfo {pages} {060501}
  (\bibinfo {year} {2015})}\BibitemShut {NoStop}%
\bibitem [{\citenamefont {Alidoust}\ \emph {et~al.}(2015)\citenamefont
  {Alidoust}, \citenamefont {Halterman},\ and\ \citenamefont
  {Valls}}]{PhysRevB.92.014508}%
  \BibitemOpen
  \bibfield  {author} {\bibinfo {author} {\bibfnamefont {M.}~\bibnamefont
  {Alidoust}}, \bibinfo {author} {\bibfnamefont {K.}~\bibnamefont {Halterman}},
  \ and\ \bibinfo {author} {\bibfnamefont {O.~T.}\ \bibnamefont {Valls}},\
  }\href {\doibase 10.1103/PhysRevB.92.014508} {\bibfield  {journal} {\bibinfo
  {journal} {Phys. Rev. B}\ }\textbf {\bibinfo {volume} {92}},\ \bibinfo
  {pages} {014508} (\bibinfo {year} {2015})}\BibitemShut {NoStop}%
\bibitem [{\citenamefont {Bernardo}\ \emph {et~al.}(2015)\citenamefont
  {Bernardo}, \citenamefont {Diesch}, \citenamefont {Gu}, \citenamefont
  {Linder}, \citenamefont {Divitini}, \citenamefont {Ducati}, \citenamefont
  {Scheer}, \citenamefont {Blamire},\ and\ \citenamefont
  {Robinson}}]{bernardo15}%
  \BibitemOpen
  \bibfield  {author} {\bibinfo {author} {\bibfnamefont {A.~D.}\ \bibnamefont
  {Bernardo}}, \bibinfo {author} {\bibfnamefont {S.}~\bibnamefont {Diesch}},
  \bibinfo {author} {\bibfnamefont {Y.}~\bibnamefont {Gu}}, \bibinfo {author}
  {\bibfnamefont {J.}~\bibnamefont {Linder}}, \bibinfo {author} {\bibfnamefont
  {G.}~\bibnamefont {Divitini}}, \bibinfo {author} {\bibfnamefont
  {C.}~\bibnamefont {Ducati}}, \bibinfo {author} {\bibfnamefont
  {E.}~\bibnamefont {Scheer}}, \bibinfo {author} {\bibfnamefont
  {M.}~\bibnamefont {Blamire}}, \ and\ \bibinfo {author} {\bibfnamefont
  {J.}~\bibnamefont {Robinson}},\ }\href
  {https://www.nature.com/articles/ncomms9053} {\bibfield  {journal} {\bibinfo
  {journal} {Nat. Commun.}\ }\textbf {\bibinfo {volume} {6}},\ \bibinfo {pages}
  {8053} (\bibinfo {year} {2015})}\BibitemShut {NoStop}%
\bibitem [{\citenamefont {Linder}\ and\ \citenamefont
  {Robinson}(2015{\natexlab{a}})}]{linder15A}%
  \BibitemOpen
  \bibfield  {author} {\bibinfo {author} {\bibfnamefont {J.}~\bibnamefont
  {Linder}}\ and\ \bibinfo {author} {\bibfnamefont {J.~W.~A.}\ \bibnamefont
  {Robinson}},\ }\href {https://www.nature.com/articles/srep15483} {\bibfield
  {journal} {\bibinfo  {journal} {Sci. Rep.}\ }\textbf {\bibinfo {volume}
  {5}},\ \bibinfo {pages} {15483} (\bibinfo {year}
  {2015}{\natexlab{a}})}\BibitemShut {NoStop}%
\bibitem [{\citenamefont {Eschrig}(2015)}]{0034-4885-78-10-104501}%
  \BibitemOpen
  \bibfield  {author} {\bibinfo {author} {\bibfnamefont {M.}~\bibnamefont
  {Eschrig}},\ }\href {http://stacks.iop.org/0034-4885/78/i=10/a=104501}
  {\bibfield  {journal} {\bibinfo  {journal} {Rep. Prog. Phys.}\ }\textbf
  {\bibinfo {volume} {78}},\ \bibinfo {pages} {104501} (\bibinfo {year}
  {2015})}\BibitemShut {NoStop}%
\bibitem [{\citenamefont {Linder}\ and\ \citenamefont
  {Robinson}(2015{\natexlab{b}})}]{LinderNat15}%
  \BibitemOpen
  \bibfield  {author} {\bibinfo {author} {\bibfnamefont {J.}~\bibnamefont
  {Linder}}\ and\ \bibinfo {author} {\bibfnamefont {J.~W.~A.}\ \bibnamefont
  {Robinson}},\ }\href {http://dx.doi.org/10.1038/nphys3242} {\bibfield
  {journal} {\bibinfo  {journal} {Nat. Phys.}\ }\textbf {\bibinfo {volume}
  {11}},\ \bibinfo {pages} {307} (\bibinfo {year}
  {2015}{\natexlab{b}})}\BibitemShut {NoStop}%
\bibitem [{\citenamefont {Hwang}\ \emph {et~al.}(2017)\citenamefont {Hwang},
  \citenamefont {Burset},\ and\ \citenamefont {Sothmann}}]{hwang17}%
  \BibitemOpen
  \bibfield  {author} {\bibinfo {author} {\bibfnamefont {S.-Y.}\ \bibnamefont
  {Hwang}}, \bibinfo {author} {\bibfnamefont {P.}~\bibnamefont {Burset}}, \
  and\ \bibinfo {author} {\bibfnamefont {B.}~\bibnamefont {Sothmann}},\ }\href
  {https://arxiv.org/abs/1712.03067} {\bibfield  {journal} {\bibinfo  {journal}
  {arXiv:1712.03067}\ } (\bibinfo {year} {2017})}\BibitemShut {NoStop}%
\bibitem [{\citenamefont {Jeon}\ \emph {et~al.}(2018)\citenamefont {Jeon},
  \citenamefont {Ciccarelli}, \citenamefont {Ferguson}, \citenamefont
  {Kurebayashi}, \citenamefont {Cohen}, \citenamefont {Montiel}, \citenamefont
  {Eschrig}, \citenamefont {Robinson},\ and\ \citenamefont {Blamire}}]{jeon18}%
  \BibitemOpen
  \bibfield  {author} {\bibinfo {author} {\bibfnamefont {K.-R.}\ \bibnamefont
  {Jeon}}, \bibinfo {author} {\bibfnamefont {C.}~\bibnamefont {Ciccarelli}},
  \bibinfo {author} {\bibfnamefont {A.~J.}\ \bibnamefont {Ferguson}}, \bibinfo
  {author} {\bibfnamefont {H.}~\bibnamefont {Kurebayashi}}, \bibinfo {author}
  {\bibfnamefont {L.~F.}\ \bibnamefont {Cohen}}, \bibinfo {author}
  {\bibfnamefont {X.}~\bibnamefont {Montiel}}, \bibinfo {author} {\bibfnamefont
  {M.}~\bibnamefont {Eschrig}}, \bibinfo {author} {\bibfnamefont {J.~W.~A.}\
  \bibnamefont {Robinson}}, \ and\ \bibinfo {author} {\bibfnamefont {M.~G.}\
  \bibnamefont {Blamire}},\ }\href
  {https://www.nature.com/articles/s41563-018-0058-9#article-info} {\bibfield
  {journal} {\bibinfo  {journal} {Nat. Mat.}\ }\textbf {\bibinfo {volume}
  {17}},\ \bibinfo {pages} {499} (\bibinfo {year} {2018})}\BibitemShut
  {NoStop}%
\bibitem [{\citenamefont {Tanaka}\ and\ \citenamefont
  {Golubov}(2007)}]{PhysRevLett.98.037003}%
  \BibitemOpen
  \bibfield  {author} {\bibinfo {author} {\bibfnamefont {Y.}~\bibnamefont
  {Tanaka}}\ and\ \bibinfo {author} {\bibfnamefont {A.~A.}\ \bibnamefont
  {Golubov}},\ }\href {\doibase 10.1103/PhysRevLett.98.037003} {\bibfield
  {journal} {\bibinfo  {journal} {Phys. Rev. Lett.}\ }\textbf {\bibinfo
  {volume} {98}},\ \bibinfo {pages} {037003} (\bibinfo {year}
  {2007})}\BibitemShut {NoStop}%
\bibitem [{\citenamefont {Tanaka}\ \emph
  {et~al.}(2007{\natexlab{a}})\citenamefont {Tanaka}, \citenamefont {Golubov},
  \citenamefont {Kashiwaya},\ and\ \citenamefont
  {Ueda}}]{PhysRevLett.99.037005}%
  \BibitemOpen
  \bibfield  {author} {\bibinfo {author} {\bibfnamefont {Y.}~\bibnamefont
  {Tanaka}}, \bibinfo {author} {\bibfnamefont {A.~A.}\ \bibnamefont {Golubov}},
  \bibinfo {author} {\bibfnamefont {S.}~\bibnamefont {Kashiwaya}}, \ and\
  \bibinfo {author} {\bibfnamefont {M.}~\bibnamefont {Ueda}},\ }\href {\doibase
  10.1103/PhysRevLett.99.037005} {\bibfield  {journal} {\bibinfo  {journal}
  {Phys. Rev. Lett.}\ }\textbf {\bibinfo {volume} {99}},\ \bibinfo {pages}
  {037005} (\bibinfo {year} {2007}{\natexlab{a}})}\BibitemShut {NoStop}%
\bibitem [{\citenamefont {Tanaka}\ \emph
  {et~al.}(2007{\natexlab{b}})\citenamefont {Tanaka}, \citenamefont {Tanuma},\
  and\ \citenamefont {Golubov}}]{PhysRevB.76.054522}%
  \BibitemOpen
  \bibfield  {author} {\bibinfo {author} {\bibfnamefont {Y.}~\bibnamefont
  {Tanaka}}, \bibinfo {author} {\bibfnamefont {Y.}~\bibnamefont {Tanuma}}, \
  and\ \bibinfo {author} {\bibfnamefont {A.~A.}\ \bibnamefont {Golubov}},\
  }\href {\doibase 10.1103/PhysRevB.76.054522} {\bibfield  {journal} {\bibinfo
  {journal} {Phys. Rev. B}\ }\textbf {\bibinfo {volume} {76}},\ \bibinfo
  {pages} {054522} (\bibinfo {year} {2007}{\natexlab{b}})}\BibitemShut
  {NoStop}%
\bibitem [{\citenamefont {Tanaka}\ \emph {et~al.}(2012)\citenamefont {Tanaka},
  \citenamefont {Sato},\ and\ \citenamefont {Nagaosa}}]{Nagaosa12}%
  \BibitemOpen
  \bibfield  {author} {\bibinfo {author} {\bibfnamefont {Y.}~\bibnamefont
  {Tanaka}}, \bibinfo {author} {\bibfnamefont {M.}~\bibnamefont {Sato}}, \ and\
  \bibinfo {author} {\bibfnamefont {N.}~\bibnamefont {Nagaosa}},\ }\href
  {\doibase http://dx.doi.org/10.1143/JPSJ.81.011013} {\bibfield  {journal}
  {\bibinfo  {journal} {J. Phys. Soc. Jpn.}\ }\textbf {\bibinfo {volume}
  {81}},\ \bibinfo {pages} {011013} (\bibinfo {year} {2012})}\BibitemShut
  {NoStop}%
\bibitem [{\citenamefont {Reeg}\ and\ \citenamefont
  {Maslov}(2015)}]{PhysRevB.92.134512}%
  \BibitemOpen
  \bibfield  {author} {\bibinfo {author} {\bibfnamefont {C.~R.}\ \bibnamefont
  {Reeg}}\ and\ \bibinfo {author} {\bibfnamefont {D.~L.}\ \bibnamefont
  {Maslov}},\ }\href {\doibase 10.1103/PhysRevB.92.134512} {\bibfield
  {journal} {\bibinfo  {journal} {Phys. Rev. B}\ }\textbf {\bibinfo {volume}
  {92}},\ \bibinfo {pages} {134512} (\bibinfo {year} {2015})}\BibitemShut
  {NoStop}%
\bibitem [{\citenamefont {Bobkova}\ and\ \citenamefont
  {Bobkov}(2017)}]{PhysRevB.95.184518}%
  \BibitemOpen
  \bibfield  {author} {\bibinfo {author} {\bibfnamefont {I.~V.}\ \bibnamefont
  {Bobkova}}\ and\ \bibinfo {author} {\bibfnamefont {A.~M.}\ \bibnamefont
  {Bobkov}},\ }\href {\doibase 10.1103/PhysRevB.95.184518} {\bibfield
  {journal} {\bibinfo  {journal} {Phys. Rev. B}\ }\textbf {\bibinfo {volume}
  {95}},\ \bibinfo {pages} {184518} (\bibinfo {year} {2017})}\BibitemShut
  {NoStop}%
\bibitem [{\citenamefont {Ebisu}\ \emph {et~al.}(2016)\citenamefont {Ebisu},
  \citenamefont {Lu}, \citenamefont {Klinovaja},\ and\ \citenamefont
  {Tanaka}}]{Ebisu16}%
  \BibitemOpen
  \bibfield  {author} {\bibinfo {author} {\bibfnamefont {H.}~\bibnamefont
  {Ebisu}}, \bibinfo {author} {\bibfnamefont {B.}~\bibnamefont {Lu}}, \bibinfo
  {author} {\bibfnamefont {J.}~\bibnamefont {Klinovaja}}, \ and\ \bibinfo
  {author} {\bibfnamefont {Y.}~\bibnamefont {Tanaka}},\ }\href {\doibase
  10.1093/ptep/ptw094} {\bibfield  {journal} {\bibinfo  {journal} {Prog. Theor.
  Exp. Phys.}\ }\textbf {\bibinfo {volume} {2016}},\ \bibinfo {pages} {083I01}
  (\bibinfo {year} {2016})}\BibitemShut {NoStop}%
\bibitem [{\citenamefont {Burset}\ \emph {et~al.}(2017)\citenamefont {Burset},
  \citenamefont {Lu}, \citenamefont {Tamura},\ and\ \citenamefont
  {Tanaka}}]{PhysRevB.95.224502}%
  \BibitemOpen
  \bibfield  {author} {\bibinfo {author} {\bibfnamefont {P.}~\bibnamefont
  {Burset}}, \bibinfo {author} {\bibfnamefont {B.}~\bibnamefont {Lu}}, \bibinfo
  {author} {\bibfnamefont {S.}~\bibnamefont {Tamura}}, \ and\ \bibinfo {author}
  {\bibfnamefont {Y.}~\bibnamefont {Tanaka}},\ }\href {\doibase
  10.1103/PhysRevB.95.224502} {\bibfield  {journal} {\bibinfo  {journal} {Phys.
  Rev. B}\ }\textbf {\bibinfo {volume} {95}},\ \bibinfo {pages} {224502}
  (\bibinfo {year} {2017})}\BibitemShut {NoStop}%
\bibitem [{\citenamefont {Balatsky}\ \emph {et~al.}(2018)\citenamefont
  {Balatsky}, \citenamefont {Pershoguba},\ and\ \citenamefont
  {Triola}}]{Triola18}%
  \BibitemOpen
  \bibfield  {author} {\bibinfo {author} {\bibfnamefont {A.~V.}\ \bibnamefont
  {Balatsky}}, \bibinfo {author} {\bibfnamefont {S.~S.}\ \bibnamefont
  {Pershoguba}}, \ and\ \bibinfo {author} {\bibfnamefont {C.}~\bibnamefont
  {Triola}},\ }\href@noop {} {\bibfield  {journal} {\bibinfo  {journal}
  {arXiv:1804.07244}\ } (\bibinfo {year} {2018})}\BibitemShut {NoStop}%
\bibitem [{\citenamefont {Black-Schaffer}\ and\ \citenamefont
  {Balatsky}(2012)}]{PhysRevB.86.144506}%
  \BibitemOpen
  \bibfield  {author} {\bibinfo {author} {\bibfnamefont {A.~M.}\ \bibnamefont
  {Black-Schaffer}}\ and\ \bibinfo {author} {\bibfnamefont {A.~V.}\
  \bibnamefont {Balatsky}},\ }\href {\doibase 10.1103/PhysRevB.86.144506}
  {\bibfield  {journal} {\bibinfo  {journal} {Phys. Rev. B}\ }\textbf {\bibinfo
  {volume} {86}},\ \bibinfo {pages} {144506} (\bibinfo {year}
  {2012})}\BibitemShut {NoStop}%
\bibitem [{\citenamefont {Black-Schaffer}\ and\ \citenamefont
  {Balatsky}(2013{\natexlab{b}})}]{PhysRevB.87.220506}%
  \BibitemOpen
  \bibfield  {author} {\bibinfo {author} {\bibfnamefont {A.~M.}\ \bibnamefont
  {Black-Schaffer}}\ and\ \bibinfo {author} {\bibfnamefont {A.~V.}\
  \bibnamefont {Balatsky}},\ }\href {\doibase 10.1103/PhysRevB.87.220506}
  {\bibfield  {journal} {\bibinfo  {journal} {Phys. Rev. B}\ }\textbf {\bibinfo
  {volume} {87}},\ \bibinfo {pages} {220506} (\bibinfo {year}
  {2013}{\natexlab{b}})}\BibitemShut {NoStop}%
\bibitem [{\citenamefont {Lu}\ and\ \citenamefont {Tanaka}(2015)}]{bo2016}%
  \BibitemOpen
  \bibfield  {author} {\bibinfo {author} {\bibfnamefont {B.}~\bibnamefont
  {Lu}}\ and\ \bibinfo {author} {\bibfnamefont {Y.}~\bibnamefont {Tanaka}},\
  }\href {https://arxiv.org/abs/1512.00916} {\bibfield  {journal} {\bibinfo
  {journal} {arXiv:1512.00916}\ } (\bibinfo {year} {2015})}\BibitemShut
  {NoStop}%
\bibitem [{\citenamefont {Burset}\ \emph {et~al.}(2015)\citenamefont {Burset},
  \citenamefont {Lu}, \citenamefont {Tkachov}, \citenamefont {Tanaka},
  \citenamefont {Hankiewicz},\ and\ \citenamefont
  {Trauzettel}}]{PhysRevB.92.205424}%
  \BibitemOpen
  \bibfield  {author} {\bibinfo {author} {\bibfnamefont {P.}~\bibnamefont
  {Burset}}, \bibinfo {author} {\bibfnamefont {B.}~\bibnamefont {Lu}}, \bibinfo
  {author} {\bibfnamefont {G.}~\bibnamefont {Tkachov}}, \bibinfo {author}
  {\bibfnamefont {Y.}~\bibnamefont {Tanaka}}, \bibinfo {author} {\bibfnamefont
  {E.~M.}\ \bibnamefont {Hankiewicz}}, \ and\ \bibinfo {author} {\bibfnamefont
  {B.}~\bibnamefont {Trauzettel}},\ }\href {\doibase
  10.1103/PhysRevB.92.205424} {\bibfield  {journal} {\bibinfo  {journal} {Phys.
  Rev. B}\ }\textbf {\bibinfo {volume} {92}},\ \bibinfo {pages} {205424}
  (\bibinfo {year} {2015})}\BibitemShut {NoStop}%
\bibitem [{\citenamefont {Cr\'epin}\ \emph {et~al.}(2015)\citenamefont
  {Cr\'epin}, \citenamefont {Burset},\ and\ \citenamefont
  {Trauzettel}}]{PhysRevB.92.100507}%
  \BibitemOpen
  \bibfield  {author} {\bibinfo {author} {\bibfnamefont {F.}~\bibnamefont
  {Cr\'epin}}, \bibinfo {author} {\bibfnamefont {P.}~\bibnamefont {Burset}}, \
  and\ \bibinfo {author} {\bibfnamefont {B.}~\bibnamefont {Trauzettel}},\
  }\href {\doibase 10.1103/PhysRevB.92.100507} {\bibfield  {journal} {\bibinfo
  {journal} {Phys. Rev. B}\ }\textbf {\bibinfo {volume} {92}},\ \bibinfo
  {pages} {100507} (\bibinfo {year} {2015})}\BibitemShut {NoStop}%
\bibitem [{\citenamefont {Cayao}\ and\ \citenamefont
  {Black-Schaffer}(2017)}]{PhysRevB.96.155426}%
  \BibitemOpen
  \bibfield  {author} {\bibinfo {author} {\bibfnamefont {J.}~\bibnamefont
  {Cayao}}\ and\ \bibinfo {author} {\bibfnamefont {A.~M.}\ \bibnamefont
  {Black-Schaffer}},\ }\href {\doibase 10.1103/PhysRevB.96.155426} {\bibfield
  {journal} {\bibinfo  {journal} {Phys. Rev. B}\ }\textbf {\bibinfo {volume}
  {96}},\ \bibinfo {pages} {155426} (\bibinfo {year} {2017})}\BibitemShut
  {NoStop}%
\bibitem [{\citenamefont {Kuzmanovski}\ and\ \citenamefont
  {Black-Schaffer}(2017)}]{PhysRevB.96.174509}%
  \BibitemOpen
  \bibfield  {author} {\bibinfo {author} {\bibfnamefont {D.}~\bibnamefont
  {Kuzmanovski}}\ and\ \bibinfo {author} {\bibfnamefont {A.~M.}\ \bibnamefont
  {Black-Schaffer}},\ }\href {\doibase 10.1103/PhysRevB.96.174509} {\bibfield
  {journal} {\bibinfo  {journal} {Phys. Rev. B}\ }\textbf {\bibinfo {volume}
  {96}},\ \bibinfo {pages} {174509} (\bibinfo {year} {2017})}\BibitemShut
  {NoStop}%
\bibitem [{\citenamefont {Keidel}\ \emph {et~al.}(2018)\citenamefont {Keidel},
  \citenamefont {Burset},\ and\ \citenamefont
  {Trauzettel}}]{PhysRevB.97.075408}%
  \BibitemOpen
  \bibfield  {author} {\bibinfo {author} {\bibfnamefont {F.}~\bibnamefont
  {Keidel}}, \bibinfo {author} {\bibfnamefont {P.}~\bibnamefont {Burset}}, \
  and\ \bibinfo {author} {\bibfnamefont {B.}~\bibnamefont {Trauzettel}},\
  }\href {\doibase 10.1103/PhysRevB.97.075408} {\bibfield  {journal} {\bibinfo
  {journal} {Phys. Rev. B}\ }\textbf {\bibinfo {volume} {97}},\ \bibinfo
  {pages} {075408} (\bibinfo {year} {2018})}\BibitemShut {NoStop}%
\bibitem [{\citenamefont {Breunig}\ \emph {et~al.}(2018)\citenamefont
  {Breunig}, \citenamefont {Burset},\ and\ \citenamefont
  {Trauzettel}}]{PhysRevLett.120.037701}%
  \BibitemOpen
  \bibfield  {author} {\bibinfo {author} {\bibfnamefont {D.}~\bibnamefont
  {Breunig}}, \bibinfo {author} {\bibfnamefont {P.}~\bibnamefont {Burset}}, \
  and\ \bibinfo {author} {\bibfnamefont {B.}~\bibnamefont {Trauzettel}},\
  }\href {\doibase 10.1103/PhysRevLett.120.037701} {\bibfield  {journal}
  {\bibinfo  {journal} {Phys. Rev. Lett.}\ }\textbf {\bibinfo {volume} {120}},\
  \bibinfo {pages} {037701} (\bibinfo {year} {2018})}\BibitemShut {NoStop}%
\bibitem [{\citenamefont {Fleckenstein}\ \emph
  {et~al.}(2018{\natexlab{a}})\citenamefont {Fleckenstein}, \citenamefont
  {Ziani},\ and\ \citenamefont {Trauzettel}}]{PhysRevB.97.134523}%
  \BibitemOpen
  \bibfield  {author} {\bibinfo {author} {\bibfnamefont {C.}~\bibnamefont
  {Fleckenstein}}, \bibinfo {author} {\bibfnamefont {N.~T.}\ \bibnamefont
  {Ziani}}, \ and\ \bibinfo {author} {\bibfnamefont {B.}~\bibnamefont
  {Trauzettel}},\ }\href {\doibase 10.1103/PhysRevB.97.134523} {\bibfield
  {journal} {\bibinfo  {journal} {Phys. Rev. B}\ }\textbf {\bibinfo {volume}
  {97}},\ \bibinfo {pages} {134523} (\bibinfo {year}
  {2018}{\natexlab{a}})}\BibitemShut {NoStop}%
\bibitem [{\citenamefont {Triola}\ and\ \citenamefont
  {Balatsky}(2016)}]{PhysRevB.94.094518}%
  \BibitemOpen
  \bibfield  {author} {\bibinfo {author} {\bibfnamefont {C.}~\bibnamefont
  {Triola}}\ and\ \bibinfo {author} {\bibfnamefont {A.~V.}\ \bibnamefont
  {Balatsky}},\ }\href {\doibase 10.1103/PhysRevB.94.094518} {\bibfield
  {journal} {\bibinfo  {journal} {Phys. Rev. B}\ }\textbf {\bibinfo {volume}
  {94}},\ \bibinfo {pages} {094518} (\bibinfo {year} {2016})}\BibitemShut
  {NoStop}%
\bibitem [{\citenamefont {Triola}\ and\ \citenamefont
  {Balatsky}(2017)}]{triola17}%
  \BibitemOpen
  \bibfield  {author} {\bibinfo {author} {\bibfnamefont {C.}~\bibnamefont
  {Triola}}\ and\ \bibinfo {author} {\bibfnamefont {A.~V.}\ \bibnamefont
  {Balatsky}},\ }\href {\doibase 10.1103/PhysRevB.95.224518} {\bibfield
  {journal} {\bibinfo  {journal} {Phys. Rev. B}\ }\textbf {\bibinfo {volume}
  {95}},\ \bibinfo {pages} {224518} (\bibinfo {year} {2017})}\BibitemShut
  {NoStop}%
\bibitem [{\citenamefont {Sigrist}\ and\ \citenamefont
  {Ueda}(1991)}]{RevModPhys.63.239}%
  \BibitemOpen
  \bibfield  {author} {\bibinfo {author} {\bibfnamefont {M.}~\bibnamefont
  {Sigrist}}\ and\ \bibinfo {author} {\bibfnamefont {K.}~\bibnamefont {Ueda}},\
  }\href {\doibase 10.1103/RevModPhys.63.239} {\bibfield  {journal} {\bibinfo
  {journal} {Rev. Mod. Phys.}\ }\textbf {\bibinfo {volume} {63}},\ \bibinfo
  {pages} {239} (\bibinfo {year} {1991})}\BibitemShut {NoStop}%
\bibitem [{\citenamefont {Higginbotham}\ \emph {et~al.}(1994)\citenamefont
  {Higginbotham}, \citenamefont {Albrecht}, \citenamefont {Kirsanskas},
  \citenamefont {Chang}, \citenamefont {Kuemmeth}, \citenamefont {Krogstrup},
  \citenamefont {Nyg{\aa}rd}, \citenamefont {Flensberg},\ and\ \citenamefont
  {Marcus}}]{Maeno94}%
  \BibitemOpen
  \bibfield  {author} {\bibinfo {author} {\bibfnamefont {A.~P.}\ \bibnamefont
  {Higginbotham}}, \bibinfo {author} {\bibfnamefont {S.~M.}\ \bibnamefont
  {Albrecht}}, \bibinfo {author} {\bibfnamefont {G.}~\bibnamefont
  {Kirsanskas}}, \bibinfo {author} {\bibfnamefont {W.}~\bibnamefont {Chang}},
  \bibinfo {author} {\bibfnamefont {F.}~\bibnamefont {Kuemmeth}}, \bibinfo
  {author} {\bibfnamefont {P.}~\bibnamefont {Krogstrup}}, \bibinfo {author}
  {\bibfnamefont {T.~S. J.~J.}\ \bibnamefont {Nyg{\aa}rd}}, \bibinfo {author}
  {\bibfnamefont {K.}~\bibnamefont {Flensberg}}, \ and\ \bibinfo {author}
  {\bibfnamefont {C.~M.}\ \bibnamefont {Marcus}},\ }\href
  {http://dx.doi.org/10.1038/372532a0} {\bibfield  {journal} {\bibinfo
  {journal} {Nature}\ }\textbf {\bibinfo {volume} {372}},\ \bibinfo {pages}
  {532} (\bibinfo {year} {1994})}\BibitemShut {NoStop}%
\bibitem [{\citenamefont {Tou}\ \emph {et~al.}(1998)\citenamefont {Tou},
  \citenamefont {Kitaoka}, \citenamefont {Ishida}, \citenamefont {Asayama},
  \citenamefont {Kimura}, \citenamefont {\={O}nuki}, \citenamefont {Yamamoto},
  \citenamefont {Haga},\ and\ \citenamefont {Maezawa}}]{PhysRevLett.80.3129}%
  \BibitemOpen
  \bibfield  {author} {\bibinfo {author} {\bibfnamefont {H.}~\bibnamefont
  {Tou}}, \bibinfo {author} {\bibfnamefont {Y.}~\bibnamefont {Kitaoka}},
  \bibinfo {author} {\bibfnamefont {K.}~\bibnamefont {Ishida}}, \bibinfo
  {author} {\bibfnamefont {K.}~\bibnamefont {Asayama}}, \bibinfo {author}
  {\bibfnamefont {N.}~\bibnamefont {Kimura}}, \bibinfo {author} {\bibfnamefont
  {Y.}~\bibnamefont {\={O}nuki}}, \bibinfo {author} {\bibfnamefont
  {E.}~\bibnamefont {Yamamoto}}, \bibinfo {author} {\bibfnamefont
  {Y.}~\bibnamefont {Haga}}, \ and\ \bibinfo {author} {\bibfnamefont
  {K.}~\bibnamefont {Maezawa}},\ }\href {\doibase 10.1103/PhysRevLett.80.3129}
  {\bibfield  {journal} {\bibinfo  {journal} {Phys. Rev. Lett.}\ }\textbf
  {\bibinfo {volume} {80}},\ \bibinfo {pages} {3129} (\bibinfo {year}
  {1998})}\BibitemShut {NoStop}%
\bibitem [{\citenamefont {Kashiwaya}\ \emph {et~al.}(2011)\citenamefont
  {Kashiwaya}, \citenamefont {Kashiwaya}, \citenamefont {Kambara},
  \citenamefont {Furuta}, \citenamefont {Yaguchi}, \citenamefont {Tanaka},\
  and\ \citenamefont {Maeno}}]{PhysRevLett.107.077003}%
  \BibitemOpen
  \bibfield  {author} {\bibinfo {author} {\bibfnamefont {S.}~\bibnamefont
  {Kashiwaya}}, \bibinfo {author} {\bibfnamefont {H.}~\bibnamefont
  {Kashiwaya}}, \bibinfo {author} {\bibfnamefont {H.}~\bibnamefont {Kambara}},
  \bibinfo {author} {\bibfnamefont {T.}~\bibnamefont {Furuta}}, \bibinfo
  {author} {\bibfnamefont {H.}~\bibnamefont {Yaguchi}}, \bibinfo {author}
  {\bibfnamefont {Y.}~\bibnamefont {Tanaka}}, \ and\ \bibinfo {author}
  {\bibfnamefont {Y.}~\bibnamefont {Maeno}},\ }\href {\doibase
  10.1103/PhysRevLett.107.077003} {\bibfield  {journal} {\bibinfo  {journal}
  {Phys. Rev. Lett.}\ }\textbf {\bibinfo {volume} {107}},\ \bibinfo {pages}
  {077003} (\bibinfo {year} {2011})}\BibitemShut {NoStop}%
\bibitem [{\citenamefont {Coleman}\ \emph {et~al.}(1993)\citenamefont
  {Coleman}, \citenamefont {Miranda},\ and\ \citenamefont
  {Tsvelik}}]{PhysRevLett.70.2960}%
  \BibitemOpen
  \bibfield  {author} {\bibinfo {author} {\bibfnamefont {P.}~\bibnamefont
  {Coleman}}, \bibinfo {author} {\bibfnamefont {E.}~\bibnamefont {Miranda}}, \
  and\ \bibinfo {author} {\bibfnamefont {A.}~\bibnamefont {Tsvelik}},\ }\href
  {\doibase 10.1103/PhysRevLett.70.2960} {\bibfield  {journal} {\bibinfo
  {journal} {Phys. Rev. Lett.}\ }\textbf {\bibinfo {volume} {70}},\ \bibinfo
  {pages} {2960} (\bibinfo {year} {1993})}\BibitemShut {NoStop}%
\bibitem [{\citenamefont {Liu}\ \emph {et~al.}(2015)\citenamefont {Liu},
  \citenamefont {Sau},\ and\ \citenamefont {Das~Sarma}}]{PhysRevB.92.014513}%
  \BibitemOpen
  \bibfield  {author} {\bibinfo {author} {\bibfnamefont {X.}~\bibnamefont
  {Liu}}, \bibinfo {author} {\bibfnamefont {J.~D.}\ \bibnamefont {Sau}}, \ and\
  \bibinfo {author} {\bibfnamefont {S.}~\bibnamefont {Das~Sarma}},\ }\href
  {\doibase 10.1103/PhysRevB.92.014513} {\bibfield  {journal} {\bibinfo
  {journal} {Phys. Rev. B}\ }\textbf {\bibinfo {volume} {92}},\ \bibinfo
  {pages} {014513} (\bibinfo {year} {2015})}\BibitemShut {NoStop}%
\bibitem [{\citenamefont {Huang}\ \emph {et~al.}(2015)\citenamefont {Huang},
  \citenamefont {W\"olfle},\ and\ \citenamefont
  {Balatsky}}]{PhysRevB.92.121404}%
  \BibitemOpen
  \bibfield  {author} {\bibinfo {author} {\bibfnamefont {Z.}~\bibnamefont
  {Huang}}, \bibinfo {author} {\bibfnamefont {P.}~\bibnamefont {W\"olfle}}, \
  and\ \bibinfo {author} {\bibfnamefont {A.~V.}\ \bibnamefont {Balatsky}},\
  }\href {\doibase 10.1103/PhysRevB.92.121404} {\bibfield  {journal} {\bibinfo
  {journal} {Phys. Rev. B}\ }\textbf {\bibinfo {volume} {92}},\ \bibinfo
  {pages} {121404} (\bibinfo {year} {2015})}\BibitemShut {NoStop}%
\bibitem [{\citenamefont {Lee}\ \emph {et~al.}(2017)\citenamefont {Lee},
  \citenamefont {Lutchyn},\ and\ \citenamefont {Maciejko}}]{lutchyn16}%
  \BibitemOpen
  \bibfield  {author} {\bibinfo {author} {\bibfnamefont {S.-P.}\ \bibnamefont
  {Lee}}, \bibinfo {author} {\bibfnamefont {R.~M.}\ \bibnamefont {Lutchyn}}, \
  and\ \bibinfo {author} {\bibfnamefont {J.}~\bibnamefont {Maciejko}},\ }\href
  {\doibase 10.1103/PhysRevB.95.184506} {\bibfield  {journal} {\bibinfo
  {journal} {Phys. Rev. B}\ }\textbf {\bibinfo {volume} {95}},\ \bibinfo
  {pages} {184506} (\bibinfo {year} {2017})}\BibitemShut {NoStop}%
\bibitem [{\citenamefont {Kashuba}\ \emph {et~al.}(2017)\citenamefont
  {Kashuba}, \citenamefont {Sothmann}, \citenamefont {Burset},\ and\
  \citenamefont {Trauzettel}}]{PhysRevB.95.174516}%
  \BibitemOpen
  \bibfield  {author} {\bibinfo {author} {\bibfnamefont {O.}~\bibnamefont
  {Kashuba}}, \bibinfo {author} {\bibfnamefont {B.}~\bibnamefont {Sothmann}},
  \bibinfo {author} {\bibfnamefont {P.}~\bibnamefont {Burset}}, \ and\ \bibinfo
  {author} {\bibfnamefont {B.}~\bibnamefont {Trauzettel}},\ }\href {\doibase
  10.1103/PhysRevB.95.174516} {\bibfield  {journal} {\bibinfo  {journal} {Phys.
  Rev. B}\ }\textbf {\bibinfo {volume} {95}},\ \bibinfo {pages} {174516}
  (\bibinfo {year} {2017})}\BibitemShut {NoStop}%
\bibitem [{\citenamefont {Linder}\ and\ \citenamefont
  {Balatsky}(2017)}]{Balatsky2017}%
  \BibitemOpen
  \bibfield  {author} {\bibinfo {author} {\bibfnamefont {J.}~\bibnamefont
  {Linder}}\ and\ \bibinfo {author} {\bibfnamefont {A.~V.}\ \bibnamefont
  {Balatsky}},\ }\href@noop {} {\bibfield  {journal} {\bibinfo  {journal}
  {arXiv:1709.03986}\ } (\bibinfo {year} {2017})}\BibitemShut {NoStop}%
\bibitem [{\citenamefont {Dresselhaus}(1955)}]{PhysRev.100.580}%
  \BibitemOpen
  \bibfield  {author} {\bibinfo {author} {\bibfnamefont {G.}~\bibnamefont
  {Dresselhaus}},\ }\href {\doibase 10.1103/PhysRev.100.580} {\bibfield
  {journal} {\bibinfo  {journal} {Phys. Rev.}\ }\textbf {\bibinfo {volume}
  {100}},\ \bibinfo {pages} {580} (\bibinfo {year} {1955})}\BibitemShut
  {NoStop}%
\bibitem [{\citenamefont {Rashba}(1960)}]{Rashba1960}%
  \BibitemOpen
  \bibfield  {author} {\bibinfo {author} {\bibfnamefont {E.}~\bibnamefont
  {Rashba}},\ }\href@noop {} {\bibfield  {journal} {\bibinfo  {journal} {Sov.
  Phys. Solid. State}\ }\textbf {\bibinfo {volume} {2}},\ \bibinfo {pages}
  {1109} (\bibinfo {year} {1960})}\BibitemShut {NoStop}%
\bibitem [{\citenamefont {Bychkov}\ and\ \citenamefont
  {Rashba}(1984)}]{rashba84a}%
  \BibitemOpen
  \bibfield  {author} {\bibinfo {author} {\bibfnamefont {Y.~A.}\ \bibnamefont
  {Bychkov}}\ and\ \bibinfo {author} {\bibfnamefont {E.~I.}\ \bibnamefont
  {Rashba}},\ }\href@noop {} {\bibfield  {journal} {\bibinfo  {journal} {Sov.
  Phys. JETP}\ }\textbf {\bibinfo {volume} {39}},\ \bibinfo {pages} {78}
  (\bibinfo {year} {1984})}\BibitemShut {NoStop}%
\bibitem [{\citenamefont {Samokhin}(2009)}]{SAMOKHIN20092385}%
  \BibitemOpen
  \bibfield  {author} {\bibinfo {author} {\bibfnamefont {K.}~\bibnamefont
  {Samokhin}},\ }\href {\doibase https://doi.org/10.1016/j.aop.2009.08.008}
  {\bibfield  {journal} {\bibinfo  {journal} {Annals of Physics}\ }\textbf
  {\bibinfo {volume} {324}},\ \bibinfo {pages} {2385 } (\bibinfo {year}
  {2009})}\BibitemShut {NoStop}%
\bibitem [{\citenamefont {Schultz}\ \emph {et~al.}(1996)\citenamefont
  {Schultz}, \citenamefont {Heinrichs}, \citenamefont {Merkt}, \citenamefont
  {Colin}, \citenamefont {Skauli},\ and\ \citenamefont
  {L{\o}vold}}]{0268-1242-11-8-009}%
  \BibitemOpen
  \bibfield  {author} {\bibinfo {author} {\bibfnamefont {M.}~\bibnamefont
  {Schultz}}, \bibinfo {author} {\bibfnamefont {F.}~\bibnamefont {Heinrichs}},
  \bibinfo {author} {\bibfnamefont {U.}~\bibnamefont {Merkt}}, \bibinfo
  {author} {\bibfnamefont {T.}~\bibnamefont {Colin}}, \bibinfo {author}
  {\bibfnamefont {T.}~\bibnamefont {Skauli}}, \ and\ \bibinfo {author}
  {\bibfnamefont {S.}~\bibnamefont {L{\o}vold}},\ }\href
  {http://stacks.iop.org/0268-1242/11/i=8/a=009} {\bibfield  {journal}
  {\bibinfo  {journal} {Semicond. Sci. Technol.}\ }\textbf {\bibinfo {volume}
  {11}},\ \bibinfo {pages} {1168} (\bibinfo {year} {1996})}\BibitemShut
  {NoStop}%
\bibitem [{\citenamefont {Nitta}\ \emph {et~al.}(1997)\citenamefont {Nitta},
  \citenamefont {Akazaki}, \citenamefont {Takayanagi},\ and\ \citenamefont
  {Enoki}}]{PhysRevLett.78.1335}%
  \BibitemOpen
  \bibfield  {author} {\bibinfo {author} {\bibfnamefont {J.}~\bibnamefont
  {Nitta}}, \bibinfo {author} {\bibfnamefont {T.}~\bibnamefont {Akazaki}},
  \bibinfo {author} {\bibfnamefont {H.}~\bibnamefont {Takayanagi}}, \ and\
  \bibinfo {author} {\bibfnamefont {T.}~\bibnamefont {Enoki}},\ }\href
  {\doibase 10.1103/PhysRevLett.78.1335} {\bibfield  {journal} {\bibinfo
  {journal} {Phys. Rev. Lett.}\ }\textbf {\bibinfo {volume} {78}},\ \bibinfo
  {pages} {1335} (\bibinfo {year} {1997})}\BibitemShut {NoStop}%
\bibitem [{\citenamefont {Liang}\ and\ \citenamefont
  {Gao}(2012)}]{doi:10.1021/nl301325h}%
  \BibitemOpen
  \bibfield  {author} {\bibinfo {author} {\bibfnamefont {D.}~\bibnamefont
  {Liang}}\ and\ \bibinfo {author} {\bibfnamefont {X.~P.}\ \bibnamefont
  {Gao}},\ }\href {\doibase 10.1021/nl301325h} {\bibfield  {journal} {\bibinfo
  {journal} {Nano Letters}\ }\textbf {\bibinfo {volume} {12}},\ \bibinfo
  {pages} {3263} (\bibinfo {year} {2012})}\BibitemShut {NoStop}%
\bibitem [{\citenamefont {Takase}\ \emph {et~al.}(2017)\citenamefont {Takase},
  \citenamefont {Ashikawa}, \citenamefont {Zhang}, \citenamefont {Tateno},\
  and\ \citenamefont {Sasaki}}]{Takase17}%
  \BibitemOpen
  \bibfield  {author} {\bibinfo {author} {\bibfnamefont {K.}~\bibnamefont
  {Takase}}, \bibinfo {author} {\bibfnamefont {Y.}~\bibnamefont {Ashikawa}},
  \bibinfo {author} {\bibfnamefont {G.}~\bibnamefont {Zhang}}, \bibinfo
  {author} {\bibfnamefont {K.}~\bibnamefont {Tateno}}, \ and\ \bibinfo {author}
  {\bibfnamefont {S.}~\bibnamefont {Sasaki}},\ }\href {\doibase
  doi:10.1038/s41598-017-01080-0} {\bibfield  {journal} {\bibinfo  {journal}
  {Sci. Rep.}\ }\textbf {\bibinfo {volume} {7}},\ \bibinfo {pages} {930}
  (\bibinfo {year} {2017})}\BibitemShut {NoStop}%
\bibitem [{\citenamefont {Triola}\ \emph {et~al.}(2016)\citenamefont {Triola},
  \citenamefont {Badiane}, \citenamefont {Balatsky},\ and\ \citenamefont
  {Rossi}}]{PhysRevLett.116.257001}%
  \BibitemOpen
  \bibfield  {author} {\bibinfo {author} {\bibfnamefont {C.}~\bibnamefont
  {Triola}}, \bibinfo {author} {\bibfnamefont {D.~M.}\ \bibnamefont {Badiane}},
  \bibinfo {author} {\bibfnamefont {A.~V.}\ \bibnamefont {Balatsky}}, \ and\
  \bibinfo {author} {\bibfnamefont {E.}~\bibnamefont {Rossi}},\ }\href
  {\doibase 10.1103/PhysRevLett.116.257001} {\bibfield  {journal} {\bibinfo
  {journal} {Phys. Rev. Lett.}\ }\textbf {\bibinfo {volume} {116}},\ \bibinfo
  {pages} {257001} (\bibinfo {year} {2016})}\BibitemShut {NoStop}%
\bibitem [{\citenamefont {Tamura}\ and\ \citenamefont
  {Tanaka}(2018)}]{Tamura18}%
  \BibitemOpen
  \bibfield  {author} {\bibinfo {author} {\bibfnamefont {S.}~\bibnamefont
  {Tamura}}\ and\ \bibinfo {author} {\bibfnamefont {Y.}~\bibnamefont
  {Tanaka}},\ }\href@noop {} {\bibfield  {journal} {\bibinfo  {journal}
  {arXiv:1804.04018}\ } (\bibinfo {year} {2018})}\BibitemShut {NoStop}%
\bibitem [{\citenamefont {Chang}\ \emph {et~al.}(2015)\citenamefont {Chang},
  \citenamefont {Albrecht}, \citenamefont {Jespersen}, \citenamefont
  {Kuemmeth}, \citenamefont {Krogstrup}, \citenamefont {Nyg{\aa}rd},\ and\
  \citenamefont {Marcus}}]{chang15}%
  \BibitemOpen
  \bibfield  {author} {\bibinfo {author} {\bibfnamefont {W.}~\bibnamefont
  {Chang}}, \bibinfo {author} {\bibfnamefont {S.~M.}\ \bibnamefont {Albrecht}},
  \bibinfo {author} {\bibfnamefont {T.~S.}\ \bibnamefont {Jespersen}}, \bibinfo
  {author} {\bibfnamefont {F.}~\bibnamefont {Kuemmeth}}, \bibinfo {author}
  {\bibfnamefont {P.}~\bibnamefont {Krogstrup}}, \bibinfo {author}
  {\bibfnamefont {J.}~\bibnamefont {Nyg{\aa}rd}}, \ and\ \bibinfo {author}
  {\bibfnamefont {C.~M.}\ \bibnamefont {Marcus}},\ }\href
  {https://www.nature.com/articles/nnano.2014.306} {\bibfield  {journal}
  {\bibinfo  {journal} {Nat. Nanotech.}\ }\textbf {\bibinfo {volume} {10}},\
  \bibinfo {pages} {232} (\bibinfo {year} {2015})}\BibitemShut {NoStop}%
\bibitem [{\citenamefont {Higginbotham}\ \emph {et~al.}(2015)\citenamefont
  {Higginbotham}, \citenamefont {Albrecht}, \citenamefont {Kirsanskas},
  \citenamefont {Chang}, \citenamefont {Kuemmeth}, \citenamefont {Krogstrup},
  \citenamefont {Nyg{\aa}rd}, \citenamefont {Flensberg},\ and\ \citenamefont
  {Marcus}}]{Higginbotham}%
  \BibitemOpen
  \bibfield  {author} {\bibinfo {author} {\bibfnamefont {A.~P.}\ \bibnamefont
  {Higginbotham}}, \bibinfo {author} {\bibfnamefont {S.~M.}\ \bibnamefont
  {Albrecht}}, \bibinfo {author} {\bibfnamefont {G.}~\bibnamefont
  {Kirsanskas}}, \bibinfo {author} {\bibfnamefont {W.}~\bibnamefont {Chang}},
  \bibinfo {author} {\bibfnamefont {F.}~\bibnamefont {Kuemmeth}}, \bibinfo
  {author} {\bibfnamefont {P.}~\bibnamefont {Krogstrup}}, \bibinfo {author}
  {\bibfnamefont {T.~S. J.~J.}\ \bibnamefont {Nyg{\aa}rd}}, \bibinfo {author}
  {\bibfnamefont {K.}~\bibnamefont {Flensberg}}, \ and\ \bibinfo {author}
  {\bibfnamefont {C.~M.}\ \bibnamefont {Marcus}},\ }\href
  {https://www.nature.com/articles/nphys3461} {\bibfield  {journal} {\bibinfo
  {journal} {Nat. Phys.}\ }\textbf {\bibinfo {volume} {11}},\ \bibinfo {pages}
  {1017} (\bibinfo {year} {2015})}\BibitemShut {NoStop}%
\bibitem [{\citenamefont {Krogstrup}\ \emph {et~al.}(2015)\citenamefont
  {Krogstrup}, \citenamefont {Ziino}, \citenamefont {Chang}, \citenamefont
  {Albrecht}, \citenamefont {Madsen}, \citenamefont {Johnson}, \citenamefont
  {Nyg{\aa}rd}, \citenamefont {Marcus},\ and\ \citenamefont
  {Jespersen}}]{Krogstrup15}%
  \BibitemOpen
  \bibfield  {author} {\bibinfo {author} {\bibfnamefont {P.}~\bibnamefont
  {Krogstrup}}, \bibinfo {author} {\bibfnamefont {N.~L.~B.}\ \bibnamefont
  {Ziino}}, \bibinfo {author} {\bibfnamefont {W.}~\bibnamefont {Chang}},
  \bibinfo {author} {\bibfnamefont {S.~M.}\ \bibnamefont {Albrecht}}, \bibinfo
  {author} {\bibfnamefont {M.~H.}\ \bibnamefont {Madsen}}, \bibinfo {author}
  {\bibfnamefont {E.}~\bibnamefont {Johnson}}, \bibinfo {author} {\bibfnamefont
  {J.}~\bibnamefont {Nyg{\aa}rd}}, \bibinfo {author} {\bibfnamefont {C.~M.}\
  \bibnamefont {Marcus}}, \ and\ \bibinfo {author} {\bibfnamefont {T.~S.}\
  \bibnamefont {Jespersen}},\ }\href {https://www.nature.com/articles/nmat4176}
  {\bibfield  {journal} {\bibinfo  {journal} {Nat. Mat.}\ }\textbf {\bibinfo
  {volume} {14}},\ \bibinfo {pages} {400} (\bibinfo {year} {2015})}\BibitemShut
  {NoStop}%
\bibitem [{\citenamefont {\"{O}nder G\"{u}l}\ \emph {et~al.}(2015)\citenamefont
  {\"{O}nder G\"{u}l}, \citenamefont {van Woerkom}, \citenamefont {van
  Weperen}, \citenamefont {Car}, \citenamefont {Plissard}, \citenamefont
  {Bakkers},\ and\ \citenamefont {Kouwenhoven}}]{0957-4484-26-21-215202}%
  \BibitemOpen
  \bibfield  {author} {\bibinfo {author} {\bibnamefont {\"{O}nder G\"{u}l}},
  \bibinfo {author} {\bibfnamefont {D.~J.}\ \bibnamefont {van Woerkom}},
  \bibinfo {author} {\bibfnamefont {I.}~\bibnamefont {van Weperen}}, \bibinfo
  {author} {\bibfnamefont {D.}~\bibnamefont {Car}}, \bibinfo {author}
  {\bibfnamefont {S.~R.}\ \bibnamefont {Plissard}}, \bibinfo {author}
  {\bibfnamefont {E.~P. A.~M.}\ \bibnamefont {Bakkers}}, \ and\ \bibinfo
  {author} {\bibfnamefont {L.~P.}\ \bibnamefont {Kouwenhoven}},\ }\href
  {http://stacks.iop.org/0957-4484/26/i=21/a=215202} {\bibfield  {journal}
  {\bibinfo  {journal} {Nanotechnology}\ }\textbf {\bibinfo {volume} {26}},\
  \bibinfo {pages} {215202} (\bibinfo {year} {2015})}\BibitemShut {NoStop}%
\bibitem [{\citenamefont {Deng}\ \emph {et~al.}(2016)\citenamefont {Deng},
  \citenamefont {Vaitiek\.{e}nas}, \citenamefont {Hansen}, \citenamefont
  {Danon}, \citenamefont {Leijnse}, \citenamefont {Flensberg}, \citenamefont
  {Nyg{\aa}rd}, \citenamefont {Krogstrup},\ and\ \citenamefont
  {Marcus}}]{Deng16}%
  \BibitemOpen
  \bibfield  {author} {\bibinfo {author} {\bibfnamefont {M.~T.}\ \bibnamefont
  {Deng}}, \bibinfo {author} {\bibfnamefont {S.}~\bibnamefont
  {Vaitiek\.{e}nas}}, \bibinfo {author} {\bibfnamefont {E.~B.}\ \bibnamefont
  {Hansen}}, \bibinfo {author} {\bibfnamefont {J.}~\bibnamefont {Danon}},
  \bibinfo {author} {\bibfnamefont {M.}~\bibnamefont {Leijnse}}, \bibinfo
  {author} {\bibfnamefont {K.}~\bibnamefont {Flensberg}}, \bibinfo {author}
  {\bibfnamefont {J.}~\bibnamefont {Nyg{\aa}rd}}, \bibinfo {author}
  {\bibfnamefont {P.}~\bibnamefont {Krogstrup}}, \ and\ \bibinfo {author}
  {\bibfnamefont {C.~M.}\ \bibnamefont {Marcus}},\ }\href {\doibase
  10.1126/science.aaf3961} {\bibfield  {journal} {\bibinfo  {journal}
  {Science}\ }\textbf {\bibinfo {volume} {354}},\ \bibinfo {pages} {1557}
  (\bibinfo {year} {2016})}\BibitemShut {NoStop}%
\bibitem [{\citenamefont {Albrecht}\ \emph {et~al.}(2016)\citenamefont
  {Albrecht}, \citenamefont {Higginbotham}, \citenamefont {Madsen},
  \citenamefont {Kuemmeth}, \citenamefont {Jespersen}, \citenamefont
  {Nyg{\aa}rd}, , \citenamefont {Krogstrup},\ and\ \citenamefont
  {Marcus}}]{Albrecht16}%
  \BibitemOpen
  \bibfield  {author} {\bibinfo {author} {\bibfnamefont {S.~M.}\ \bibnamefont
  {Albrecht}}, \bibinfo {author} {\bibfnamefont {A.~P.}\ \bibnamefont
  {Higginbotham}}, \bibinfo {author} {\bibfnamefont {M.}~\bibnamefont
  {Madsen}}, \bibinfo {author} {\bibfnamefont {F.}~\bibnamefont {Kuemmeth}},
  \bibinfo {author} {\bibfnamefont {T.~S.}\ \bibnamefont {Jespersen}}, \bibinfo
  {author} {\bibfnamefont {J.}~\bibnamefont {Nyg{\aa}rd}}, , \bibinfo {author}
  {\bibfnamefont {P.}~\bibnamefont {Krogstrup}}, \ and\ \bibinfo {author}
  {\bibfnamefont {C.~M.}\ \bibnamefont {Marcus}},\ }\href
  {http://dx.doi.org/10.1038/nature17162} {\bibfield  {journal} {\bibinfo
  {journal} {Nature}\ }\textbf {\bibinfo {volume} {531}},\ \bibinfo {pages}
  {206} (\bibinfo {year} {2016})}\BibitemShut {NoStop}%
\bibitem [{\citenamefont {Zhang}\ \emph {et~al.}(2017)\citenamefont {Zhang},
  \citenamefont {\"{O}nder G\"{u}l}, \citenamefont {Conesa-Boj}, \citenamefont
  {Zuo}, \citenamefont {Mourik}, \citenamefont {de~Vries}, \citenamefont {van
  Veen}, \citenamefont {van Woerkom}, \citenamefont {Nowak}, \citenamefont
  {Wimmer}, \citenamefont {Car}, \citenamefont {Plissard}, \citenamefont
  {Bakkers}, \citenamefont {Quintero-P\'{e}rez}, \citenamefont {Goswami},
  \citenamefont {Watanabe}, \citenamefont {Taniguchi},\ and\ \citenamefont
  {Kouwenhoven}}]{zhang16}%
  \BibitemOpen
  \bibfield  {author} {\bibinfo {author} {\bibfnamefont {H.}~\bibnamefont
  {Zhang}}, \bibinfo {author} {\bibnamefont {\"{O}nder G\"{u}l}}, \bibinfo
  {author} {\bibfnamefont {S.}~\bibnamefont {Conesa-Boj}}, \bibinfo {author}
  {\bibfnamefont {K.}~\bibnamefont {Zuo}}, \bibinfo {author} {\bibfnamefont
  {V.}~\bibnamefont {Mourik}}, \bibinfo {author} {\bibfnamefont {F.~K.}\
  \bibnamefont {de~Vries}}, \bibinfo {author} {\bibfnamefont {J.}~\bibnamefont
  {van Veen}}, \bibinfo {author} {\bibfnamefont {D.~J.}\ \bibnamefont {van
  Woerkom}}, \bibinfo {author} {\bibfnamefont {M.~P.}\ \bibnamefont {Nowak}},
  \bibinfo {author} {\bibfnamefont {M.}~\bibnamefont {Wimmer}}, \bibinfo
  {author} {\bibfnamefont {D.}~\bibnamefont {Car}}, \bibinfo {author}
  {\bibfnamefont {S.}~\bibnamefont {Plissard}}, \bibinfo {author}
  {\bibfnamefont {E.~P. A.~M.}\ \bibnamefont {Bakkers}}, \bibinfo {author}
  {\bibfnamefont {M.}~\bibnamefont {Quintero-P\'{e}rez}}, \bibinfo {author}
  {\bibfnamefont {S.}~\bibnamefont {Goswami}}, \bibinfo {author} {\bibfnamefont
  {K.}~\bibnamefont {Watanabe}}, \bibinfo {author} {\bibfnamefont
  {T.}~\bibnamefont {Taniguchi}}, \ and\ \bibinfo {author} {\bibfnamefont
  {L.~P.}\ \bibnamefont {Kouwenhoven}},\ }\href
  {http://www.nature.com/articles/ncomms16025} {\bibfield  {journal} {\bibinfo
  {journal} {Nat. Commun.}\ }\textbf {\bibinfo {volume} {8}},\ \bibinfo {pages}
  {16025} (\bibinfo {year} {2017})}\BibitemShut {NoStop}%
\bibitem [{\citenamefont {G\"{u}l}\ \emph {et~al.}(2017)\citenamefont
  {G\"{u}l}, \citenamefont {Zhang}, \citenamefont {de~Vries}, \citenamefont
  {van Veen}, \citenamefont {Zuo}, \citenamefont {Mourik}, \citenamefont
  {Conesa-Boj}, \citenamefont {Nowak}, \citenamefont {van Woerkom},
  \citenamefont {Quintero-P\'{e}rez}, \citenamefont {Cassidy}, \citenamefont
  {Geresdi}, \citenamefont {Koelling}, \citenamefont {Car}, \citenamefont
  {Plissard}, \citenamefont {Bakkers},\ and\ \citenamefont
  {Kouwenhoven}}]{gulonder}%
  \BibitemOpen
  \bibfield  {author} {\bibinfo {author} {\bibfnamefont {O.}~\bibnamefont
  {G\"{u}l}}, \bibinfo {author} {\bibfnamefont {H.}~\bibnamefont {Zhang}},
  \bibinfo {author} {\bibfnamefont {F.~K.}\ \bibnamefont {de~Vries}}, \bibinfo
  {author} {\bibfnamefont {J.}~\bibnamefont {van Veen}}, \bibinfo {author}
  {\bibfnamefont {K.}~\bibnamefont {Zuo}}, \bibinfo {author} {\bibfnamefont
  {V.}~\bibnamefont {Mourik}}, \bibinfo {author} {\bibfnamefont
  {S.}~\bibnamefont {Conesa-Boj}}, \bibinfo {author} {\bibfnamefont {M.~P.}\
  \bibnamefont {Nowak}}, \bibinfo {author} {\bibfnamefont {D.~J.}\ \bibnamefont
  {van Woerkom}}, \bibinfo {author} {\bibfnamefont {M.}~\bibnamefont
  {Quintero-P\'{e}rez}}, \bibinfo {author} {\bibfnamefont {M.~C.}\ \bibnamefont
  {Cassidy}}, \bibinfo {author} {\bibfnamefont {A.}~\bibnamefont {Geresdi}},
  \bibinfo {author} {\bibfnamefont {S.}~\bibnamefont {Koelling}}, \bibinfo
  {author} {\bibfnamefont {D.}~\bibnamefont {Car}}, \bibinfo {author}
  {\bibfnamefont {S.~R.}\ \bibnamefont {Plissard}}, \bibinfo {author}
  {\bibfnamefont {E.~P. A.~M.}\ \bibnamefont {Bakkers}}, \ and\ \bibinfo
  {author} {\bibfnamefont {L.~P.}\ \bibnamefont {Kouwenhoven}},\ }\href
  {\doibase 10.1021/acs.nanolett.7b00540} {\bibfield  {journal} {\bibinfo
  {journal} {Nano Letters}\ }\textbf {\bibinfo {volume} {17}},\ \bibinfo
  {pages} {2690} (\bibinfo {year} {2017})}\BibitemShut {NoStop}%
\bibitem [{\citenamefont {Gazibegovic}\ \emph {et~al.}(2017)\citenamefont
  {Gazibegovic}, \citenamefont {Car}, \citenamefont {Zhang}, \citenamefont
  {Balk}, \citenamefont {Logan}, \citenamefont {de~Moor}, \citenamefont
  {Cassidy}, \citenamefont {Schmits}, \citenamefont {Xu}, \citenamefont {Wang},
  \citenamefont {Krogstrup}, \citenamefont {het Veld}, \citenamefont {Shen},
  \citenamefont {Bouman}, \citenamefont {Shojaei}, \citenamefont {Pennachio},
  \citenamefont {Lee}, \citenamefont {van Veldhoven}, \citenamefont {Koelling},
  \citenamefont {Verheijen}, \citenamefont {Kouwenhoven}, \citenamefont
  {Palmstr{\o}m},\ and\ \citenamefont {Bakkers}}]{Gazibegovic17}%
  \BibitemOpen
  \bibfield  {author} {\bibinfo {author} {\bibfnamefont {S.}~\bibnamefont
  {Gazibegovic}}, \bibinfo {author} {\bibfnamefont {D.}~\bibnamefont {Car}},
  \bibinfo {author} {\bibfnamefont {H.}~\bibnamefont {Zhang}}, \bibinfo
  {author} {\bibfnamefont {S.~C.}\ \bibnamefont {Balk}}, \bibinfo {author}
  {\bibfnamefont {J.~A.}\ \bibnamefont {Logan}}, \bibinfo {author}
  {\bibfnamefont {M.~W.~A.}\ \bibnamefont {de~Moor}}, \bibinfo {author}
  {\bibfnamefont {M.~C.}\ \bibnamefont {Cassidy}}, \bibinfo {author}
  {\bibfnamefont {R.}~\bibnamefont {Schmits}}, \bibinfo {author} {\bibfnamefont
  {D.}~\bibnamefont {Xu}}, \bibinfo {author} {\bibfnamefont {G.}~\bibnamefont
  {Wang}}, \bibinfo {author} {\bibfnamefont {P.}~\bibnamefont {Krogstrup}},
  \bibinfo {author} {\bibfnamefont {R.~L. M.~O.}\ \bibnamefont {het Veld}},
  \bibinfo {author} {\bibfnamefont {J.}~\bibnamefont {Shen}}, \bibinfo {author}
  {\bibfnamefont {D.}~\bibnamefont {Bouman}}, \bibinfo {author} {\bibfnamefont
  {B.}~\bibnamefont {Shojaei}}, \bibinfo {author} {\bibfnamefont
  {D.}~\bibnamefont {Pennachio}}, \bibinfo {author} {\bibfnamefont {J.~S.}\
  \bibnamefont {Lee}}, \bibinfo {author} {\bibfnamefont {P.~J.}\ \bibnamefont
  {van Veldhoven}}, \bibinfo {author} {\bibfnamefont {S.}~\bibnamefont
  {Koelling}}, \bibinfo {author} {\bibfnamefont {M.~A.}\ \bibnamefont
  {Verheijen}}, \bibinfo {author} {\bibfnamefont {L.~P.}\ \bibnamefont
  {Kouwenhoven}}, \bibinfo {author} {\bibfnamefont {C.~J.}\ \bibnamefont
  {Palmstr{\o}m}}, \ and\ \bibinfo {author} {\bibfnamefont {E.~P.}\
  \bibnamefont {Bakkers}},\ }\href
  {https://www.nature.com/articles/nature23468} {\bibfield  {journal} {\bibinfo
   {journal} {Nature}\ }\textbf {\bibinfo {volume} {548}},\ \bibinfo {pages}
  {434} (\bibinfo {year} {2017})}\BibitemShut {NoStop}%
\bibitem [{\citenamefont {Sestoft}\ \emph {et~al.}(2018)\citenamefont
  {Sestoft}, \citenamefont {Kanne}, \citenamefont {Gejl}, \citenamefont {von
  Soosten}, \citenamefont {Yodh}, \citenamefont {Sherman}, \citenamefont
  {Tarasinski}, \citenamefont {Wimmer}, \citenamefont {Johnson}, \citenamefont
  {Deng}, \citenamefont {Nyg\aa{}rd}, \citenamefont {Jespersen}, \citenamefont
  {Marcus},\ and\ \citenamefont {Krogstrup}}]{PhysRevMaterials.2.044202}%
  \BibitemOpen
  \bibfield  {author} {\bibinfo {author} {\bibfnamefont {J.~E.}\ \bibnamefont
  {Sestoft}}, \bibinfo {author} {\bibfnamefont {T.}~\bibnamefont {Kanne}},
  \bibinfo {author} {\bibfnamefont {A.~N.}\ \bibnamefont {Gejl}}, \bibinfo
  {author} {\bibfnamefont {M.}~\bibnamefont {von Soosten}}, \bibinfo {author}
  {\bibfnamefont {J.~S.}\ \bibnamefont {Yodh}}, \bibinfo {author}
  {\bibfnamefont {D.}~\bibnamefont {Sherman}}, \bibinfo {author} {\bibfnamefont
  {B.}~\bibnamefont {Tarasinski}}, \bibinfo {author} {\bibfnamefont
  {M.}~\bibnamefont {Wimmer}}, \bibinfo {author} {\bibfnamefont
  {E.}~\bibnamefont {Johnson}}, \bibinfo {author} {\bibfnamefont
  {M.}~\bibnamefont {Deng}}, \bibinfo {author} {\bibfnamefont {J.}~\bibnamefont
  {Nyg\aa{}rd}}, \bibinfo {author} {\bibfnamefont {T.~S.}\ \bibnamefont
  {Jespersen}}, \bibinfo {author} {\bibfnamefont {C.~M.}\ \bibnamefont
  {Marcus}}, \ and\ \bibinfo {author} {\bibfnamefont {P.}~\bibnamefont
  {Krogstrup}},\ }\href {\doibase 10.1103/PhysRevMaterials.2.044202} {\bibfield
   {journal} {\bibinfo  {journal} {Phys. Rev. Materials}\ }\textbf {\bibinfo
  {volume} {2}},\ \bibinfo {pages} {044202} (\bibinfo {year}
  {2018})}\BibitemShut {NoStop}%
\bibitem [{\citenamefont {Vaitiek\.{e}nas}\ \emph {et~al.}(2017)\citenamefont
  {Vaitiek\.{e}nas}, \citenamefont {Deng}, \citenamefont {Nyg{\aa}rd},
  \citenamefont {Krogstrup},\ and\ \citenamefont {Marcus}}]{vaitienkenas17}%
  \BibitemOpen
  \bibfield  {author} {\bibinfo {author} {\bibfnamefont {S.}~\bibnamefont
  {Vaitiek\.{e}nas}}, \bibinfo {author} {\bibfnamefont {M.~T.}\ \bibnamefont
  {Deng}}, \bibinfo {author} {\bibfnamefont {J.}~\bibnamefont {Nyg{\aa}rd}},
  \bibinfo {author} {\bibfnamefont {P.}~\bibnamefont {Krogstrup}}, \ and\
  \bibinfo {author} {\bibfnamefont {C.~M.}\ \bibnamefont {Marcus}},\
  }\href@noop {} {\bibfield  {journal} {\bibinfo  {journal} {arXiv:1710.04300}\
  } (\bibinfo {year} {2017})}\BibitemShut {NoStop}%
\bibitem [{\citenamefont {Zhang}\ \emph {et~al.}(2018)\citenamefont {Zhang},
  \citenamefont {Liu}, \citenamefont {Gazibegovic}, \citenamefont {Xu},
  \citenamefont {Logan}, \citenamefont {Wang}, \citenamefont {van Loo},
  \citenamefont {Bommer}, \citenamefont {de~Moor}, \citenamefont {Car},
  \citenamefont {het Veld}, \citenamefont {van Veldhoven}, \citenamefont
  {Koelling}, \citenamefont {Verheijen}, \citenamefont {Pendharkar},
  \citenamefont {Pennachio}, \citenamefont {Shojaei}, \citenamefont {Lee},
  \citenamefont {Palmstrom}, \citenamefont {Bakkers}, \citenamefont {Sarma},\
  and\ \citenamefont {Kouwenhoven}}]{zhang18}%
  \BibitemOpen
  \bibfield  {author} {\bibinfo {author} {\bibfnamefont {H.}~\bibnamefont
  {Zhang}}, \bibinfo {author} {\bibfnamefont {C.-X.}\ \bibnamefont {Liu}},
  \bibinfo {author} {\bibfnamefont {S.}~\bibnamefont {Gazibegovic}}, \bibinfo
  {author} {\bibfnamefont {D.}~\bibnamefont {Xu}}, \bibinfo {author}
  {\bibfnamefont {J.~A.}\ \bibnamefont {Logan}}, \bibinfo {author}
  {\bibfnamefont {G.}~\bibnamefont {Wang}}, \bibinfo {author} {\bibfnamefont
  {N.}~\bibnamefont {van Loo}}, \bibinfo {author} {\bibfnamefont {J.~D.}\
  \bibnamefont {Bommer}}, \bibinfo {author} {\bibfnamefont {M.~W.}\
  \bibnamefont {de~Moor}}, \bibinfo {author} {\bibfnamefont {D.}~\bibnamefont
  {Car}}, \bibinfo {author} {\bibfnamefont {R.~L. M.~O.}\ \bibnamefont {het
  Veld}}, \bibinfo {author} {\bibfnamefont {P.~J.}\ \bibnamefont {van
  Veldhoven}}, \bibinfo {author} {\bibfnamefont {S.}~\bibnamefont {Koelling}},
  \bibinfo {author} {\bibfnamefont {M.~A.}\ \bibnamefont {Verheijen}}, \bibinfo
  {author} {\bibfnamefont {M.}~\bibnamefont {Pendharkar}}, \bibinfo {author}
  {\bibfnamefont {D.~J.}\ \bibnamefont {Pennachio}}, \bibinfo {author}
  {\bibfnamefont {B.}~\bibnamefont {Shojaei}}, \bibinfo {author} {\bibfnamefont
  {J.~S.}\ \bibnamefont {Lee}}, \bibinfo {author} {\bibfnamefont {C.~J.}\
  \bibnamefont {Palmstrom}}, \bibinfo {author} {\bibfnamefont {E.~P.}\
  \bibnamefont {Bakkers}}, \bibinfo {author} {\bibfnamefont {S.~D.}\
  \bibnamefont {Sarma}}, \ and\ \bibinfo {author} {\bibfnamefont {L.~P.}\
  \bibnamefont {Kouwenhoven}},\ }\href {http://dx.doi.org/10.1038/nature26142}
  {\bibfield  {journal} {\bibinfo  {journal} {Nature}\ }\textbf {\bibinfo
  {volume} {556}},\ \bibinfo {pages} {74} (\bibinfo {year} {2018})}\BibitemShut
  {NoStop}%
\bibitem [{\citenamefont {Deng}\ \emph {et~al.}(2018)\citenamefont {Deng},
  \citenamefont {Vaitiek\ifmmode~\dot{e}\else \.{e}\fi{}nas}, \citenamefont
  {Prada}, \citenamefont {San-Jose}, \citenamefont {Nyg\aa{}rd}, \citenamefont
  {Krogstrup}, \citenamefont {Aguado},\ and\ \citenamefont {Marcus}}]{deng18}%
  \BibitemOpen
  \bibfield  {author} {\bibinfo {author} {\bibfnamefont {M.-T.}\ \bibnamefont
  {Deng}}, \bibinfo {author} {\bibfnamefont {S.}~\bibnamefont
  {Vaitiek\ifmmode~\dot{e}\else \.{e}\fi{}nas}}, \bibinfo {author}
  {\bibfnamefont {E.}~\bibnamefont {Prada}}, \bibinfo {author} {\bibfnamefont
  {P.}~\bibnamefont {San-Jose}}, \bibinfo {author} {\bibfnamefont
  {J.}~\bibnamefont {Nyg\aa{}rd}}, \bibinfo {author} {\bibfnamefont
  {P.}~\bibnamefont {Krogstrup}}, \bibinfo {author} {\bibfnamefont
  {R.}~\bibnamefont {Aguado}}, \ and\ \bibinfo {author} {\bibfnamefont {C.~M.}\
  \bibnamefont {Marcus}},\ }\href {\doibase 10.1103/PhysRevB.98.085125}
  {\bibfield  {journal} {\bibinfo  {journal} {Phys. Rev. B}\ }\textbf {\bibinfo
  {volume} {98}},\ \bibinfo {pages} {085125} (\bibinfo {year}
  {2018})}\BibitemShut {NoStop}%
\bibitem [{\citenamefont {Gor'kov}\ and\ \citenamefont
  {Rashba}(2001)}]{PhysRevLett.87.037004}%
  \BibitemOpen
  \bibfield  {author} {\bibinfo {author} {\bibfnamefont {L.~P.}\ \bibnamefont
  {Gor'kov}}\ and\ \bibinfo {author} {\bibfnamefont {E.~I.}\ \bibnamefont
  {Rashba}},\ }\href {\doibase 10.1103/PhysRevLett.87.037004} {\bibfield
  {journal} {\bibinfo  {journal} {Phys. Rev. Lett.}\ }\textbf {\bibinfo
  {volume} {87}},\ \bibinfo {pages} {037004} (\bibinfo {year}
  {2001})}\BibitemShut {NoStop}%
\bibitem [{\citenamefont {Bauer}\ \emph {et~al.}(2004)\citenamefont {Bauer},
  \citenamefont {Hilscher}, \citenamefont {Michor}, \citenamefont {Paul},
  \citenamefont {Scheidt}, \citenamefont {Gribanov}, \citenamefont {Seropegin},
  \citenamefont {No\"el}, \citenamefont {Sigrist},\ and\ \citenamefont
  {Rogl}}]{PhysRevLett.92.027003}%
  \BibitemOpen
  \bibfield  {author} {\bibinfo {author} {\bibfnamefont {E.}~\bibnamefont
  {Bauer}}, \bibinfo {author} {\bibfnamefont {G.}~\bibnamefont {Hilscher}},
  \bibinfo {author} {\bibfnamefont {H.}~\bibnamefont {Michor}}, \bibinfo
  {author} {\bibfnamefont {C.}~\bibnamefont {Paul}}, \bibinfo {author}
  {\bibfnamefont {E.~W.}\ \bibnamefont {Scheidt}}, \bibinfo {author}
  {\bibfnamefont {A.}~\bibnamefont {Gribanov}}, \bibinfo {author}
  {\bibfnamefont {Y.}~\bibnamefont {Seropegin}}, \bibinfo {author}
  {\bibfnamefont {H.}~\bibnamefont {No\"el}}, \bibinfo {author} {\bibfnamefont
  {M.}~\bibnamefont {Sigrist}}, \ and\ \bibinfo {author} {\bibfnamefont
  {P.}~\bibnamefont {Rogl}},\ }\href {\doibase 10.1103/PhysRevLett.92.027003}
  {\bibfield  {journal} {\bibinfo  {journal} {Phys. Rev. Lett.}\ }\textbf
  {\bibinfo {volume} {92}},\ \bibinfo {pages} {027003} (\bibinfo {year}
  {2004})}\BibitemShut {NoStop}%
\bibitem [{\citenamefont {Frigeri}\ \emph {et~al.}(2004)\citenamefont
  {Frigeri}, \citenamefont {Agterberg}, \citenamefont {Koga},\ and\
  \citenamefont {Sigrist}}]{PhysRevLett.92.097001}%
  \BibitemOpen
  \bibfield  {author} {\bibinfo {author} {\bibfnamefont {P.~A.}\ \bibnamefont
  {Frigeri}}, \bibinfo {author} {\bibfnamefont {D.~F.}\ \bibnamefont
  {Agterberg}}, \bibinfo {author} {\bibfnamefont {A.}~\bibnamefont {Koga}}, \
  and\ \bibinfo {author} {\bibfnamefont {M.}~\bibnamefont {Sigrist}},\ }\href
  {\doibase 10.1103/PhysRevLett.92.097001} {\bibfield  {journal} {\bibinfo
  {journal} {Phys. Rev. Lett.}\ }\textbf {\bibinfo {volume} {92}},\ \bibinfo
  {pages} {097001} (\bibinfo {year} {2004})}\BibitemShut {NoStop}%
\bibitem [{\citenamefont {Reyren}\ \emph {et~al.}(2007)\citenamefont {Reyren},
  \citenamefont {Thiel}, \citenamefont {Caviglia}, \citenamefont {Kourkoutis},
  \citenamefont {Hammerl}, \citenamefont {Richter}, \citenamefont {Schneider},
  \citenamefont {Kopp}, \citenamefont {R\"{u}etschi}, \citenamefont {Jaccard},
  \citenamefont {Gabay}, \citenamefont {Muller}, \citenamefont {Triscone},\
  and\ \citenamefont {Mannhart}}]{Reyren07}%
  \BibitemOpen
  \bibfield  {author} {\bibinfo {author} {\bibfnamefont {N.}~\bibnamefont
  {Reyren}}, \bibinfo {author} {\bibfnamefont {S.}~\bibnamefont {Thiel}},
  \bibinfo {author} {\bibfnamefont {A.~D.}\ \bibnamefont {Caviglia}}, \bibinfo
  {author} {\bibfnamefont {L.~F.}\ \bibnamefont {Kourkoutis}}, \bibinfo
  {author} {\bibfnamefont {G.}~\bibnamefont {Hammerl}}, \bibinfo {author}
  {\bibfnamefont {C.}~\bibnamefont {Richter}}, \bibinfo {author} {\bibfnamefont
  {C.~W.}\ \bibnamefont {Schneider}}, \bibinfo {author} {\bibfnamefont
  {T.}~\bibnamefont {Kopp}}, \bibinfo {author} {\bibfnamefont {A.-S.}\
  \bibnamefont {R\"{u}etschi}}, \bibinfo {author} {\bibfnamefont
  {D.}~\bibnamefont {Jaccard}}, \bibinfo {author} {\bibfnamefont
  {M.}~\bibnamefont {Gabay}}, \bibinfo {author} {\bibfnamefont {D.~A.}\
  \bibnamefont {Muller}}, \bibinfo {author} {\bibfnamefont {J.-M.}\
  \bibnamefont {Triscone}}, \ and\ \bibinfo {author} {\bibfnamefont
  {J.}~\bibnamefont {Mannhart}},\ }\href
  {http://science.sciencemag.org/content/317/5842/1196} {\bibfield  {journal}
  {\bibinfo  {journal} {Science}\ }\textbf {\bibinfo {volume} {317}},\ \bibinfo
  {pages} {1196} (\bibinfo {year} {2007})}\BibitemShut {NoStop}%
\bibitem [{\citenamefont {Fujimoto}(2007)}]{doi:10.1143/JPSJ.76.051008}%
  \BibitemOpen
  \bibfield  {author} {\bibinfo {author} {\bibfnamefont {S.}~\bibnamefont
  {Fujimoto}},\ }\href {\doibase 10.1143/JPSJ.76.051008} {\bibfield  {journal}
  {\bibinfo  {journal} {J. Phys. Soc. Jpn.}\ }\textbf {\bibinfo {volume}
  {76}},\ \bibinfo {pages} {051008} (\bibinfo {year} {2007})}\BibitemShut
  {NoStop}%
\bibitem [{\citenamefont {Sato}\ and\ \citenamefont
  {Fujimoto}(2009)}]{PhysRevB.79.094504}%
  \BibitemOpen
  \bibfield  {author} {\bibinfo {author} {\bibfnamefont {M.}~\bibnamefont
  {Sato}}\ and\ \bibinfo {author} {\bibfnamefont {S.}~\bibnamefont
  {Fujimoto}},\ }\href {\doibase 10.1103/PhysRevB.79.094504} {\bibfield
  {journal} {\bibinfo  {journal} {Phys. Rev. B}\ }\textbf {\bibinfo {volume}
  {79}},\ \bibinfo {pages} {094504} (\bibinfo {year} {2009})}\BibitemShut
  {NoStop}%
\bibitem [{\citenamefont {Liu}\ \emph {et~al.}(2014)\citenamefont {Liu},
  \citenamefont {Jain},\ and\ \citenamefont {Liu}}]{PhysRevLett.113.227002}%
  \BibitemOpen
  \bibfield  {author} {\bibinfo {author} {\bibfnamefont {X.}~\bibnamefont
  {Liu}}, \bibinfo {author} {\bibfnamefont {J.~K.}\ \bibnamefont {Jain}}, \
  and\ \bibinfo {author} {\bibfnamefont {C.-X.}\ \bibnamefont {Liu}},\ }\href
  {\doibase 10.1103/PhysRevLett.113.227002} {\bibfield  {journal} {\bibinfo
  {journal} {Phys. Rev. Lett.}\ }\textbf {\bibinfo {volume} {113}},\ \bibinfo
  {pages} {227002} (\bibinfo {year} {2014})}\BibitemShut {NoStop}%
\bibitem [{\citenamefont {Smidman}\ \emph {et~al.}(2017)\citenamefont
  {Smidman}, \citenamefont {Salamon}, \citenamefont {Yuan},\ and\ \citenamefont
  {Agterberg}}]{0034-4885-80-3-036501}%
  \BibitemOpen
  \bibfield  {author} {\bibinfo {author} {\bibfnamefont {M.}~\bibnamefont
  {Smidman}}, \bibinfo {author} {\bibfnamefont {M.~B.}\ \bibnamefont
  {Salamon}}, \bibinfo {author} {\bibfnamefont {H.~Q.}\ \bibnamefont {Yuan}}, \
  and\ \bibinfo {author} {\bibfnamefont {D.~F.}\ \bibnamefont {Agterberg}},\
  }\href {http://stacks.iop.org/0034-4885/80/i=3/a=036501} {\bibfield
  {journal} {\bibinfo  {journal} {Rep. Prog. Phys.}\ }\textbf {\bibinfo
  {volume} {80}},\ \bibinfo {pages} {036501} (\bibinfo {year}
  {2017})}\BibitemShut {NoStop}%
\bibitem [{Note1()}]{Note1}%
  \BibitemOpen
  \bibinfo {note} {Spatial variations of $\Delta (x)$ at the NS interfaces\cite
  {Fer18} do not alter the main conclusions of this work.}\BibitemShut {Stop}%
\bibitem [{\citenamefont {McMillan}(1968)}]{PhysRev.175.559}%
  \BibitemOpen
  \bibfield  {author} {\bibinfo {author} {\bibfnamefont {W.~L.}\ \bibnamefont
  {McMillan}},\ }\href {\doibase 10.1103/PhysRev.175.559} {\bibfield  {journal}
  {\bibinfo  {journal} {Phys. Rev.}\ }\textbf {\bibinfo {volume} {175}},\
  \bibinfo {pages} {559} (\bibinfo {year} {1968})}\BibitemShut {NoStop}%
\bibitem [{\citenamefont {Pannetier}\ and\ \citenamefont
  {Courtois}(2000)}]{Pannetier2000}%
  \BibitemOpen
  \bibfield  {author} {\bibinfo {author} {\bibfnamefont {B.}~\bibnamefont
  {Pannetier}}\ and\ \bibinfo {author} {\bibfnamefont {H.}~\bibnamefont
  {Courtois}},\ }\href {\doibase 10.1023/A:1004635226825} {\bibfield  {journal}
  {\bibinfo  {journal} {J. Low Temp. Phys.}\ }\textbf {\bibinfo {volume}
  {118}},\ \bibinfo {pages} {599} (\bibinfo {year} {2000})}\BibitemShut
  {NoStop}%
\bibitem [{\citenamefont {Klapwijk}(2004)}]{Klapwijk2004}%
  \BibitemOpen
  \bibfield  {author} {\bibinfo {author} {\bibfnamefont {T.~M.}\ \bibnamefont
  {Klapwijk}},\ }\href {\doibase 10.1007/s10948-004-0773-0} {\bibfield
  {journal} {\bibinfo  {journal} {Journal of Superconductivity}\ }\textbf
  {\bibinfo {volume} {17}},\ \bibinfo {pages} {593} (\bibinfo {year}
  {2004})}\BibitemShut {NoStop}%
\bibitem [{\citenamefont {Covaci}\ and\ \citenamefont
  {Marsiglio}(2006)}]{PhysRevB.73.014503}%
  \BibitemOpen
  \bibfield  {author} {\bibinfo {author} {\bibfnamefont {L.}~\bibnamefont
  {Covaci}}\ and\ \bibinfo {author} {\bibfnamefont {F.}~\bibnamefont
  {Marsiglio}},\ }\href {\doibase 10.1103/PhysRevB.73.014503} {\bibfield
  {journal} {\bibinfo  {journal} {Phys. Rev. B}\ }\textbf {\bibinfo {volume}
  {73}},\ \bibinfo {pages} {014503} (\bibinfo {year} {2006})}\BibitemShut
  {NoStop}%
\bibitem [{\citenamefont {Black-Schaffer}\ and\ \citenamefont
  {Doniach}(2010)}]{PhysRevB.81.014517}%
  \BibitemOpen
  \bibfield  {author} {\bibinfo {author} {\bibfnamefont {A.~M.}\ \bibnamefont
  {Black-Schaffer}}\ and\ \bibinfo {author} {\bibfnamefont {S.}~\bibnamefont
  {Doniach}},\ }\href {\doibase 10.1103/PhysRevB.81.014517} {\bibfield
  {journal} {\bibinfo  {journal} {Phys. Rev. B}\ }\textbf {\bibinfo {volume}
  {81}},\ \bibinfo {pages} {014517} (\bibinfo {year} {2010})}\BibitemShut
  {NoStop}%
\bibitem [{\citenamefont {B\'eri}\ \emph {et~al.}(2008)\citenamefont {B\'eri},
  \citenamefont {Bardarson},\ and\ \citenamefont
  {Beenakker}}]{PhysRevB.77.045311}%
  \BibitemOpen
  \bibfield  {author} {\bibinfo {author} {\bibfnamefont {B.}~\bibnamefont
  {B\'eri}}, \bibinfo {author} {\bibfnamefont {J.~H.}\ \bibnamefont
  {Bardarson}}, \ and\ \bibinfo {author} {\bibfnamefont {C.~W.~J.}\
  \bibnamefont {Beenakker}},\ }\href {\doibase 10.1103/PhysRevB.77.045311}
  {\bibfield  {journal} {\bibinfo  {journal} {Phys. Rev. B}\ }\textbf {\bibinfo
  {volume} {77}},\ \bibinfo {pages} {045311} (\bibinfo {year}
  {2008})}\BibitemShut {NoStop}%
\bibitem [{\citenamefont {Dimitrova}\ and\ \citenamefont
  {Feigel'man}(2006)}]{Dimitrova2006}%
  \BibitemOpen
  \bibfield  {author} {\bibinfo {author} {\bibfnamefont {O.~V.}\ \bibnamefont
  {Dimitrova}}\ and\ \bibinfo {author} {\bibfnamefont {M.~V.}\ \bibnamefont
  {Feigel'man}},\ }\href {\doibase 10.1134/S1063776106040157} {\bibfield
  {journal} {\bibinfo  {journal} {JETP}\ }\textbf {\bibinfo {volume} {102}},\
  \bibinfo {pages} {652} (\bibinfo {year} {2006})}\BibitemShut {NoStop}%
\bibitem [{\citenamefont {San-Jos\'{e}}\ \emph {et~al.}(2013)\citenamefont
  {San-Jos\'{e}}, \citenamefont {Cayao}, \citenamefont {Prada},\ and\
  \citenamefont {Aguado}}]{SanJoseNJP:13}%
  \BibitemOpen
  \bibfield  {author} {\bibinfo {author} {\bibfnamefont {P.}~\bibnamefont
  {San-Jos\'{e}}}, \bibinfo {author} {\bibfnamefont {J.}~\bibnamefont {Cayao}},
  \bibinfo {author} {\bibfnamefont {E.}~\bibnamefont {Prada}}, \ and\ \bibinfo
  {author} {\bibfnamefont {R.}~\bibnamefont {Aguado}},\ }\href
  {http://stacks.iop.org/1367-2630/15/i=7/a=075019} {\bibfield  {journal}
  {\bibinfo  {journal} {New J. of Phys.}\ }\textbf {\bibinfo {volume} {15}},\
  \bibinfo {pages} {075019} (\bibinfo {year} {2013})}\BibitemShut {NoStop}%
\bibitem [{\citenamefont {Nakhmedov}\ \emph {et~al.}(2017)\citenamefont
  {Nakhmedov}, \citenamefont {Alekperov}, \citenamefont {Tatardar},
  \citenamefont {Shukrinov}, \citenamefont {Rahmonov},\ and\ \citenamefont
  {Sengupta}}]{PhysRevB.96.014519}%
  \BibitemOpen
  \bibfield  {author} {\bibinfo {author} {\bibfnamefont {E.}~\bibnamefont
  {Nakhmedov}}, \bibinfo {author} {\bibfnamefont {O.}~\bibnamefont
  {Alekperov}}, \bibinfo {author} {\bibfnamefont {F.}~\bibnamefont {Tatardar}},
  \bibinfo {author} {\bibfnamefont {Y.~M.}\ \bibnamefont {Shukrinov}}, \bibinfo
  {author} {\bibfnamefont {I.}~\bibnamefont {Rahmonov}}, \ and\ \bibinfo
  {author} {\bibfnamefont {K.}~\bibnamefont {Sengupta}},\ }\href {\doibase
  10.1103/PhysRevB.96.014519} {\bibfield  {journal} {\bibinfo  {journal} {Phys.
  Rev. B}\ }\textbf {\bibinfo {volume} {96}},\ \bibinfo {pages} {014519}
  (\bibinfo {year} {2017})}\BibitemShut {NoStop}%
\bibitem [{\citenamefont {van Heck}\ \emph {et~al.}(2017)\citenamefont {van
  Heck}, \citenamefont {V\"ayrynen},\ and\ \citenamefont
  {Glazman}}]{PhysRevB.96.075404}%
  \BibitemOpen
  \bibfield  {author} {\bibinfo {author} {\bibfnamefont {B.}~\bibnamefont {van
  Heck}}, \bibinfo {author} {\bibfnamefont {J.~I.}\ \bibnamefont {V\"ayrynen}},
  \ and\ \bibinfo {author} {\bibfnamefont {L.~I.}\ \bibnamefont {Glazman}},\
  }\href {\doibase 10.1103/PhysRevB.96.075404} {\bibfield  {journal} {\bibinfo
  {journal} {Phys. Rev. B}\ }\textbf {\bibinfo {volume} {96}},\ \bibinfo
  {pages} {075404} (\bibinfo {year} {2017})}\BibitemShut {NoStop}%
\bibitem [{\citenamefont {Cayao}\ \emph {et~al.}(2017)\citenamefont {Cayao},
  \citenamefont {San-Jose}, \citenamefont {Black-Schaffer}, \citenamefont
  {Aguado},\ and\ \citenamefont {Prada}}]{Cayao17b}%
  \BibitemOpen
  \bibfield  {author} {\bibinfo {author} {\bibfnamefont {J.}~\bibnamefont
  {Cayao}}, \bibinfo {author} {\bibfnamefont {P.}~\bibnamefont {San-Jose}},
  \bibinfo {author} {\bibfnamefont {A.~M.}\ \bibnamefont {Black-Schaffer}},
  \bibinfo {author} {\bibfnamefont {R.}~\bibnamefont {Aguado}}, \ and\ \bibinfo
  {author} {\bibfnamefont {E.}~\bibnamefont {Prada}},\ }\href {\doibase
  10.1103/PhysRevB.96.205425} {\bibfield  {journal} {\bibinfo  {journal} {Phys.
  Rev. B}\ }\textbf {\bibinfo {volume} {96}},\ \bibinfo {pages} {205425}
  (\bibinfo {year} {2017})}\BibitemShut {NoStop}%
\bibitem [{\citenamefont {Cayao}\ \emph {et~al.}(2018)\citenamefont {Cayao},
  \citenamefont {Black-Schaffer}, \citenamefont {Prada},\ and\ \citenamefont
  {Aguado}}]{cayao18a}%
  \BibitemOpen
  \bibfield  {author} {\bibinfo {author} {\bibfnamefont {J.}~\bibnamefont
  {Cayao}}, \bibinfo {author} {\bibfnamefont {A.~M.}\ \bibnamefont
  {Black-Schaffer}}, \bibinfo {author} {\bibfnamefont {E.}~\bibnamefont
  {Prada}}, \ and\ \bibinfo {author} {\bibfnamefont {R.}~\bibnamefont
  {Aguado}},\ }\href {https://doi.org/10.3762/bjnano.9.127} {\bibfield
  {journal} {\bibinfo  {journal} {Beilstein J. Nanotechnol.}\ }\textbf
  {\bibinfo {volume} {9}},\ \bibinfo {pages} {1339} (\bibinfo {year}
  {2018})}\BibitemShut {NoStop}%
\bibitem [{\citenamefont {Gill}\ \emph {et~al.}(2016)\citenamefont {Gill},
  \citenamefont {Damasco}, \citenamefont {Car}, \citenamefont {Bakkers},\ and\
  \citenamefont {Mason}}]{doi:10.1063/1.4971394}%
  \BibitemOpen
  \bibfield  {author} {\bibinfo {author} {\bibfnamefont {S.~T.}\ \bibnamefont
  {Gill}}, \bibinfo {author} {\bibfnamefont {J.}~\bibnamefont {Damasco}},
  \bibinfo {author} {\bibfnamefont {D.}~\bibnamefont {Car}}, \bibinfo {author}
  {\bibfnamefont {E.~P. A.~M.}\ \bibnamefont {Bakkers}}, \ and\ \bibinfo
  {author} {\bibfnamefont {N.}~\bibnamefont {Mason}},\ }\href {\doibase
  10.1063/1.4971394} {\bibfield  {journal} {\bibinfo  {journal} {Applied
  Physics Letters}\ }\textbf {\bibinfo {volume} {109}},\ \bibinfo {pages}
  {233502} (\bibinfo {year} {2016})}\BibitemShut {NoStop}%
\bibitem [{\citenamefont {Fleckenstein}\ \emph
  {et~al.}(2018{\natexlab{b}})\citenamefont {Fleckenstein}, \citenamefont
  {Dom\'{\i}nguez}, \citenamefont {Traverso~Ziani},\ and\ \citenamefont
  {Trauzettel}}]{Fer18}%
  \BibitemOpen
  \bibfield  {author} {\bibinfo {author} {\bibfnamefont {C.}~\bibnamefont
  {Fleckenstein}}, \bibinfo {author} {\bibfnamefont {F.}~\bibnamefont
  {Dom\'{\i}nguez}}, \bibinfo {author} {\bibfnamefont {N.}~\bibnamefont
  {Traverso~Ziani}}, \ and\ \bibinfo {author} {\bibfnamefont {B.}~\bibnamefont
  {Trauzettel}},\ }\href {\doibase 10.1103/PhysRevB.97.155425} {\bibfield
  {journal} {\bibinfo  {journal} {Phys. Rev. B}\ }\textbf {\bibinfo {volume}
  {97}},\ \bibinfo {pages} {155425} (\bibinfo {year}
  {2018}{\natexlab{b}})}\BibitemShut {NoStop}%
\bibitem [{\citenamefont {Furusaki}\ and\ \citenamefont
  {Tsukada}(1991)}]{FURUSAKI1991299}%
  \BibitemOpen
  \bibfield  {author} {\bibinfo {author} {\bibfnamefont {A.}~\bibnamefont
  {Furusaki}}\ and\ \bibinfo {author} {\bibfnamefont {M.}~\bibnamefont
  {Tsukada}},\ }\href {\doibase https://doi.org/10.1016/0038-1098(91)90201-6}
  {\bibfield  {journal} {\bibinfo  {journal} {Solid State Commun.}\ }\textbf
  {\bibinfo {volume} {78}},\ \bibinfo {pages} {299 } (\bibinfo {year}
  {1991})}\BibitemShut {NoStop}%
\bibitem [{\citenamefont {Furusaki}\ \emph {et~al.}(1991)\citenamefont
  {Furusaki}, \citenamefont {Takayanagi},\ and\ \citenamefont
  {Tsukada}}]{PhysRevLett.67.132}%
  \BibitemOpen
  \bibfield  {author} {\bibinfo {author} {\bibfnamefont {A.}~\bibnamefont
  {Furusaki}}, \bibinfo {author} {\bibfnamefont {H.}~\bibnamefont
  {Takayanagi}}, \ and\ \bibinfo {author} {\bibfnamefont {M.}~\bibnamefont
  {Tsukada}},\ }\href {\doibase 10.1103/PhysRevLett.67.132} {\bibfield
  {journal} {\bibinfo  {journal} {Phys. Rev. Lett.}\ }\textbf {\bibinfo
  {volume} {67}},\ \bibinfo {pages} {132} (\bibinfo {year} {1991})}\BibitemShut
  {NoStop}%
\end{thebibliography}%
%\onecolumngrid
%\clearpage
%\twocolumngrid

% -------------------------------------- %
% APPENDIX:
% -------------------------------------- %
\appendix

% GREEN'S FUNCTIONS:
\section{Retarded and advanced Green's functions}
\label{AppG}
In this appendix we briefly outline the method we use to calculate the pairing amplitudes. We follow Ref.~[\onlinecite{PhysRevB.96.155426}] and first construct the retarded Green's function $G^{r}(x,x',\omega)$ with outgoing boundary conditions in each region  from  the scattering processes at the interface.\cite{PhysRev.175.559}
Thus, the retarded Green's function reads as
\begin{widetext}
\begin{equation}
\label{RGF}
\begin{split}
G^{r}(x,x',\omega)=
\begin{cases}
\Psi_{1}(x)[\alpha_{11}\tilde{\Psi}_{5}^{T}(x')+\alpha_{12}\tilde{\Psi}_{6}^{T}(x')+\alpha_{13}\tilde{\Psi}_{7}^{T}(x')+\alpha_{14}\tilde{\Psi}_{8}^{T}(x')]\\
+
\Psi_{2}(x)[\alpha_{21}\tilde{\Psi}_{5}^{T}(x')+\alpha_{22}\tilde{\Psi}_{6}^{T}(x')+\alpha_{23}\tilde{\Psi}_{7}^{T}(x')+\alpha_{24}\tilde{\Psi}_{8}^{T}(x')]\\
+
\Psi_{3}(x)[\alpha_{31}\tilde{\Psi}_{5}^{T}(x')+\alpha_{32}\tilde{\Psi}_{6}^{T}(x')+\alpha_{33}\tilde{\Psi}_{7}^{T}(x')+\alpha_{34}\tilde{\Psi}_{8}^{T}(x')]\\
+
\Psi_{4}(x)[\alpha_{41}\tilde{\Psi}_{5}^{T}(x')+\alpha_{42}\tilde{\Psi}_{6}^{T}(x')+\alpha_{43}\tilde{\Psi}_{7}^{T}(x')+\alpha_{44}\tilde{\Psi}_{8}^{T}(x')]
\,,\quad x>x'&\\
\Psi_{5}(x)[\beta_{11}\tilde{\Psi}_{1}^{T}(x')+\beta_{12}\tilde{\Psi}_{2}^{T}(x')+\beta_{13}\tilde{\Psi}_{3}^{T}(x')+\beta_{14}\tilde{\Psi}_{4}^{T}(x')]\\
+\Psi_{6}(x)[\beta_{21}\tilde{\Psi}_{1}^{T}(x')+\beta_{22}\tilde{\Psi}_{2}^{T}(x')+\beta_{23}\tilde{\Psi}_{3}^{T}(x')+\beta_{24}\tilde{\Psi}_{4}^{T}(x')]\\
+\Psi_{7}(x)[\beta_{31}\tilde{\Psi}_{1}^{T}(x')+\beta_{32}\tilde{\Psi}_{2}^{T}(x')+\beta_{33}\tilde{\Psi}_{3}^{T}(x')+\beta_{34}\tilde{\Psi}_{4}^{T}(x')]\\
+\Psi_{8}(x)[\beta_{41}\tilde{\Psi}_{1}^{T}(x')+\beta_{42}\tilde{\Psi}_{2}^{T}(x')+\beta_{43}\tilde{\Psi}_{3}^{T}(x')+\beta_{44}\tilde{\Psi}_{4}^{T}(x')]\,, \quad x<x'&
\end{cases}
\end{split}
\end{equation}
\end{widetext}
where $\Psi_{i}$ represent the scattering processes at the interface of the junction under investigation and they are found after solving the BdG equations given by $H_{\rm BdG}(k)$; their specific form for NS and SNS junctions is given in subsequent appendices. Moreover, $\tilde{\Psi}_{i}$ correspond to the conjugated processes found after solving the BdG equations using $H_{\rm BdG}(-k)^{*}$ instead $H_{\rm BdG}(k)$.
The Green's function $G^{r}$ includes eight scattering processes $\Psi_{i}$ assuming that spin-up and -down particles are involved in the problem. 
The first four processes $\Psi_{1,2, 3,4}$ account for right moving particles  (up and down electrons, up and down holes) from the left region towards the interface. The last four processes $\Psi_{5,6, 7,8}$ correspond to left moving particles from the right region towards the interface (up and down electrons, up and down holes).
If spin is not an active degree of freedom, half of the scattering states drop out of the problem. 
Equation \eqref{RGF} is a generalization of the expression given by \citet{PhysRev.175.559} and later by Furusaki and Tsukada\cite{FURUSAKI1991299} and the method allows the calculation of the Green's function in the left and right regions separately. We thus do not consider a situation that accounts for the total Green's function of the left region coupled to the right one.

The coefficients $\alpha_{ij}$ and $\beta_{mn}$ in Eq.~\eqref{RGF} are found from the continuity of the Green's function
\begin{equation}
[\omega-H_{\rm BdG}(x)]G^{r}(x,x',\omega)=\delta(x-x')\,,
\end{equation}
where $H_{\rm BdG}$ is the BdG Hamiltonian of the system defined in Eq.\,(\ref{HBdG}).
Then, by integrating around $x=x'$ we obtain
\begin{equation}
\label{conditionGRSO}
\begin{split}
&[G^{r}(x>x')]_{x=x'}=[G^{r}(x<x')]_{x=x'}\,,\\
&[\partial_{x}G^{r}(x>x')]_{x=x'}-[\partial_{x}G^{r}(x<x')]_{x=x'}=\eta\sigma_{0}\tau_{z}\,,%=\eta
%\begin{pmatrix}
%1&0&0&0\\
%0&1&0&0\\
%0&0&-1&0\\
%0&0&0&-1
%\end{pmatrix}\,,
\end{split}
\end{equation}
where $\eta=2m/\hbar^{2}$ and $\sigma_{i}$ and $\tau_{i}$ are $i$-Pauli matrices in spin and electron-hole spaces, respectively. 

In general, the Green's function, either in the left or right region, is a $2\times2$ matrix in electron-hole space,
\begin{equation}
\label{GF}
G^{r}(x,x',\omega)=
\begin{pmatrix}
G^{r}_{ee}&G^{r}_{eh}\\
G^{r}_{he}&G^{r}_{hh}
\end{pmatrix}\,,
\end{equation}
where each element is a matrix. If spin is an active degree of freedom, and using the basis of $H_{\rm BdG}$ they individual Green's function components read  as
\begin{equation}
\begin{split}
G_{ee}^{r}(x,x',\omega)&=
\begin{pmatrix}
[G^{r}_{ee}]_{\uparrow\uparrow}&[G^{r}_{ee}]_{\uparrow\downarrow}\\
[G^{r}_{ee}]_{\downarrow\uparrow}&[G^{r}_{ee}]_{\downarrow\downarrow}
\end{pmatrix}\,,\\ 
G_{eh}^{r}(x,x',\omega)&=
\begin{pmatrix}
[G^{r}_{eh}]_{\uparrow\uparrow}&[G^{r}_{eh}]_{\uparrow\downarrow}\\
[G^{r}_{eh}]_{\downarrow\uparrow}&[G^{r}_{eh}]_{\downarrow\downarrow}
\end{pmatrix}\,.
\end{split}
\end{equation}
Electron-hole symmetry connects the electron-electron and hole-hole blocks and also the electron-hole and hole-electron blocks. Thus, it is enough to calculate $G_{ee}^{r}$ and $G^{r}_{eh}$. Notice that if spin is not active, then the Green's function in Eq.~\eqref{GF} is a $2\times2$ matrix in electron-hole space but the electron-electron and electron-hole components are just numbers.

We are here interested in the pairing amplitudes, which are obtained from the anomalous electron-hole element $G_{eh}^{r}$.
The spin symmetry is decomposed according to Eq.~\eqref{decomposeSPIN} in the main text, where we obtain the pairing amplitudes as
\begin{equation}
\label{pairingfunctions}
\begin{split}
f_{0}^{r}(x,x',\omega)&= \frac{[G^{r}_{eh}]_{\uparrow\downarrow}-[G^{r}_{eh}]_{\downarrow\uparrow}}{2}\,,\\
f_{1}^{r}(x,x',\omega)&= \frac{[G^{r}_{eh}]_{\downarrow\downarrow}-[G^{r}_{eh}]_{\uparrow\uparrow}}{2}\,,\\
f_{2}^{r}(x,x',\omega)&= \frac{[G^{r}_{eh}]_{\downarrow\downarrow}+[G^{r}_{eh}]_{\uparrow\uparrow}}{2i}\,,\\
f_{3}^{r}(x,x',\omega)&= \frac{[G^{r}_{eh}]_{\uparrow\downarrow}+[G^{r}_{eh}]_{\downarrow\uparrow}}{2}\,.\\
\end{split}
\end{equation}
Here $f_{0}^{r}$ corresponds to spin-singlet ($\uparrow\downarrow-\downarrow\uparrow$), $f_{1,2}^{r}$ equal spin-triplet ($\downarrow\downarrow\pm\uparrow\uparrow$), and $f_{3}^{r}$ mixed spin-triplet ($\uparrow\downarrow+\downarrow\uparrow$) amplitudes.

%%%%%%%%%%%%%%%%%%%%%%%%
%           ZERO SPIN-ORBIT COUPLING         %
%%%%%%%%%%%%%%%%%%%% %%%%
\section{Zero spin-orbit coupling}
\label{zeroSOC}
In this appendix we revisit the emergence of odd-frequency superconducting pairing in junctions without SO coupling. Although the induced  odd-frequency pairing in NS junctions is well established within the quasiclassical Usadel and Eilenberger frameworks,\cite{PhysRevB.76.054522,PhysRevLett.98.037003,Eschrig2007} a detailed scattering approach has not yet been carried out. 
We believe our approach is useful, yet simple to visualize the emergence of odd-frequency superconducting pairing and especially establish its relation with the scattering processes at the interfaces.
As explained above, we first calculate the Green's functions, a $2\times2$ matrix in either the N or S regions in electron-hole subspace as spin is now not an active degree of freedom, from scattering states and then obtain the pairing amplitudes. 

With spin is not actively involved in the problem, the system's Hamiltonian is given by 
\begin{equation}
\label{BdG2}
\begin{split}
H_{\rm BdG}(x)&=\begin{pmatrix}
H_{0}&\Delta(x)\\
\Delta^{\dagger}(x)&-H_{0}
\end{pmatrix}\,,
\end{split}
\end{equation}
where $H_{0}(x)=\frac{\hbar^{2}k^{2}}{2m}-\mu(x)$. For NS junctions 
\begin{equation}
\label{NSDelta}
\Delta(x)=
\begin{cases}
     0\,, & x<0, \\
     \Delta\,, & x>0,
\end{cases}
\end{equation} 
while for short SNS junctions 
\begin{equation}
\label{SNSDelta}
\Delta(x)=
\begin{cases}
     \Delta\,, & x<0, \\
     \Delta{\rm e}^{i\phi}\,, & x>0.
\end{cases}
\end{equation} 
The chemical potential $\mu(x)$ can, in principle, take different values in N and S.
First, we discuss semi-infinite NS junctions with the interface located at $x=0$ and then short SNS junctions

%%%%%%%%%%%%%%%%%%%%%%%%%%%%%
%            NS JUNCTIONS AT ZERO SOC                       %
%%%%%%%%%%%%%%%%%%%%%%%%%%%%%
\subsection{NS junction}
\label{NSApp}
Here, we discuss NS junctions, whose interface is located at $x=0$ in the limit of vanishing SO coupling. In this case, the scattering processes at the  interface read as
\begin{equation}
\label{normalw}
\begin{split}
\Psi_{1}(x)&=
\begin{cases}
     \phi_{1}^{N}\,{\rm e}^{ik_{e}x}+a_{1}\phi_{3}^{N}\,{\rm e}^{ik_{h}x} +b_{1}\phi_{2}^{N}\,{\rm e}^{-ik_{e}x},\,x<0&  \\
    c_{1}  \phi_{1}^{S}\,{\rm e}^{ik_{e}^{S}x} +d_{1}  \phi_{4}^{S}\,{\rm e}^{-ik_{h}^{S}x},\,x>0 & 
\end{cases}\\
\Psi_{2}(x)&=
\begin{cases}
     \phi_{4}^{N}\,{\rm e}^{-ik_{h}x}+a_{2}\phi_{2}^{N}\,{\rm e}^{-ik_{e}x} +b_{2}\phi_{3}^{N}\,{\rm e}^{ik_{h}x},\,x<0& \\
    c_{2}  \phi_{4}^{S}\,{\rm e}^{-ik_{h}^{S}x} +    d_{2}  \phi_{1}^{S}\,{\rm e}^{ik_{e}^{S}x}, \,x>0 & 
\end{cases}\\
\Psi_{3}(x)&=
\begin{cases}
    c_{3}  \phi_{2}^{N}\,{\rm e}^{-ik_{e}x} +d_{3}  \phi_{3}^{N}\,{\rm e}^{ik_{h}x},\, x<0&\\
     \phi_{2}^{S}\,{\rm e}^{-ik_{e}^{S}x}+a_{3}\phi_{4}^{S}\,{\rm e}^{-ik_{h}^{S}x}+b_{3}\phi_{1}^{S}\,{\rm e}^{ik_{e}^{S}x},\,x>0&  
\end{cases}\\
\Psi_{4}(x)&=
\begin{cases}
    c_{4}  \phi_{3}^{N}\,{\rm e}^{ik_{h}x} +    d_{4}  \phi_{2}^{N}\,{\rm e}^{-ik_{e}x},\,x<0  & \\
\phi_{3}^{S}\,{\rm e}^{ik_{h}^{S}x}+a_{4}\phi_{1}^{S}\,{\rm e}^{ik_{e}^{S}x}+b_{4}\phi_{4}^{S}\,{\rm e}^{-ik_{h}^{S}x},\,x>0&
\end{cases}
\end{split}
\end{equation}
where 
\begin{equation}
\begin{split}
\phi_{1,2}^{N}&=
\begin{pmatrix}
1\\
0
\end{pmatrix}\,,\quad\phi_{3,4}^{N}=
\begin{pmatrix}
0\\
1
\end{pmatrix}\,,\\
\phi_{1,2}^{S}&=
\begin{pmatrix}
u\\
v
\end{pmatrix}\,,\quad \phi_{3,4}^{S}=
\begin{pmatrix}
v\\
u
\end{pmatrix}\,,
\end{split}
\end{equation}
and
\begin{equation}
\begin{split}
k_{e,h}&= \sqrt{\frac{2m}{\hbar^{2}}(\mu_{N}\pm \omega)}\,,\\
q_{e,h}&=\sqrt{\frac{2m}{\hbar^{2}}\bigg[\mu_{S}\pm \sqrt{\omega^{2}-\Delta^{2}}\bigg]}\,,\\
u&=\sqrt{\frac{1}{2}\bigg[1+\frac{\sqrt{\omega^{2}+\Delta^{2}}}{\omega}\bigg]}\,,\\
v&=\sqrt{\frac{1}{2}\bigg[1-\frac{\sqrt{\omega^{2}+\Delta^{2}}}{\omega}\bigg]}\,.
\end{split}
\end{equation}
The conjugated processes needed for the Green's functions are found after solving for $H_{\rm BdG}^{*}(-k)$. In this special case without SO coupling, however, the solutions are the same as previous equations, namely $\tilde{\Psi}_{i}=\Psi_{i}$ because $\tilde{\phi}_{i}^{N(S)}=\phi_{i}^{N(S)}$, also resulting in the same coefficients, namely, $a_{i}=\tilde{a}_{i}$ and so on.
The coefficients in the scattering states are found from the conditions established when integrating the BdG equations,
\begin{equation}
\begin{split}
\label{conditions}
\Big[\partial_{x}\Psi(x>0)] -[\partial_{x}\Psi(x<0)]&=Z\Psi(x=0)\,,\\
[\Psi_{i}(x<0)]&=[\Psi_{i}(x>0)]\,,
\end{split}
\end{equation}
where $Z=2mV/\hbar^{2}$ is the interface transparency, where for generality we consider a delta potential $V(x)=Z\delta(x)$ at $x=0$. 

\subsubsection{Green's function in N}
The next step consists on finding the Green's function $G^{r}$ following Eqs.\,(\ref{RGF}), where we only include four scattering processes as spin is not involved in the problem. For doing so we need to find the coefficients $\alpha$ and $\beta$ from the continuity and discontinuity of $G^{r}$ at $x=x'$ given by Eq.\,(\ref{conditionGRSO}), but without $\sigma_{0}$. After some algebra, we finally obtain the elements of the Green's function
\begin{equation}
\begin{split}
G^{r}_{ee}(x,x',\omega)&=\frac{\eta}{2ik_{e}}{\rm e}^{ik_{e}|x-x'|}+\frac{\eta b_{1}}{2ik_{e}}{\rm e}^{-i(x+x')k_{e}}\,,\\
G^{r}_{eh}(x,x',\omega)&=\frac{\eta}{2ik_{h}}\bar{A}{\rm e}^{-i(k_{e}x-k_{h}x')}\,,\\
G^{r}_{he}(x,x',\omega)&=\frac{\eta}{2ik_{e}}\bar{A}{\rm e}^{i(k_{h}x-k_{e}x')}\,,\\
G^{r}_{hh}(x,x',\omega)&=\frac{\eta}{2ik_{h}}{\rm e}^{-ik_{h}|x-x'|}+\frac{\eta b_{2}}{2ik_{h}}{\rm e}^{i(x+x')k_{h}}\,,
\end{split}
\end{equation}
where $\bar{A}=(a_{1}/k_{e})=(a_{2}/k_{h})$ depends on the Andreev reflection coefficient  for a right moving electron from N, $a_1$, where
\begin{equation}
\begin{split}
a_{1}&=\frac{2k_{e}(k_{e}^{S}+k_{h}^{S})uv}{\bar{D}}\,,\quad 
a_{2}=\frac{2k_{h}(k_{e}^{S}+k_{h}^{S})uv}{\bar{D}}\,,\\
b_{1}&=\frac{P}{\bar{D}}\,,
\quad b_{2}=\frac{Q}{\bar{D}}\,,\\
\bar{D}&=u^{2}\Big[(k_{e}+k_{e}^{S}+iZ)(k_{h}+k_{h}^{S}-iZ)\Big]\\
&+v^{2}\Big[(k_{e}^{S}-k_{h}+iZ)(k_{e}-k_{h}^{S}+iZ)\Big]\,,\\
P&=u^{2}\Big[(k_{e}-k_{e}^{S}-iZ)(k_{h}+k_{h}^{S}-iZ)\Big]\\
&+v^{2}\Big[(k_{e}^{S}-k_{h}+iZ)(k_{e}+k_{h}^{S}-iZ)\Big]\,,\\
Q&=u^{2}\Big[(k_{e}+k_{h}^{S}+iZ)(k_{h}-k_{h}^{S}+iZ)\Big]\\
&+v^{2}\Big[(k_{e}^{S}+k_{h}+iZ)(k_{h}^{S}-k_{e}-iZ)\Big]\,.
\end{split}
\end{equation}
Note that in these expressions we have omitted the normalization constants in the Andreev coefficients $a_{i}$ as they simplify out in the general expression of the Green's functions.

The pairing amplitudes are determined by the anomalous terms, which in this case are not matrices, but just numbers due to the absence of spin.
Thus, the electron-hole Green's function is the pairing amplitude and reads as
\begin{equation}
\label{eq0}
f_{0}^{r}(x,x'\omega)=\frac{\eta}{2i}\bar{A}\,{\rm e}^{-i(k_{e}x-k_{h}x')}\,,
\end{equation} 
where $k_{e,h}=k_{\mu_{N}}\sqrt{1\pm \frac{\omega}{\mu_{N}}}$ with $k_{\mu_{N}}=\sqrt{2m\mu_{N}/\hbar^{2}}$.

Notice that the pairing amplitude is proportional to the Andreev reflection coefficient through $\bar{A}$. The exponential term mixes electron and hole wave vectors at different positions, introducing a mixing of spatial parity. In this case, the simple expression given above describes the effect of the superconducting region on the normal region. Since there is no active spin mechanism in our assumption, previous expression has the same spin-singlet symmetry as the initial superconductor before contacting with the normal region. 

Further insight is obtained by writing  the even and odd-frequency pairing components  for large $\mu_{N}$, where we approximate $k_{e,h}\approx k_{\mu}\big(1\pm \frac{\omega}{2\mu_{N}}\big)$,
\begin{equation}
\label{EQ0}
\begin{split}
f_{0}^{r,{\rm O}}(x,x',\omega)&=-\frac{\eta}{2}\bar{A}\,{\rm e}^{-ik^{N}(x+x')}{\rm sin}[k_{\mu_{N}}(x-x')]\,,\\
f_{0}^{r,{\rm E}}(x,x',\omega)&=\frac{\eta}{2i}\bar{A}\,{\rm e}^{-ik^{N}(x+x')}{\rm cos}[k_{\mu_{N}}(x-x')]
\end{split}
\end{equation}
where, $k^{N}=\omega k_{\mu_{N}}/(2\mu_{N})$. Since we consider a spin-singlet $\Delta$, the only possibilities for the pairing classes are the spin-singlet ESE and OSO symmetries, respectively. 
These are proportional to the Andreev reflection, through the coefficient $a_{1}$, an effect which is at the core of the proximity effect\cite{Pannetier2000,Klapwijk2004} and, quite interesting,  both even- and odd-frequency pairing coexist with a dominant behavior of one or the other depending on the modulation factors, ${\rm sine}$ or ${\rm cosine}$. Locally, at $x=x'$, however, only even-frequency pairing exists and is maximum, while odd-frequency can dominate non-locally.
In fact, it is hard to think about pair formation at the same position ($x=x'$) after Andreev reflection at an interface. Thus, we believe the most common formation of pairs occurs in fact when $x\neq x'$, a condition that directly enables the existence of odd-frequency component as we then always have a contribution from OSO. In particular, when $k_{\mu_{N}}(x-x')=\frac{\pi}{2}+\pi n$ for $n=0,1,2,\cdots$, OSO becomes maximum and ESE zero. This latter situation might arise when considering Copper pairs formed from electrons at different positions.
Notice that these results arise purely due to the NS interface breaking spatial invariance and thus mixing even and odd spatial parities, which is a fundamental effect directly connected to Andreev reflection. This information is shown in the exponential part of Eq.\,(\ref{eq0}), which mixes electron and hole wave vectors with $x$ and $x'$ and acts as the generator of even- and odd-frequency components. We close this part by pointing out that in order to observe the decay of superconducting correlations in the normal metal we needs to incorporate a finite temperature by going to the Matsubara representation where $\omega\rightarrow i\omega$.

\subsubsection{Green's function in S}
The Green's function in the S region is obtained similarly to in the N region and we arrive at the following expression
\begin{widetext}
\begin{equation}
\label{GSNS}
\begin{split}
G^{r}(x,x',\omega)&=
\frac{\eta}{2ik_{e}^{S}}\frac{1}{u^{2}-v^{2}}
\bigg[{\rm e}^{ik_{e}^{S}|x-x'|}
\begin{pmatrix}
u^{2}& uv\\
uv & v^{2}
\end{pmatrix}
+
b_{3}{\rm e}^{ik_{e}^{S}(x+x')}
\begin{pmatrix}
u^{2}& uv\\
uv & v^{2}
\end{pmatrix}
+a_{3}{\rm e}^{i(k_{e}^{S}x-k_{h}^{S}x')}
\begin{pmatrix}
uv& u^{2}\\
v^{2} & uv
\end{pmatrix}
\bigg]\\
&+\frac{\eta}{2ik_{h}^{S}}\frac{1}{u^{2}-v^{2}}\bigg[{\rm e}^{-ik_{h}^{S}|x-x'|}
\begin{pmatrix}
v^{2}& uv\\
uv & u^{2}
\end{pmatrix}
+
b_{4}{\rm e}^{-ik_{h}^{S}(x+x')}
\begin{pmatrix}
v^{2}& uv\\
uv & u^{2}
\end{pmatrix}
+a_{4}{\rm e}^{i(k_{e}^{S}x'-k_{h}^{S}x)}
\begin{pmatrix}
uv&  v^{2}\\
u^{2} & uv
\end{pmatrix}
\bigg]\,,
\end{split}
\end{equation}
where $(a_{4}/k_{h}^{S})=(a_{3}/k_{e}^{S})$ with 
\begin{equation}
\begin{split}
a_{3}&=-\frac{2k_{e}^{S}(k_{e}+k_{h})uv}{\bar{D}}\,,\quad
b_{3}=\frac{R}{\bar{D}}\,,\quad
b_{4}=\frac{S}{\bar{D}}\,,\\
S&=u^{2}\Big[(k_{e}+k_{e}^{S}+iZ)(k_{h}^{S}-k_{h}+iZ)\Big]+
v^{2}\Big[(k_{h}-k_{e}^{S}-iZ)(k_{e}+k_{h}^{S}+iZ)\Big]
\end{split}
\end{equation}
are the Andreev and normal coefficients found from wave-matching. 
Then, from Eq.\,(\ref{GSNS}) the electron-hole term is found to be
\begin{equation}
\label{GNSzeroSOC}
\begin{split}
G_{eh}^{r}(x,x',\omega)&=
\frac{\eta}{2i}\frac{uv}{u^{2}-v^{2}}
\bigg\{
\frac{{\rm e}^{ik_{e}^{S}|x-x'|}}{k_{e}^{S}}
+
\frac{{\rm e}^{-ik_{h}^{S}|x-x'|}}{k_{h}^{S}}+
\frac{b_{3}}{k_{e}^{S}}{\rm e}^{ik_{e}^{S}(x+x')}+
\frac{b_{4}}{k_{h}^{S}}{\rm e}^{-ik_{h}^{S}(x+x')}\\
&+
\frac{a_{3}}{k_{e}^{S}}\Big[ \frac{u}{v}
{\rm e}^{i(k_{e}^{S}x-k_{h}^{S}x')}+
\frac{v}{u}
{\rm e}^{i(k_{e}^{S}x'-k_{h}^{S}x)}\Big]
\bigg\}
\,,
\end{split}
\end{equation}
which again corresponds to the pairing amplitude only of spin-singlet nature $f^{r}_{0}$.
Next, we can write the wave vectors in the large chemical potential limit:
$k_{e,h}^{S}=k_{\mu_{S}}\pm i\kappa$, where $\kappa=\sqrt{\Delta^{2}-\omega^{2}}[k_{\mu_{S}}/(2\mu_{S})]$ to to arrive at the simpler expressions
\begin{equation}\
\label{fNSZEROCO}
\begin{split}
f_{0}^{r}(x,x',\omega)&=
\frac{\eta}{2i}\frac{uv}{u^{2}-v^{2}}
\bigg\{
{\rm e}^{-\kappa|x-x'|}
\bigg[
\frac{{\rm e}^{ik_{\mu_{S}}|x-x'|}}{k_{e}^{S}}
+
\frac{{\rm e}^{-ik_{\mu_{\mu}}|x-x'|}}{k_{h}^{S}}\bigg]+
{\rm e}^{-\kappa(x+x')}
\bigg[
\frac{b_{3}}{k_{e}^{S}}{\rm e}^{ik_{\mu_{S}}(x+x')}+
\frac{b_{4}}{k_{h}^{S}}{\rm e}^{-ik_{\mu_{S}}(x+x')}\bigg]\\
&+
\frac{a_{3}{\rm e}^{-\kappa(x+x')}}{k_{e}^{S}}\Big[ \frac{u}{v}
{\rm e}^{ik_{\mu_{S}}(x-x')}+
\frac{v}{u}
{\rm e}^{-ik_{\mu_{S}}(x-x)}\Big]
\bigg\}
\,.
\end{split}
\end{equation}
Here, the pairing amplitude is formed out from correlations deep in the bulk (first square bracket) and contributions from the NS interface (second and third square brackets). The interface contributions correspond to normal reflection ($b_{i}$ in second square bracket) and Andreev reflection ($a_{3}$ in third square bracket). At this level, we observe that the bulk and normal reflection contributions exhibit an even in space contribution without mixing the spatial parity, while the bulk becomes space independent at $x=x'$.
At the interface, however, Andreev processes add a very interesting feature.
The Andreev term is proportional to  ${\rm e}^{i(k_{e}^{S}x'-k_{h}^{S}x)}$ in Eq.\,(\ref{GNSzeroSOC}), which mixes spatial coordinates with electron and holes  wave vectors. This leads to ${\rm e}^{ik_{\mu_{S}}(x-x')}={\rm cos}[k_{\mu_{S}}(x-x')]+i{\rm sin}[k_{\mu_{S}}(x-x')]$ in Eq.\,(\ref{fNSZEROCO}). The first term is even in space, while the second is odd, showing directly that the Andreev reflection is responsible for
spatial parity mixing. The Andreev reflection thus generates even- and odd-parity components, which in turn gives rise to even and odd-frequency dependence due to the antisymmetry condition. 
This discussion can be further clarify by writing  the even-frequency (ESE) pairing
\begin{equation}
\begin{split}
f_{0}^{r,{\rm E}}(x,x',\omega)&=\frac{\eta uv}{2i(u^{2}-v^{2})}{\rm e}^{-\kappa|x-x'|}\Big[\frac{{\rm e}^{ik_{\mu_{S}}|x-x'|}}{k_{e}^{S}}+\frac{{\rm e}^{-ik_{\mu_{S}}|x-x'|}}{k_{h}^{S}}\Big]\\
&+\frac{\eta uv}{2i(u^{2}-v^{2})}{\rm e}^{-\kappa(x+x')}
\Big[\frac{b_{3}{\rm e}^{ik_{\mu_{S}}(x+x')}}{k_{e}^{S}}+\frac{b_{4}{\rm e}^{-ik_{\mu_{S}}(x+x')}}{k_{h}^{S}}\Big]\\
&+\frac{\eta a_{3}}{2i k_{e}^{S}} \frac{u^{2}+v^{2}}{u^{2}-v^{2}}{\rm cos}[k_{\mu_{S}}(x-x')] {\rm e}^{-\kappa(x+x')}\,,
\end{split}
\end{equation}
and odd-frequency (OSO) pairing
\begin{equation}
f_{0}^{r,{\rm O}}(x,x',\omega)=\frac{\eta a_{3}}{2 k_{e}^{S}}{\rm sin}[k_{\mu_{S}}(x-x')] {\rm e}^{-\kappa(x+x')}\,.
\end{equation}
The odd-frequency component is purely  proportional to the Andreev reflection coefficient $a_{3}$. The even-frequency component, however, has contributions from the bulk (first term in square brackets), normal reflection (second term in square brackets), and Andreev reflection (last term in square brackets). However, normal reflection coefficients $b_{i}$ are very small (negligible) if the interface is transparent ($Z$) and if there is not mismatch of large chemical potentials, leaving only large contributions due to Andreev reflection $a_{3}$.
Locally, at $x=x'$ OSO is zero while ESE is maximum. On the other hand, when $k_{\mu_{S}}(x-x')=\frac{\pi}{2}+\pi n$ for $n=0,1,2,\cdots$ the opposite case happens: OSO dominates over the completely reduced ESE. 
The odd-frequency term is thus generated at the interface and exhibits an exponential decay into the bulk of S. Previous expression together with Eq.\,(\ref{EQ0}) shows on a very a strong relation between odd-frequency pairing and Andreev reflection.  
In summary, it is the Andreev reflection process that mixes spatial parities and is responsible for the coexistence for the odd-frequency components at the interface.

%%%%%%%%%%%%%%%%%%%%%%%%%%%%%
%            Short SNS JUNCTIONS AT ZERO SOC           %
%%%%%%%%%%%%%%%%%%%%%%%%%%%%%

\subsection{Short SNS junction}
Next we treat a SNS junction, where for analytical tractability we restrict ourselves to a very short N region. 
The scattering processes are constructed in a similar way as for NS junctions.  
Thus,
\begin{equation}
\label{SNS1}
\begin{split}
\Psi_{1}(x)&=
\begin{cases}
     \phi_{1}^{S_{L}}\,{\rm e}^{ik_{e}^{S}x}+a_{1}\phi_{3}^{S_{L}}\,{\rm e}^{ik_{h}^{S}x} +b_{1}\phi_{2}^{S_{L}}\,{\rm e}^{-ik_{e}^{S}x},\,x<0&  \\
   %p_{1}  \phi_{1}^{N}\,{\rm e}^{ik_{e}x}+q_{1}  \phi_{2}^{N}\,{\rm e}^{-ik_{e}x} +r_{1}  \phi_{3}^{N}\,{\rm e}^{ik_{h}x}+s_{1}  \phi_{4}^{N}\,{\rm e}^{-ik_{h}x},\,0<x<L_{\rm N}\\
    c_{1}  \phi_{1}^{S_{R}}\,{\rm e}^{ik_{e}^{S}x} +d_{1}  \phi_{4}^{S_{R}}\,{\rm e}^{-ik_{h}^{S}x},\,x>0 & 
\end{cases}\\
\Psi_{2}(x)&=
\begin{cases}
     \phi_{4}^{S_{L}}\,{\rm e}^{-ik_{h}^{S}x}+a_{2}\phi_{2}^{S_{L}}\,{\rm e}^{-ik_{e}^{S}x} +b_{2}\phi_{3}^{S_{L}}\,{\rm e}^{ik_{h}^{S}x},\,x<0& \\
    %  p_{2}  \phi_{1}^{N}\,{\rm e}^{ik_{e}x}+q_{2}  \phi_{2}^{N}\,{\rm e}^{-ik_{e}x} +r_{2}  \phi_{3}^{N}\,{\rm e}^{ik_{h}x}+s_{2}  \phi_{4}^{N}\,{\rm e}^{-ik_{h}x},\,0<x<L_{\rm N}\\
    c_{2}  \phi_{4}^{S_{R}}\,{\rm e}^{-ik_{h}^{S}x} +    d_{2}  \phi_{1}^{S_{R}}\,{\rm e}^{ik_{e}^{S}x}, \,x>0 & 
\end{cases}\\
\Psi_{3}(x)&=
\begin{cases}
    c_{3}  \phi_{2}^{S_{L}}\,{\rm e}^{-ik_{e}^{S}x} +d_{3}  \phi_{3}^{S_{L}}\,{\rm e}^{ik_{h}^{S}x},\, x<0&\\
   %  p_{3}  \phi_{1}^{N}\,{\rm e}^{ik_{e}x}+q_{3}  \phi_{2}^{N}\,{\rm e}^{-ik_{e}x} +r_{3}  \phi_{3}^{N}\,{\rm e}^{ik_{h}x}+s_{3}  \phi_{4}^{N}\,{\rm e}^{-ik_{h}x},\,0<x<L_{\rm N}\\
     \phi_{2}^{S}\,{\rm e}^{-ik_{e}^{S}x}+a_{3}\phi_{4}^{S}\,{\rm e}^{-ik_{h}^{S}x}+b_{3}\phi_{1}^{S_{R}}\,{\rm e}^{ik_{e}^{S}x},\,x>0&  
\end{cases}\\
\Psi_{4}(x)&=
\begin{cases}
    c_{4}  \phi_{3}^{S_{L}}\,{\rm e}^{ik_{h}x} +    d_{4}  \phi_{2}^{S_{L}}\,{\rm e}^{-ik_{e}x},\,x<0  & \\
    % p_{4}  \phi_{1}^{N}\,{\rm e}^{ik_{e}x}+q_{4}  \phi_{2}^{N}\,{\rm e}^{-ik_{e}x} +r_{4}  \phi_{3}^{N}\,{\rm e}^{ik_{h}x}+s_{4}  \phi_{4}^{N}\,{\rm e}^{-ik_{h}x},\,0<x<L_{\rm N}\\
\phi_{3}^{S}\,{\rm e}^{ik_{h}^{S}x}+a_{4}\phi_{1}^{S_{R}}\,{\rm e}^{ik_{e}^{S}x}+b_{4}\phi_{4}^{S_{R}}\,{\rm e}^{-ik_{h}^{S}x},\,x>0&
\end{cases}
\end{split}
\end{equation}
where 
\begin{equation}
\begin{split}
\phi_{1,2}^{N}&=
\begin{pmatrix}
1\\
0
\end{pmatrix}\,,\quad\phi_{3,4}^{N}=
\begin{pmatrix}
0\\
1
\end{pmatrix}\,,\quad
\phi_{1,2}^{S_{i}}=
\begin{pmatrix}
u\,{\rm e}^{i\phi_{i}/2}\\
v\,{\rm e}^{-i\phi_{i}/2}
\end{pmatrix}\,,
\quad 
\phi_{3,4}^{S_{L}}=
\begin{pmatrix}
v\,{\rm e}^{i\phi_{i}/2}\\
u\,{\rm e}^{-i\phi_{i}/2}
\end{pmatrix}
\end{split}
\end{equation}
where $i=L,R$ denote the left and right S regions.
The conjugated processes in this special case without SOC are the same as previous equations.
The coefficients in the scattering states are again found from 
\begin{equation}
\label{discontPsi}
\begin{split}
\Big\{[\partial_{x}\Psi(x>0)] - [\partial_{x}\Psi(x<0)]\Big\}_{x=0}&=Z[\Psi(x<0)]_{x=0}\,,\\
[\Psi_{i}(x<0)]_{x=0}&=[\Psi_{i}(x>0)]_{x=0}\,,\\
\end{split}
\end{equation}
with similar conditions applying for $\tilde{\Psi}_{i}$

\subsubsection{Green's function}
The Greens function in the left S region is
%\begin{widetext}
\begin{equation}
\begin{split}
G^{r}(x,x',\omega)&=
\frac{\eta}{2ik_{e}^{S}}\frac{1}{u^{2}-v^{2}}
\bigg\{
\Big[{\rm e}^{ik_{e}^{S}|x-x'|}+b_{1}{\rm e}^{-ik_{e}^{S}(x+x')} \Big]
\begin{pmatrix}
u^{2}&uv {\rm e}^{i\phi_{L}}\\
uv {\rm e}^{-i\phi_{L}}&v^{2}
\end{pmatrix}
+a_{1}{\rm e}^{i(k_{h}^{S}x-k_{e}^{S}x')}
\begin{pmatrix}
uv&v^{2} {\rm e}^{i\phi_{L}}\\
u^{2} {\rm e}^{-i\phi_{L}}&uv
\end{pmatrix}
\bigg\}
\\
&+
\frac{\eta}{2ik_{h}^{S}}\frac{1}{u^{2}-v^{2}}
\bigg\{
\Big[{\rm e}^{-ik_{h}^{S}|x-x'|}+b_{2}{\rm e}^{ik_{h}^{S}(x+x')} \Big]
\begin{pmatrix}
v^{2}&uv {\rm e}^{i\phi_{L}}\\
uv {\rm e}^{-i\phi_{L}}&u^{2}
\end{pmatrix}+
a_{2}{\rm e}^{i(k_{h}^{S}x'-k_{e}^{S}x)}
\begin{pmatrix}
uv&u^{2} {\rm e}^{i\phi_{L}}\\
v^{2} {\rm e}^{-i\phi_{L}}&uv
\end{pmatrix}
\bigg\}
\end{split}
\end{equation}
while in the right S region we obtain
\begin{equation}
\begin{split}
G^{r}(x,x',\omega)&=
\frac{\eta}{2ik_{e}^{S}}\frac{1}{u^{2}-v^{2}}
\bigg\{
\Big[{\rm e}^{ik_{e}^{S}|x-x'|}+b_{3}{\rm e}^{ik_{e}^{S}(x+x')} \Big]
\begin{pmatrix}
u^{2}&uv {\rm e}^{i\phi_{R}}\\
uv {\rm e}^{-i\phi_{R}}&v^{2}
\end{pmatrix}
+
\tilde{a}_{3}{\rm e}^{i(k_{e}^{S}x-k_{h}^{S}x')}
\begin{pmatrix}
uv&v^{2} {\rm e}^{i\phi_{R}}\\
u^{2} {\rm e}^{-i\phi_{R}}&uv
\end{pmatrix}
\bigg\}
\\
&+
\frac{\eta}{2ik_{h}^{S}}\frac{1}{u^{2}-v^{2}}
\bigg\{
\Big[{\rm e}^{-ik_{h}^{S}|x-x'|}+b_{4}{\rm e}^{-ik_{h}^{S}(x+x')} \Big]
\begin{pmatrix}
v^{2}&uv {\rm e}^{i\phi_{R}}\\
uv {\rm e}^{-i\phi_{R}}&u^{2}
\end{pmatrix}
+
\tilde{a}_{4}{\rm e}^{i(k_{e}^{S}x'-k_{h}^{S}x)}
\begin{pmatrix}
uv&u^{2} {\rm e}^{i\phi_{R}}\\
v^{2} {\rm e}^{-i\phi_{R}}&uv
\end{pmatrix}\bigg\}
\end{split}
\end{equation}
where $a_{i}$ and $b_{i}$ in this case are the coefficients corresponding to the SNS geometry.
Since both Green's functions provide the same information, such as LDOS, supercurrents, and pairing amplitudes, we only need to analyze the one in the right S region. Thus, we write  the anomalous electron-hole component  
\begin{equation}
\begin{split}
G_{eh}^{r}(x,x',\omega)&=
\frac{\eta}{2i}\frac{uv\,{\rm e}^{i\phi_{R}}}{u^{2}-v^{2}}
\bigg\{
\frac{{\rm e}^{ik_{e}^{S}|x-x'|}}{k_{e}^{S}}
+
\frac{{\rm e}^{-ik_{h}^{S}|x-x'|}}{k_{h}^{S}}+
\frac{b_{1}}{k_{e}^{S}}{\rm e}^{ik_{e}^{S}(x+x')}+
\frac{b_{2}}{k_{h}^{S}}{\rm e}^{-ik_{h}^{S}(x+x')}\\
&+
\frac{a_{1}}{k_{e}^{S}}\frac{u}{v}
{\rm e}^{i(k_{e}^{S}x-k_{h}^{S}x')}+
\frac{a_{2}}{k_{h}^{S}}\frac{v}{u}
{\rm e}^{i(k_{e}^{S}x'-k_{h}^{S}x)}\Big]
\bigg\}
\,,
\end{split}
\end{equation}
where we have used that $b_{3(4)}=b_{1(2)}$ and $\tilde{a}_{3(4)}=a_{1(2)}$, and
\begin{equation}
\label{ARSNS}
\begin{split}
a_{1}(\phi_{L},\phi_{R})&=\frac{2k_{e}^{S}(k_{e}^{S}+k_{h}^{S})uv(u^{2}{\rm e}^{i\phi_{L}}-v^{2}{\rm e}^{i\phi_{R}})({\rm e}^{i\phi_{R}}-{\rm e}^{i\phi_{L}})}{D}\,,\\
a_{2}(\phi_{L},\phi_{R})&=\frac{k_{h}^{S}}{k_{e}^{S}}a_{1}(\phi_{R},\phi_{L})\,,\\
b_{1}(\phi_{L},\phi_{R})&={\rm e}^{i(\phi_{R}+\phi_{L})}\frac{(u^{4}+v^{4})Z(2ik_{h}^{S}+Z)-2u^{2}v^{2}\Big[\big({k^{S}_{e}}^{2}-{k_{h}^{S}}^{2}\big)(1-{\rm cos}(\phi_{L}+\phi_{R}))+Z(Z+2ik_{h}^{S})\Big]}{D}\,,\\
b_{2}(\phi_{L},\phi_{R})&=-b_{1}(\phi_{L},\phi_{R})\,,
\end{split}
\end{equation}
where the denominator is given by $D=-(u^{4}+v^{4}){\rm e}^{\phi_{L}+\phi_{R}}\big(2k_{h}^{S}-iZ\big)\big(2k_{e}^{S}+iZ\big)+u^{2}v^{2}\Big[(k_{e}^{S}+k_{h}^{S})^{2}({\rm e}^{2i\phi_{L}}+{\rm e}^{2i\phi_{R}})-2{\rm e}^{i(\phi_{L}+\phi_{R})}(k_{e}^{S}-k_{h}^{S}+iZ)^{2}\Big]$. Notice how Andreev reflection is fully determined by a finite phase difference between the two superconducting regions, where at zero phase difference the Andreev coefficient $a_{1}=0$. On the other hand, normal reflection is allowed even if there is no  phase difference, but in the full transparent regime ($Z=0$) we still have $b_{1}=0$ at zero phase difference. These coefficients play an important role in the pairing amplitudes as they fully determine their existence at the interface.

Then, by using the wave vectors in the large chemical potential limit and energies within $\Delta$, $k_{e,h}^{S}=k_{\mu_{S}}\pm i\kappa$, where $\kappa=\sqrt{\Delta^{2}-\omega^{2}}[k_{\mu_{S}}/(2\mu_{S})]$, we obtain
\begin{equation}
\begin{split}
G_{eh}^{r}(x,x',\omega)&=
\frac{\eta}{2i}\frac{uv\,{\rm e}^{i\phi_{R}}}{u^{2}-v^{2}}
\bigg\{
{\rm e}^{-\kappa|x-x'|}
\bigg[
\frac{{\rm e}^{ik_{\mu_{S}}|x-x'|}}{k_{e}^{S}}
+
\frac{{\rm e}^{-ik_{\mu_{\mu}}|x-x'|}}{k_{h}^{S}}\bigg]+
{\rm e}^{-\kappa(x+x')}
\bigg[
\frac{b_{1}}{k_{e}^{S}}{\rm e}^{ik_{\mu_{S}}(x+x')}+
\frac{b_{2}}{k_{h}^{S}}{\rm e}^{-ik_{\mu_{S}}(x+x')}\bigg]\\
&+
{\rm e}^{-\kappa(x+x')}\bigg[
\frac{a_{1}}{k_{e}^{S}} \frac{u}{v}
{\rm e}^{ik_{\mu_{S}}(x-x')}+
\frac{a_{2}}{k_{h}^{S}}\frac{v}{u}
{\rm e}^{-ik_{\mu_{S}}(x-x)}\bigg]
\bigg\}
\,.
\end{split}
\end{equation}
Observe that in this case the anomalous electron-hole component, being the pairing amplitude in this case, acquires an overall factor that is dependent on the superconducting phase of the right S region. As in the NS case, such amplitude contains contributions from the bulk (first term) and interface (second and third terms). The interface contributes through normal (terms with $b_{i}$) and Andreev ($a_{i}$) reflections. The interface contributions exponentially decay into the bulk of the S region. By writing the even and odd-frequency amplitudes separately we obtain
\begin{equation}
\label{FSNSO}
\begin{split}
f_{0}^{r,{\rm E}}(x,x',\omega)&=\frac{\eta}{2i}\frac{uv\,{\rm e}^{i\phi_{R}}}{u^{2}-v^{2}}
\bigg\{
{\rm e}^{-\kappa|x-x'|}
\bigg[
\frac{{\rm e}^{ik_{\mu_{S}}|x-x'|}}{k_{e}^{S}}
+
\frac{{\rm e}^{-ik_{\mu_{\mu}}|x-x'|}}{k_{h}^{S}}\bigg]\\
&+
{\rm e}^{-\kappa(x+x')}
\bigg[
\frac{b_{1}}{k_{e}^{S}}{\rm e}^{ik_{\mu_{S}}(x+x')}+
\frac{b_{2}}{k_{h}^{S}}{\rm e}^{-ik_{\mu_{S}}(x+x')}\bigg]
+{\rm e}^{-\kappa(x+x')}{\rm cos}[k_{\mu_{S}}(x-x')]\bigg[
\frac{a_{1}}{k_{e}^{S}} \frac{u}{v}+
\frac{a_{2}}{k_{h}^{S}}\frac{v}{u}
\bigg]
\bigg\}\,,\\
f_{0}^{r,{\rm O}}(x,x',\omega)&=\frac{\eta}{2i}\frac{uv\,{\rm e}^{i\phi_{R}}}{u^{2}-v^{2}}{\rm e}^{-\kappa(x+x')}i{\rm sin}[k_{\mu_{S}}(x-x')]\bigg[
\frac{a_{1}}{k_{e}^{S}} \frac{u}{v}-
\frac{a_{2}}{k_{h}^{S}}\frac{v}{u}
\bigg]\,,
\end{split}
\end{equation}
\end{widetext}
where we notice that bulk and normal reflection contributions induce parity even terms, while the Andreev reflection mixes spatial parity giving rise to both even and odd components in the spatial coordinates and therefore even- and odd-frequency terms. These two pairing components correspond to ESE and OSO classes, respectively. Observe that the odd-frequency component is solely proportional to the Andreev reflection coefficients; normal reflections do not generate odd-frequency pairs. 

The very first observation we make is that at $\phi=0$, the Andreev coefficients are zero, as seen directly in Eqs.\,(\ref{ARSNS}), and therefore the Andreev contribution is zero, leaving only normal reflection and bulk terms.  This indicates a relation between phase-dependent properties at the junction interface, namely, Andreev bound states and supercurrents. In fact, in the large chemical potential limit  we can directly obtain  from $f^{r,{\rm O}}$ the well-known expression for the energy for the Andreev bound states: $\omega_{\pm}=\pm\Delta\sqrt{1-\tau {\rm sin}^{2}(\phi/2)}$, with $\tau=1/(1+\bar{Z}^{2})$ and $\bar{Z}=Z/(2k_{\mu_{S}})$, since the bound state show up as poles in the Andreev reflection and thus also as poles in $f^{r,{\rm O}}$.
Thus, the interface  pairing functions capture the formation of Andreev bound states, which in turn fully determine the supercurrent in short junctions. Indeed, the supercurrent across a short SNS junction is proportional to the integral over frequency of $(a_{1}/k_{e}^{S})-(a_{2}/k_{h}^{S})$ as was reported already long time ago.\cite{PhysRevLett.67.132,FURUSAKI1991299} The supercurrent is thus described by an  expression very similar to the odd-frequency term given by Eq.\,(\ref{FSNSO}).  In the fully transparent regime, the normal coefficients are much smaller than the Andreev coefficients and therefore the interface pairing amplitudes are all approximately determined by Andreev reflections. 

We conclude this section by pointing out that within a scattering approach we have explained the well-established induced odd-frequency pairing in NS and in SNS junctions in terms of Andreev reflections and have also been able to directly relate  odd-frequency pairing to the supercurrent in short SNS junctions.

\begin{widetext}
\section{Finite spin-orbit coupling}
\label{finiteSOC}
In this appendix, we treat the case of finite SO coupling, i.e.,~the junction is modeled by Eq.\,(\ref{HBdG}) in the main text. The construction of the Green's functions follows the same recipe as in the previous appendix, with the sole difference that now we need to account for the spin degree of freedom. Under spin-orbit coupling the spin becomes an active degree of freedom and the problem gets notably more complicated than the case discussed in the previous appendix. Here, we provide the detailed equations and results underlying the results in the main text for NS and SNS junctions with Rashba SO coupling in the main text.

\subsection{NS junction}
\label{NSAppSOC}
For the NS junction there are at finite SO coupling eight scattering processes, four particles coming from the left region and four from the right one, and they read as
\begin{equation}
\label{PsiNSSOC}
\begin{split}
\Psi_{1}(x)&=
\begin{cases}
    \phi_{1}^{N}{\rm e}^{ik_{2}x}+b_{11}\phi_{2}^{N}{\rm e}^{-ik_{1}x}+b_{12}\phi_{4}^{N}{\rm e}^{-ik_{2}x}+a_{11}\phi_{6}^{N}{\rm e}^{i\bar{k}_{1}x}+a_{12}\phi_{8}^{N}{\rm e}^{i\bar{k}_{2}x}\,,  & x<0, \\
     t_{11}\phi_{1}^{S}{\rm e}^{ik^{S}_{e_{2}}x}+   t_{12}\phi_{3}^{S}{\rm e}^{ik^{S}_{e_{1}}x}+   c_{11}\phi_{5}^{S}{\rm e}^{-ik^{S}_{h_{2}}x}+c_{12}\phi_{7}^{S}{\rm e}^{-ik^{S}_{h_{1}}x}
 \,,& x>0.
\end{cases}
\\
\Psi_{2}(x)&=
\begin{cases}
    \phi_{3}^{N}{\rm e}^{ik_{1}x}+b_{21}\phi_{2}^{N}{\rm e}^{-ik_{1}x}+b_{22}\phi_{4}^{N}{\rm e}^{-ik_{2}x}+a_{21}\phi_{6}^{N}{\rm e}^{i\bar{k}_{1}x}+a_{22}\phi_{8}^{N}{\rm e}^{i\bar{k}_{2}x}\,,  & x<0, \\   
    t_{21}\phi_{1}^{S}{\rm e}^{ik^{S}_{e_{2}}x}+   t_{22}\phi_{3}^{S}{\rm e}^{ik^{S}_{e_{1}}x}+   c_{21}\phi_{5}^{S}{\rm e}^{-ik^{S}_{h_{2}}x}+c_{22}\phi_{7}^{S}{\rm e}^{-ik^{S}_{h_{1}}x}  \,,& x>0.
\end{cases}\\
\Psi_{3}(x)&=
\begin{cases}
    \phi_{5}^{N}{\rm e}^{-i\bar{k}_{2}x}+b_{31}\phi_{6}^{N}{\rm e}^{i\bar{k}_{1}x}+b_{32}\phi_{8}^{N}{\rm e}^{i\bar{k}_{2}x}+a_{31}\phi_{4}^{N}{\rm e}^{-ik_{2}x}+a_{32}\phi_{2}^{N}{\rm e}^{-ik_{1}x}\,,  & x<0, \\
    t_{31}\phi_{5}^{S}{\rm e}^{-ik^{S}_{h_{2}}x}+t_{32}\phi_{7}^{S}{\rm e}^{-ik^{S}_{h_{1}}x} +  c_{31}\phi_{1}^{S}{\rm e}^{ik^{S}_{e_{2}}x}+   c_{32}\phi_{3}^{S}{\rm e}^{ik^{S}_{e_{1}}x}\,,& x>0.
\end{cases}\\
\Psi_{4}(x)&=
\begin{cases}
    \phi_{7}^{N}{\rm e}^{-i\bar{k}_{1}x}+b_{41}\phi_{8}^{N}{\rm e}^{i\bar{k}_{2}x}+b_{42}\phi_{6}^{N}{\rm e}^{i\bar{k}_{1}x}+a_{41}\phi_{4}^{N}{\rm e}^{-ik_{2}x}+a_{42}\phi_{2}^{N}{\rm e}^{-ik_{1}x}\,,  & x<0, \\
    t_{41}\phi_{5}^{S}{\rm e}^{-ik^{S}_{h_{2}}x}+t_{42}\phi_{7}^{S}{\rm e}^{-ik^{S}_{h_{1}}x} +  c_{41}\phi_{1}^{S}{\rm e}^{ik^{S}_{e_{2}}x}+   c_{42}\phi_{3}^{S}{\rm e}^{ik^{S}_{e_{1}}x}\,,& x>0.
\end{cases}
\\
\Psi_{5}(x)&=
\begin{cases}
  t_{51}\phi_{4}^{N}{\rm e}^{-ik_{2}x}+t_{52}\phi_{2}^{N}{\rm e}^{-ik_{1}x}+c_{51}\phi_{6}^{N}{\rm e}^{i\bar{k}_{1}x}+c_{52}\phi_{8}^{N}{\rm e}^{i\bar{k}_{2}x}\,,  & x<0, \\
   \phi_{2}^{S}{\rm e}^{-ik^{S}_{e_{1}}x}+  b_{51}\phi_{1}^{S}{\rm e}^{ik^{S}_{e_{2}}x}+   b_{52}\phi_{3}^{S}{\rm e}^{ik^{S}_{e_{1}}x}+  a_{51}\phi_{5}^{S}{\rm e}^{-ik^{S}_{h_{2}}x}+a_{52}\phi_{7}^{S}{\rm e}^{-ik^{S}_{h_{1}}x}  \,,& x>0.
\end{cases}
\\
\Psi_{6}(x)&=
\begin{cases}
  t_{61}\phi_{4}^{N}{\rm e}^{-ik_{2}x}+t_{62}\phi_{2}^{N}{\rm e}^{-ik_{1}x}+c_{61}\phi_{6}^{N}{\rm e}^{i\bar{k}_{1}x}+c_{62}\phi_{8}^{N}{\rm e}^{i\bar{k}_{2}x}\,,  & x<0, \\
   \phi_{4}^{S}{\rm e}^{-ik^{S}_{e_{2}}x}+  b_{61}\phi_{1}^{S}{\rm e}^{ik^{S}_{e_{2}}x}+   b_{62}\phi_{3}^{S}{\rm e}^{ik^{S}_{e_{1}}x}+  a_{61}\phi_{5}^{S}{\rm e}^{-ik^{S}_{h_{2}}x}+a_{62}\phi_{7}^{S}{\rm e}^{-ik^{S}_{h_{1}}x}  \,,& x>0.
\end{cases}\\
\Psi_{7}(x)&=
\begin{cases}
  t_{71}\phi_{6}^{N}{\rm e}^{i\bar{k}_{1}x}+t_{72}\phi_{8}^{N}{\rm e}^{i\bar{k}_{2}x}++c_{71}\phi_{4}^{N}{\rm e}^{-ik_{2}x}+c_{72}\phi_{2}^{N}{\rm e}^{-ik_{1}x}\,,  & x<0, \\
   \phi_{6}^{S}{\rm e}^{ik^{S}_{h_{1}}x} +  b_{71}\phi_{5}^{S}{\rm e}^{-ik^{S}_{h_{2}}x}+b_{72}\phi_{7}^{S}{\rm e}^{-ik^{S}_{h_{1}}x} +a_{71}\phi_{1}^{S}{\rm e}^{ik^{S}_{e_{2}}x}+   a_{72}\phi_{3}^{S}{\rm e}^{ik^{S}_{e_{1}}x} \,,& x>0.
\end{cases}
\\
\Psi_{8}(x)&=
\begin{cases}
  t_{81}\phi_{6}^{N}{\rm e}^{i\bar{k}_{1}x}+t_{82}\phi_{8}^{N}{\rm e}^{i\bar{k}_{2}x}++c_{81}\phi_{4}^{N}{\rm e}^{-ik_{2}x}+c_{82}\phi_{2}^{N}{\rm e}^{-ik_{1}x}\,,  & x<0, \\
   \phi_{8}^{S}{\rm e}^{ik^{S}_{h_{2}}x} +  b_{81}\phi_{5}^{S}{\rm e}^{-ik^{S}_{h_{2}}x}+b_{82}\phi_{7}^{S}{\rm e}^{-ik^{S}_{h_{1}}x} +a_{81}\phi_{1}^{S}{\rm e}^{ik^{S}_{e_{2}}x}+   a_{82}\phi_{3}^{S}{\rm e}^{ik^{S}_{e_{1}}x} \,,& x>0.
\end{cases}
\end{split}
\end{equation}
 where $k_{1,2}=k_{e_{1,2}}$, $\bar{k}_{1,2}=k_{h_{1,2}}$ and
 \begin{equation}
 \begin{split}
 \phi_{1,2}^{N}=
 \begin{pmatrix}
 1\\
 0\\
 0\\
 0
 \end{pmatrix},\, 
 \phi_{3,4}^{N}=
 \begin{pmatrix}
 0\\
 1\\
 0\\
 0
 \end{pmatrix},\, 
 \phi_{5,6}^{N}=
 \begin{pmatrix}
 0\\
 0\\
 1\\
 0
 \end{pmatrix},\, 
 \phi_{7,8}^{N}=
 \begin{pmatrix}
 0\\
 0\\
 0\\
 1
 \end{pmatrix},\, 
 \phi_{1,2}^{S}=
 \begin{pmatrix}
 u\\
 0\\
 0\\
v
 \end{pmatrix},\, 
 \phi_{3,4}^{S}=
 \begin{pmatrix}
 0\\
 -u\\
 v\\
0
 \end{pmatrix},\, 
 \phi_{5,6}^{S}=
 \begin{pmatrix}
0\\
 -v\\
 u\\
0
 \end{pmatrix},\, 
 \phi_{7,8}^{S}=
 \begin{pmatrix}
 v\\
 0\\
 0\\
u
 \end{pmatrix}
 \end{split}
 \end{equation}
The conjugated processes $\tilde{\Psi}_{i}$ have the same form but instead of previous vectors we obtain $\tilde{\phi}_{1,2}^{N}=\phi_{3}^{N}$, $\tilde{\phi}_{3,4}^{N}=\phi_{1}^{N}$, $\tilde{\phi}_{5,6}^{N}=\phi_{7}^{N}$, $\tilde{\phi}_{7,8}^{N}=\phi_{5}^{N}$, $\tilde{\phi}_{1,2}^{S}=\phi_{3}^{S}$, $\tilde{\phi}_{3,4}^{S}=\phi_{1}^{S}$, $\tilde{\phi}_{5,6}^{S}=\phi_{7}^{S}$, $\tilde{\phi}_{7,8}^{S}=\phi_{5}^{S}$.
The coefficients of these scattering processes are found by matching them at the interface $x=0$ as outlined in Eq.\,(\ref{conditions}).
Then, the retarded Green's function is found by plugging all scattering functions into Eq.\,\eqref{RGF}, where Eqs.\,(\ref{conditionGRSO}) are also employed to find the coefficients. 

\subsubsection{Green's function in N}
After some tedious but straightforward algebra, we obtain in the N region the electron-electron and electron-hole components
\begin{equation}
\begin{split}
[G_{ee}^{r}]_{\uparrow\uparrow}(x,x',\omega)&=\frac{\eta}{i(k_{e_{1}}+k_{e_{2}})}
\Big\{
\Big[\theta(x-x'){\rm e}^{ik_{e_{2}}(x-x')}+\theta(x'-x){\rm e}^{-ik_{e_{1}}(x-x')}\Big]\\
&+{\rm e}^{-i(k_{e_{1}}x+k_{e_{2}}x')}[b_{11}\theta(x-x')+b_{22}\theta(x'-x)]\,,\\
[G_{ee}^{r}]_{\downarrow\downarrow}(x,x',\omega)&=\frac{\eta}{i(k_{e_{1}}+k_{e_{2}})}
\Big\{
\Big[\theta(x-x'){\rm e}^{ik_{e_{1}}(x-x')}+\theta(x'-x){\rm e}^{-ik_{e_{2}}(x-x')}\Big]\\
&+{\rm e}^{-i(k_{e_{2}}x+k_{e_{1}}x')}[b_{22}\theta(x-x')+b_{11}\theta(x'-x)]
\Big\}\,,\\
[G_{ee}^{r}]_{\uparrow\downarrow}(x,x',\omega)&=0\,,\quad [G_{ee}^{r}]_{\downarrow\uparrow}(x,x',\omega)=0\\
[G_{eh}^{r}]_{\uparrow\uparrow}(x,x',\omega)&=0\,,\quad [G_{eh}^{r}]_{\downarrow\downarrow}(x,x',\omega)=0\,,\\
[G_{eh}^{r}]_{\uparrow\downarrow}(x,x',\omega)&=\frac{\eta}{i}{\rm e}^{i(-k_{e_{1}}x+k_{h_{1}}x')}[a_{42}\theta(x-x')+\tilde{a}_{21}\theta(x'-x)]\,,\\
[G_{eh}^{r}]_{\downarrow\uparrow}(x,x',\omega)&=\frac{\eta}{i}{\rm e}^{i(-k_{e_{2}}x+k_{h_{2}}x')}[a_{31}\theta(x-x')+\tilde{a}_{12}\theta(x'-x)]\,,
\end{split}
\end{equation}
where $\tilde{a}_{12}=-a_{42}$, $\tilde{a}_{21}=-a_{31}$,
\begin{equation}
\begin{split}
a_{42}&=\frac{(k_{e_{2}}^{S}+k_{h_{1}}^{S})uv}{u^{2}(k_{h_{1}}^{S}+k_{h_{2}}-iZ)(k_{e_{2}}^{S}+k_{e_{1}}+iZ)+v^{2}(k_{e_{1}}-k_{h_{1}}^{S}+iZ)(k_{e_{2}}^{S}-k_{h_{2}}+iZ)}\,,\\
a_{31}&=-\frac{(k_{e_{1}}^{S}+k_{h_{2}}^{S})uv}{u^{2}(k_{h_{2}}^{S}+k_{h_{1}}-iZ)(k_{e_{1}}^{S}+k_{e_{2}}+iZ)+v^{2}(k_{e_{1}}^{S}-k_{h_{1}}+iZ)(k_{e_{2}}-k_{h_{2}}^{S}+iZ)}\,,\\
b_{11}&=\frac{u^{2}(k_{h_{1}}^{S}+k_{h_{2}}-iZ)(k_{e_{2}}-k_{e_{2}}^{S}-iZ)+v^{2}(k_{e_{2}}^{S}-k_{h_{2}}+iZ)(k_{e_{2}}+k_{h_{1}}^{S}-iZ)}{u^{2}(k_{h_{1}}^{S}+k_{h_{2}}-iZ)(k_{e_{2}}^{S}+k_{e_{1}}+iZ)+v^{2}(k_{e_{1}}-k_{h_{1}}^{S}+iZ)(k_{e_{2}}^{S}-k_{h_{2}}+iZ)}\,,\\
b_{22}&=\frac{u^{2}(k_{h_{2}}^{S}+k_{h_{1}}-iZ)(k_{e_{1}}-k_{e_{1}}^{S}-iZ)+v^{2}(k_{e_{1}}^{S}-k_{h_{1}}+iZ)(k_{e_{1}}+k_{h_{2}}^{S}-iZ)}{u^{2}(k_{h_{2}}^{S}+k_{h_{1}}-iZ)(k_{e_{1}}^{S}+k_{e_{2}}+iZ)+v^{2}(k_{e_{1}}^{S}-k_{h_{1}}+iZ)(k_{e_{2}}-k_{h_{2}}^{S}+iZ)}\,.
\end{split}
\end{equation}
The spin-singlet and triplet pairing amplitudes are then found using Eqs.\,(\ref{decomposeSPIN}) and \,(\ref{pairingfunctions}), resulting in
\begin{equation}
\begin{split}
f_{0}^{r}(x,x',\omega)&=\frac{\eta}{2i}\Big\{
a_{42}\Big[ \theta(x-x'){\rm e}^{i(-k_{e_{1}}x+k_{h_{1}}x')}+\theta(x'-x){\rm e}^{i(-k_{e_{2}}x+k_{h_{2}}x')}\Big]\\
&-
a_{31}\Big[ \theta(x-x'){\rm e}^{i(-k_{e_{2}}x+k_{h_{2}}x')}+\theta(x'-x){\rm e}^{i(-k_{e_{1}}x+k_{h_{1}}x')}\Big]
\Big\}
\,,\\
f_{3}^{r}(x,x',\omega)&=\frac{\eta}{2i}\Big\{
a_{42}\Big[ \theta(x-x'){\rm e}^{i(-k_{e_{1}}x+k_{h_{1}}x')}-\theta(x'-x){\rm e}^{i(-k_{e_{2}}x+k_{h_{2}}x')}\Big]\\
&+
a_{31}\Big[ \theta(x-x'){\rm e}^{i(-k_{e_{2}}x+k_{h_{2}}x')}-\theta(x'-x){\rm e}^{i(-k_{e_{1}}x+k_{h_{1}}x')}\Big]
\Big\}\,.
\end{split}
\end{equation}
In these pairing amplitudes we can  introduce the wave vectors defined in Eqs.\,(\ref{kmomenta}) and demonstrate that $f_{3}(x,x,\omega)=0$. 
To visualize this result, we further simplify the pairing amplitudes in the limit of large chemical potential and at the same time we  isolate the even- and odd-frequency components. Then, we obtain
\begin{equation}
\begin{split}
f_{0}^{r,{\rm E}}&=\frac{\eta}{2i}{\rm e}^{-i\kappa_{\omega}^{N}(x+x')}{\rm cos}[\bar{k}(x-x')]\Big(a_{42}{\rm e}^{-ik_{SO}|x-x'|}-a_{31}{\rm e}^{ik_{SO}|x-x'|}\Big)\,,\\
f_{0}^{r,{\rm O}}&=\frac{\eta}{2i}{\rm e}^{-i\kappa_{\omega}^{N}(x+x')}(-i){\rm sin}[\bar{k}(x-x')]\Big(a_{42}{\rm e}^{-ik_{SO}|x-x'|}-a_{31}{\rm e}^{ik_{SO}|x-x'|}\Big)\,,\\
f_{3}^{r,{\rm E}}&=\frac{\eta}{2i}{\rm e}^{-i\kappa_{\omega}^{N}(x+x')}{\rm sgn}(x-x'){\rm cos}[\bar{k}(x-x')]\Big(a_{42}{\rm e}^{-ik_{SO}|x-x'|}+a_{31}{\rm e}^{ik_{SO}|x-x'|}\Big)\,,\\
f_{3}^{r,{\rm O}}&=\frac{\eta}{2i}{\rm e}^{-i\kappa_{\omega}^{N}(x+x')}{\rm sgn}(x-x')(-i){\rm cos}[\bar{k}(x-x')]\Big(a_{42}{\rm e}^{-ik_{SO}|x-x'|}+a_{31}{\rm e}^{ik_{SO}|x-x'|}\Big)\,,
\end{split}
\end{equation}
which, in this large chemical potential limit, can be further simplified using $a_{31}=-a_{42}$,
\begin{equation}
\begin{split}
f_{0}^{r,{\rm E}}&=\frac{\eta}{2i}{\rm e}^{-i\kappa_{\omega}^{N}(x+x')}{\rm cos}[\bar{k}(x-x')]2a_{42}{\rm cos}[k_{SO}|x-x'|]\,,\\
f_{0}^{r,{\rm O}}&=\frac{\eta}{2i}{\rm e}^{-i\kappa_{\omega}^{N}(x+x')}(-i){\rm sin}[\bar{k}(x-x')]2a_{42}{\rm cos}[k_{SO}|x-x'|]\,,\\
f_{3}^{r,{\rm E}}&=\frac{\eta}{2i}{\rm e}^{-i\kappa_{\omega}^{N}(x+x')}{\rm sgn}(x-x'){\rm cos}[\bar{k}(x-x')](-2i)a_{42}{\rm sin}[k_{SO}|x-x'|]\,,\\
f_{3}^{r,{\rm O}}&=\frac{\eta}{2i}{\rm e}^{-i\kappa_{\omega}^{N}(x+x')}{\rm sgn}(x-x')(-i){\rm sin}[\bar{k}(x-x')](-2i)a_{42}{\rm sin}[k_{SO}|x-x'|]\,.
\end{split}
\end{equation}
This is the final result given by Eqs.\,(\ref{PairinNS_N}) in the main text, where the result is also further analyzed.
%The first observation suggests that at $x=x'$ only the first amplitude (ESE) is finite. The rest (OSO, ETO, OTE) are zero, indicating that in the normal metal with Rashba SO coupling only non-local spin-triplet and non-local odd-frequency pairing amplitudes are allowed.
%Second, the amplitudes are proportional to the Andreev reflection coefficient $a_{42}$, suggesting a strong relation between them. We emphasise here that the vanishing of spin-triplet and odd-frequency correlations is not subjected to any approximation; the large chemical potential limit was chosen only to write down simple expressions.

\subsubsection{Green's function in S}
In the S region we obtain for the electron-electron part
\begin{equation}
\label{NSSOCAPPgee}
\begin{split}
[G_{ee}^{r}]_{\uparrow\uparrow}(x,x',\omega)&=\frac{\eta u^{2}}{i(k_{{e}_{1}}^{S}+k_{{e}_{2}}^{S})(u^{2}-v^{2})}\bigg\{\theta(x-x'){\rm e}^{ik_{{e}_{2}}^{S}(x-x')}+
\theta(x'-x){\rm e}^{-ik_{{e}_{1}}^{S}(x-x')}\\
&+{\rm e}^{i(k_{{e}_{2}}^{S}x+k_{{e}_{1}}^{S}x')}[\theta(x-x')b_{62}+\theta(x'-x)b_{51}]\bigg\}
+\frac{\eta uv a_{52}\theta(x'-x)}{i(u^{2}-v^{2})}\Big[{\rm e}^{i(k_{{e}_{2}}^{S}x-k_{{h}_{2}}^{S}x')} +{\rm e}^{i(k_{{e}_{1}}^{S}x'-k_{{h}_{1}}^{S}x)}\Big]\\
&+
\frac{\eta v^{2}}{i(k_{{h}_{1}}^{S}+k_{{h}_{2}}^{S})(u^{2}-v^{2})}
\bigg\{\theta(x-x'){\rm e}^{-ik_{{h}_{1}}^{S}(x-x')}+
\theta(x'-x){\rm e}^{ik_{{h}_{2}}^{S}(x-x')}\\
&+{\rm e}^{-i(k_{{h}_{1}}^{S}x+k_{{h}_{2}}^{S}x')}[\theta(x-x')b_{71}+\theta(x'-x)b_{82}]\bigg\}
+\frac{\eta uv \tilde{a}_{72}\theta(x-x')}{i(u^{2}-v^{2})}
\Big[{\rm e}^{i(k_{{e}_{2}}^{S}x-k_{{h}_{2}}^{S}x')} +{\rm e}^{i(k_{{e}_{1}}^{S}x'-k_{{h}_{1}}^{S}x)}\Big]\,,\\
[G_{ee}^{r}]_{\downarrow\downarrow}(x,x',\omega)&=
\frac{\eta u^{2}}{i(k_{{e}_{1}}^{S}+k_{{e}_{2}}^{S})(u^{2}-v^{2})}
\bigg\{
\theta(x-x'){\rm e}^{ik_{{e}_{1}}^{S}(x-x')}+
\theta(x'-x){\rm e}^{-ik_{{e}_{2}}^{S}(x-x')}\\
&+{\rm e}^{i(k_{{e}_{1}}^{S}x+k_{{e}_{2}}^{S}x')}[\theta(x-x')b_{51}+\theta(x'-x)b_{62}]\bigg\}
+\frac{\eta uv a_{61}\theta(x'-x)}{i(u^{2}-v^{2})}
\Big[{\rm e}^{i(k_{{e}_{1}}^{S}x-k_{{h}_{1}}^{S}x')} +{\rm e}^{i(k_{{e}_{2}}^{S}x'-k_{{h}_{2}}^{S}x)}\Big]\\
&+
\frac{\eta v^{2}}{i(k_{{h}_{1}}^{S}+k_{{h}_{2}}^{S})(u^{2}-v^{2})}
\bigg\{\theta(x-x'){\rm e}^{-ik_{{h}_{2}}^{S}(x-x')}+
\theta(x'-x){\rm e}^{ik_{{h}_{1}}^{S}(x-x')}\\
&+{\rm e}^{-i(k_{{h}_{2}}^{S}x+k_{{h}_{1}}^{S}x')}[\theta(x-x')b_{82}+\theta(x'-x)b_{71}]\bigg\}
+\frac{\eta uv \tilde{a}_{81}\theta(x-x')}{i(u^{2}-v^{2})}
\Big[{\rm e}^{i(k_{{e}_{1}}^{S}x-k_{{h}_{1}}^{S}x')} +{\rm e}^{i(k_{{e}_{2}}^{S}x'-k_{{h}_{2}}^{S}x)}\Big]\,,\\
[G_{ee}^{r}]_{\uparrow\downarrow}(x,x',\omega)&=0\,,
[G_{ee}^{r}]_{\downarrow\uparrow}(x,x',\omega)=0\,,
\end{split}
\end{equation}
which contain elements from bulk, normal (terms proportional to $b_{ij}$), and Andreev reflections (terms proportional to $a_{ij}$). %
Here we find that $\tilde{a}_{72}=a_{61}$, $a_{52}=\tilde{a}_{81}$, $\tilde{a}_{72}=a_{52}$, $b_{51}=b_{62}$, $b_{71}=b_{82}$, and
\begin{equation}
\begin{split}
%\tilde{a}_{72}&=-\frac{(k_{e_{2}}+k_{h_{1}})uv}{
%u^{2}(k_{h_{1}}+k_{h_{2}}^{S}-iZ)(k_{e_{1}}^{S}+k_{e_{2}}+iZ)+
%v^{2}(k_{e_{1}}^{S}+k_{h_{1}}+iZ)(k_{e_{2}}-k_{h_{2}}^{S}+iZ)}\,,\\
a_{52}&=-\frac{(k_{e_{1}}+k_{h_{2}})uv}{
u^{2}(k_{h_{1}}^{S}+k_{h_{2}}^{S}-iZ)(k_{e_{2}}^{S}+k_{e_{1}}+iZ)+
v^{2}(k_{e_{1}}-k_{h_{1}}^{S}+iZ)(k_{e_{2}}^{S}-k_{h_{2}}+iZ)}\,,\\
b_{51}&=\frac{u^{2}(k_{e_{1}}^{S}-k_{e_{1}}-iZ)(k_{h_{1}}^{S}+k_{h_{2}}-iZ)+
v^{2}(k_{e_{1}}^{S}+k_{h_{2}}-iZ)(k_{e_{1}}-k_{h_{1}}^{S}+iZ)}{
u^{2}(k_{h_{1}}^{S}+k_{h_{2}}-iZ)(k_{e_{2}}^{S}+k_{e_{1}}+iZ)+
v^{2}(k_{e_{1}}-k_{h_{1}}^{S}+iZ)(k_{e_{2}}^{S}-k_{h_{2}}+iZ)}\,,\\
%b_{62}&=\frac{u^{2}(k_{e_{2}}^{S}-k_{e_{2}}-iZ)(k_{h_{2}}^{S}+k_{h_{1}}-iZ)+
%v^{2}(k_{e_{2}}^{S}+k_{h_{1}}-iZ)(k_{e_{2}}-k_{h_2}^{S}+iZ)}{
%u^{2}(k_{h_{2}}^{S}+k_{h_{1}}-iZ)(k_{e_{1}}^{S}+k_{e_{2}}+iZ)+
%v^{2}(k_{e_{1}}^{S}-k_{h_{1}}+iZ)(k_{e_{2}}-k_{h_{2}}^{S}+iZ)}\,,\\
b_{71}&=\frac{u^{2}(k_{h_{1}}^{S}-k_{h_{1}}+iZ)(k_{e_{1}}^{S}+k_{e_{2}}+iZ)+
v^{2}(k_{h_{1}}-k_{e_{1}}^{S}-iZ)(k_{e_{2}}+k_{h_1}^{S}+iZ)}{
u^{2}(k_{h_{2}}^{S}+k_{h_{1}}-iZ)(k_{e_{1}}^{S}+k_{e_{2}}+iZ)+
v^{2}(k_{e_{1}}^{S}-k_{h_{1}}+iZ)(k_{e_{2}}-k_{h_{2}}^{S}+iZ)}\,,\\
%b_{82}&=
%\frac{u^{2}(k_{e_{2}}^{S}+k_{e_{1}}+iZ)(k_{h_{2}}^{S}-k_{h_{2}}+iZ)
%+
%v^{2}(k_{h_{2}}-k_{e_{2}}^{S}-iZ)(k_{h_{2}}^{S}+k_{e_{1}}+iZ)
%}{
%u^{2}(k_{e_{2}}^{S}+k_{e_{1}}+iZ)(k_{h_{1}}^{S}+k_{h_{2}}-iZ)+
%v^{2}(k_{e_{1}}-k_{h_{1}}^{S}+iZ)(k_{e_{2}}^{S}-k_{h_{2}}+iZ)}\,.
\end{split}
\end{equation}
For the electron-hole component we get
\begin{equation}
\begin{split}
[G_{eh}^{r}]_{\uparrow\downarrow}(x,x',\omega)&=
\frac{\eta uv}{i(k_{{e}_{1}}^{S}+k_{{e}_{2}}^{S})(u^{2}-v^{2})}
\bigg\{
\theta(x-x'){\rm e}^{ik_{{e}_{2}}^{S}(x-x')}+
\theta(x'-x){\rm e}^{-ik_{{e}_{1}}^{S}(x-x')}+
b_{51}{\rm e}^{i(k_{{e}_{2}}^{S}x+k_{{e}_{1}}^{S}x')}
%[\theta(x-x')b_{62}+\theta(x'-x)b_{51}]
\bigg\}\\
&+\frac{\eta uv}{i(k_{{h}_{1}}^{S}+k_{{h}_{2}}^{S})(u^{2}-v^{2})}
\bigg\{
\theta(x-x'){\rm e}^{-ik_{{h}_{1}}^{S}(x-x')}+
\theta(x'-x){\rm e}^{ik_{{h}_{2}}^{S}(x-x')}
+b_{71}{\rm e}^{-i(k_{{h}_{2}}^{S}x'+k_{{h}_{1}}^{S}x)}
%[\theta(x-x')b_{71}+\theta(x'-x)b_{82}]
\bigg\}\\
&+\frac{\eta a_{52}}{i(u^{2}-v^{2})}
\bigg[u^{2}{\rm e}^{i(k_{{e}_{2}}^{S}x-k_{{h}_{2}}^{S}x')} +v^{2}{\rm e}^{i(k_{{e}_{1}}^{S}x'-k_{{h}_{1}}^{S}x)}\bigg]
\,,\\
[G_{eh}^{r}]_{\downarrow\uparrow}(x,x',\omega)&=
-\frac{\eta uv}{i(k_{{e}_{1}}^{S}+k_{{e}_{2}}^{S})(u^{2}-v^{2})}
\bigg\{
\theta(x-x'){\rm e}^{ik_{{e}_{1}}^{S}(x-x')}+
\theta(x'-x){\rm e}^{-ik_{{e}_{2}}^{S}(x-x')}+b_{51}{\rm e}^{i(k_{{e}_{2}}^{S}x'+k_{{e}_{1}}^{S}x)}
%[\theta(x-x')b_{51}+\theta(x'-x)b_{62}]
\bigg\}\\
%&-\frac{\eta a_{61}\theta(x'-x)}{i(u^{2}-v^{2})}
%\bigg[\frac{u}{v}{\rm e}^{i(k_{{e}_{1}}^{S}x-k_{{h}_{1}}^{S}x')} +\frac{v}{u}{\rm e}^{i(k_{{e}_{2}}^{S}x'-k_{{h}_{2}}^{S}x)}\bigg]\\
&-
\frac{\eta uv}{i(k_{{h}_{1}}^{S}+k_{{h}_{2}}^{S})(u^{2}-v^{2})}
\bigg\{
\theta(x-x'){\rm e}^{-ik_{{h}_{2}}^{S}(x-x')}+
\theta(x'-x){\rm e}^{ik_{{h}_{1}}^{S}(x-x')}
+b_{71}{\rm e}^{-i(k_{{h}_{2}}^{S}x+k_{{h}_{1}}^{S}x')}
%[\theta(x-x')b_{82}+\theta(x'-x)b_{71}]
\bigg\}\\
&-\frac{\eta a_{61}}{i(u^{2}-v^{2})}
\bigg[u^{2}{\rm e}^{i(k_{{e}_{1}}^{S}x-k_{{h}_{1}}^{S}x')} +v^{2}{\rm e}^{i(k_{{e}_{2}}^{S}x'-k_{{h}_{2}}^{S}x)}\bigg]\,\\
[G_{eh}^{r}]_{\uparrow\uparrow}(x,x',\omega)&=0\,,\\
[G_{eh}^{r}]_{\downarrow\downarrow}(x,x',\omega)&=0\,.
\end{split}
\end{equation}

From the diagonal elements of  $G_{ee}^{r}$ given by Eqs.\,(\ref{NSSOCAPPgee}) we calculate the LDOS as $\rho_{S}(x,\omega)=(-1/\pi){\rm Im}{\rm Tr}[G^{r}_{ee}(x,x,\omega)]$, which in the large chemical potential limit and for energies within $\Delta$ read as $\rho_{S}(x,\omega)=(-1/\pi){\rm Im}[\bar{\rho}(x,\omega)]$, where
\begin{equation}
\label{LDOSNSS}
\begin{split}
\bar{\rho}(x,\omega)
&=\frac{2\eta}{i(u^{2}-v^{2})}\bigg[\frac{u^{2}}{k_{e_{1}}^{S}+k_{e_{2}}^{S}}+\frac{v^{2}}{k_{h_{1}}^{S}+k_{h_{2}}^{S}} \bigg]+
\frac{2\eta {\rm e}^{-2\kappa x}}{i(u^{2}-v^{2})}\bigg[\frac{u^{2}{\rm e}^{2i\bar{k} x}b_{51}}{k_{e_{1}}^{S}+k_{e_{2}}^{S}}+\frac{v^{2}{\rm e}^{-2i\bar{k} x}b_{71}}{k_{h_{1}}^{S}+k_{h_{2}}^{S}} \bigg]\,
+\frac{4\eta uv a_{52}{\rm e}^{-2\kappa x}}{i(u^{2}-v^{2})}.
\end{split}
\end{equation}
Thus the LDOS includes contributions from the bulk $\bar{\rho}_{B}$ (first term in square brackets) and interface $\bar{\rho}_{I}$ through normal (second term) and Andreev reflections (third term). The interface terms are discussed in the main text and its relation to odd-frequency pairing is there clearly pointed out.

From the anomalous Green's function we obtain the pairing amplitudes,  decomposed according to Eq.\,(\ref{decomposeSPIN}) and in the large chemical potential limit are given by
\begin{equation}
\begin{split}
f_{0}^{r}(x,x',\omega)&=
\frac{\eta uv}{2i(u^{2}-v^{2})}
\bigg\{
2{\rm cos}[k_{SO}|x-x'|]{\rm e}^{-\kappa|x-x'|}
\bigg[
\frac{{\rm e}^{i\bar{k}|x-x'|}}{k_{e_1}^{S}+k_{e_2}^{S}}
+
\frac{{\rm e}^{-i\bar{k}|x-x'|}}{k_{h_1}^{S}+k_{h_2}^{S}}
\bigg]\,,\\
&+
2{\rm cos}[k_{\rm SO}|x-x'|]{\rm e}^{-\kappa(x+x')}
\bigg[
\frac{{\rm e}^{i\bar{k}|x-x'|}}{k_{e_1}^{S}+k_{e_2}^{S}}
b_{51}
%\bigg[
%b_{51}{\rm e}^{ik_{SO}|x-x'|}+b_{62}{\rm e}^{-ik_{SO}|x-x'|}
%\bigg]
+
\frac{{\rm e}^{-i\bar{k}|x-x'|}}{k_{h_1}^{S}+k_{h_2}^{S}}
b_{71}
%\bigg[
%b_{82}{\rm e}^{ik_{SO}|x-x'|}+b_{71}{\rm e}^{-ik_{SO}|x-x'|}
%\bigg]
\bigg]\\
&+{\rm e}^{-\kappa(x+x')}
\Big(\frac{u}{v}{\rm e}^{i\bar{k}(x-x')}+\frac{v}{u}{\rm e}^{-i\bar{k}(x-x')}  \Big)
a_{52}2{\rm cos}[k_{SO}|x-x'|]
%\Big(\tilde{a}_{72}{\rm e}^{-ik_{SO}|x-x'|}+a_{52}{\rm e}^{ik_{SO}|x-x'|}\Big)
\bigg\}\,,
\\
f_{1}^{r}(x,x',\omega)&=0\,,\quad f_{2}^{r}(x,x',\omega)=0\,,\\
f_{3}^{r}(x,x',\omega)&=
\frac{\eta uv}{2i(u^{2}-v^{2})}
\bigg\{
(-2i){\rm sin}[k_{SO}|x-x'|]{\rm e}^{-\kappa|x-x'|}
\bigg[
\frac{{\rm e}^{i\bar{k}|x-x'|}}{k_{e_1}^{S}+k_{e_2}^{S}}
+
\frac{{\rm e}^{-i\bar{k}|x-x'|}}{k_{h_1}^{S}+k_{h_2}^{S}}
\bigg]\,,\\
&+
{\rm sgn}(x-x')(-2i){\rm sin}[k_{SO}|x-x'|]{\rm e}^{-\kappa(x+x')}
\bigg[
\frac{{\rm e}^{i\bar{k}|x-x'|}}{k_{e_1}^{S}+k_{e_2}^{S}}
b_{51}
%\big(
%b_{62}{\rm e}^{-ik_{SO}|x-x'|}-b_{51}{\rm e}^{ik_{SO}|x-x'|}
%\bigg)
+
\frac{{\rm e}^{-i\bar{k}|x-x'|}}{k_{h_1}^{S}+k_{h_2}^{S}}
b_{71}
%\big(
%b_{71}{\rm e}^{-ik_{SO}|x-x'|}-b_{82}{\rm e}^{ik_{SO}|x-x'|}
%\big)
\bigg]\\
&+{\rm sgn}(x-x'){\rm e}^{-\kappa(x+x')}
\Big(\frac{u}{v}{\rm e}^{i\bar{k}(x-x')}+\frac{v}{u}{\rm e}^{-i\bar{k}(x-x')}  \Big)
a_{52}(-2i){\rm sin}[k_{SO}|x-x'|]
%\Big(\tilde{a}_{72}{\rm e}^{-ik_{SO}|x-x'|}-a_{52}{\rm e}^{ik_{SO}|x-x'|}\Big)
\bigg\}\,.
\end{split}
\end{equation}
Further isolating the even- and odd-frequency components we arrive at
\begin{equation}
\begin{split}
f^{r,{\rm E}}_{0}(x,x',\omega)&=
\frac{\eta uv}{2i(u^{2}-v^{2})}
\bigg\{
2{\rm cos}[k_{SO}|x-x'|]{\rm e}^{-\kappa|x-x'|}
\bigg[
\frac{{\rm e}^{i\bar{k}|x-x'|}}{k_{e_1}^{S}+k_{e_2}^{S}}
+
\frac{{\rm e}^{-i\bar{k}|x-x'|}}{k_{h_1}^{S}+k_{h_2}^{S}}
\bigg]\,,\\
&+
2{\rm cos}[k_{SO}|x-x'|]{\rm e}^{-\kappa(x+x')}
\bigg[
\frac{{\rm e}^{i\bar{k}|x-x'|}}{k_{e_1}^{S}+k_{e_2}^{S}}
b_{51}
%\bigg[
%b_{51}{\rm e}^{ik_{SO}|x-x'|}+b_{62}{\rm e}^{-ik_{SO}|x-x'|}
%\bigg]
+
\frac{{\rm e}^{-i\bar{k}|x-x'|}}{k_{h_1}^{S}+k_{h_2}^{S}}
b_{71}
%\bigg[
%b_{82}{\rm e}^{ik_{SO}|x-x'|}+b_{71}{\rm e}^{-ik_{SO}|x-x'|}
%\bigg]
\bigg]\\
&+{\rm e}^{-\kappa(x+x')}{\rm cos}[\bar{k}(x-x')]
\Big(\frac{u}{v}+\frac{v}{u}\Big)
a_{52}2{\rm cos}[k_{SO}|x-x'|]
\bigg\}
\,,\\
f^{r,{\rm O}}_{0}(x,x',\omega)&=\frac{\eta}{2i}{\rm e}^{-\kappa(x+x')}(i){\rm sin}[\bar{k}(x-x')]
a_{52}2{\rm cos}[k_{SO}|x-x'|]
%\Big(\tilde{a}_{72}{\rm e}^{-ik_{SO}|x-x'|}+a_{52}{\rm e}^{ik_{SO}|x-x'|}\Big)
\,,\\
f^{r,{\rm E}}_{3}(x,x',\omega)&=
\frac{\eta uv}{2i(u^{2}-v^{2})}
\bigg\{
(-2i){\rm sin}[k_{SO}|x-x'|]{\rm e}^{-\kappa|x-x'|}
\bigg[
\frac{{\rm e}^{i\bar{k}|x-x'|}}{k_{e_1}^{S}+k_{e_2}^{S}}
+
\frac{{\rm e}^{-i\bar{k}|x-x'|}}{k_{h_1}^{S}+k_{h_2}^{S}}
\bigg]\,,\\
&+
{\rm sgn}(x-x')(-2i){\rm sin}[k_{SO}|x-x'|]{\rm e}^{-\kappa(x+x')}
\bigg[
\frac{{\rm e}^{i\bar{k}|x-x'|}}{k_{e_1}^{S}+k_{e_2}^{S}}
b_{51}
%\bigg[
%b_{62}{\rm e}^{-ik_{SO}|x-x'|}-b_{51}{\rm e}^{ik_{SO}|x-x'|}
%\bigg]
+
\frac{{\rm e}^{-i\bar{k}|x-x'|}}{k_{h_1}^{S}+k_{h_2}^{S}}
b_{71}
%\bigg[
%b_{71}{\rm e}^{-ik_{SO}|x-x'|}-b_{82}{\rm e}^{ik_{SO}|x-x'|}
%\bigg]
\bigg]\\
&+{\rm sgn}(x-x'){\rm cos}[\bar{k}(x-x')]{\rm e}^{-\kappa(x+x')}
\Big(\frac{u}{v}+\frac{v}{u}\Big)
a_{52}(-2i){\rm sin}[k_{SO}|x-x'|]
%\Big(\tilde{a}_{72}{\rm e}^{-ik_{SO}|x-x'|}-a_{52}{\rm e}^{ik_{SO}|x-x'|}\Big)
\bigg\}\,,\\
f^{r,{\rm O}}_{3}(x,x',\omega)&=
\frac{\eta}{2i}
{\rm sgn}(x-x')i{\rm sin}[\bar{k}(x-x')]{\rm e}^{-\kappa(x+x')}
a_{52}(-2i){\rm sin}[k_{SO}|x-x'|]
%\Big(\tilde{a}_{72}{\rm e}^{-ik_{SO}|x-x'|}-a_{52}{\rm e}^{ik_{SO}|x-x'|}\Big)
\,.
\end{split}
\end{equation}
These expressions correspond to the pairing amplitudes given in the main text by Eqs.\,(\ref{PairinNS_S}).

\subsection{Short SNS junction}
\label{SNSAppSOC}
Finally, we treat a short SNS junctions located at $x=0$ with finite Rashba SO coupling. The solution method is the same as above but here we also have to keep track of a  finite phase difference across the junction, as given by Eq.\,(\ref{SNSDelta}).
The scattering states are defined in the left (zero phase) and right (finite phase $\phi$) superconducting regions and acquire the same form as in Eqs.\,(\ref{PsiNSSOC}), with the eigenvectors in the left region labelled by $S_{L}$ and of the the same form as in NS junctions, and in the right region labeled by $S_{R}$ and reading as
\begin{equation}
\begin{split}
\phi_{1,2}^{S_{R}}&=
 \begin{pmatrix}
 u{\rm e}^{i\phi/2}\\
 0\\
 0\\
v{\rm e}^{-i\phi/2}
 \end{pmatrix},\, 
 \phi_{3,4}^{S_{R}}=
 \begin{pmatrix}
 0\\
 -u{\rm e}^{i\phi/2}\\
 v{\rm e}^{-i\phi/2}\\
0
 \end{pmatrix},\, 
 \phi_{5,6}^{S_{R}}=
 \begin{pmatrix}
0\\
 -v{\rm e}^{i\phi/2}\\
 u{\rm e}^{-i\phi/2}\\
0
 \end{pmatrix},\, 
 \phi_{7,8}^{S_{R}}=
 \begin{pmatrix}
 v{\rm e}^{i\phi/2}\\
 0\\
 0\\
u{\rm e}^{-i\phi/2}
 \end{pmatrix}\,,\\
 \tilde{\phi}_{1,2}^{S_{R}}&=
 \begin{pmatrix}
 0\\
 -u{\rm e}^{-i\phi/2}\\
 v{\rm e}^{i\phi/2}\\
0
 \end{pmatrix}\,,
 \tilde{\phi}_{3,4}^{S_{R}}=
 \begin{pmatrix}
 u{\rm e}^{-i\phi/2}\\
 0\\
 0\\
v{\rm e}^{i\phi/2}
 \end{pmatrix}\,,
  \tilde{\phi}_{5,6}^{S_{R}}=
 \begin{pmatrix}
v{\rm e}^{-i\phi/2}\\
 0\\
 0\\
u{\rm e}^{i\phi/2}
 \end{pmatrix}\,,
  \tilde{\phi}_{7,8}^{S_{R}}=
 \begin{pmatrix}
 0\\
  -v{\rm e}^{-i\phi/2}\\
u{\rm e}^{i\phi/2}\\
 0
 \end{pmatrix}
 \end{split}
 \end{equation} 
 The scattering states have the same form as in previous section for NS junctions 
and are fully determined after matching them at the interface $x=0$. Then, the Green's functions are constructed following Eqs.\,(\ref{RGF}) and (\ref{conditionGRSO}).
The left and right S regions provide the same information and we can only focus on the right region. The electron-electron component is given by
\begin{equation}
\begin{split}
[G_{ee}^{r}]_{\uparrow\uparrow}&=
\frac{\eta u^{2}}{i(k_{e_{1}}^{S}+k_{e_{2}}^{S})(u^{2}-v^{2})}
\bigg\{
\theta(x-x'){\rm e}^{ik_{e_{2}}^{S}(x-x')}+\theta(x'-x){\rm e}^{-ik_{e_{1}}^{S}(x-x')}\\
&+
{\rm e}^{i(k_{e_{2}}^{S}x+k_{e_{1}}^{S}x')}[\tilde{b}_{62}\theta(x-x')+b_{51}\theta(x'-x)]
\bigg\}
+
\frac{\eta uv }{i(u^{2}-v^{2})}\Big[\theta(x-x')\tilde{a}_{61}{\rm e}^{i(k_{e_{2}}^{S}x-k_{h_{2}}^{S}x')}+
\theta(x'-x)a_{52}{\rm e}^{i(k_{e_{1}}^{S}x'-k_{h_{1}}^{S}x)}\Big]\\
&+
\frac{\eta v^{2}}{i(k_{h_{1}}^{S}+k_{h_{2}}^{S})(u^{2}-v^{2})}
\bigg\{
\theta(x-x'){\rm e}^{-ik_{h_{1}}^{S}(x-x')}+\theta(x'-x){\rm e}^{ik_{h_{2}}^{S}(x-x')}\\
&+
{\rm e}^{-i(k_{h_{1}}^{S}x+k_{h_{2}}^{S}x')}[\tilde{b}_{71}\theta(x-x')+b_{82}\theta(x'-x)]
\bigg\}
+
\frac{\eta uv }{i(u^{2}-v^{2})}\Big[\theta(x-x')\tilde{a}_{72}{\rm e}^{i(k_{e_{1}}^{S}x'-k_{h_{1}}^{S}x)}+
\theta(x'-x)a_{81}{\rm e}^{i(k_{e_{2}}^{S}x-k_{h_{2}}^{S}x')}\Big]
\,,\\
[G_{ee}^{r}]_{\downarrow\downarrow}&=
\frac{\eta u^{2}}{i(k_{e_{1}}^{S}+k_{e_{2}}^{S})(u^{2}-v^{2})}
\bigg\{
\theta(x-x'){\rm e}^{ik_{e_{1}}^{S}(x-x')}+\theta(x'-x){\rm e}^{-ik_{e_{2}}^{S}(x-x')}\\
&+
{\rm e}^{i(k_{e_{1}}^{S}x+k_{e_{2}}^{S}x')}[\tilde{b}_{51}\theta(x-x')+b_{62}\theta(x'-x)]
\bigg\}
+
\frac{\eta uv }{i(u^{2}-v^{2})}\Big[\theta(x-x')\tilde{a}_{52}{\rm e}^{i(k_{e_{1}}^{S}x-k_{h_{1}}^{S}x')}+
\theta(x'-x)a_{61}{\rm e}^{i(k_{e_{2}}^{S}x'-k_{h_{2}}^{S}x)}\Big]\\
&+
\frac{\eta v^{2}}{i(k_{h_{1}}^{S}+k_{h_{2}}^{S})(u^{2}-v^{2})}
\bigg\{
\theta(x-x'){\rm e}^{-ik_{h_{2}}^{S}(x-x')}+\theta(x'-x){\rm e}^{ik_{h_{1}}^{S}(x-x')}\\
&+
{\rm e}^{-i(k_{h_{2}}^{S}x+k_{h_{1}}^{S}x')}[\tilde{b}_{82}\theta(x-x')+b_{71}\theta(x'-x)]
\bigg\}
+
\frac{\eta uv }{i(u^{2}-v^{2})}\Big[\theta(x-x')\tilde{a}_{81}{\rm e}^{i(k_{e_{2}}^{S}x'-k_{h_{2}}^{S}x)}+
\theta(x'-x)a_{72}{\rm e}^{i(k_{e_{1}}^{S}x-k_{h_{1}}^{S}x')}\Big]
\,,\\
[G_{ee}^{r}]_{\uparrow\downarrow}&=0\,,\\
 [G_{ee}^{r}]_{\downarrow\uparrow}&=0\,,\\
\end{split}
\end{equation}
and the electron-hole component is given by
\begin{equation}
\begin{split}
[G_{eh}^{r}]_{\uparrow\uparrow}&=0\,,\\
 [G_{eh}^{r}]_{\downarrow\downarrow}&=0\,,\\
[G_{eh}^{r}]_{\uparrow\downarrow}&=
\frac{\eta uv {\rm e}^{i\phi}}{i(k_{e_{1}}^{S}+k_{e_{2}}^{S})(u^{2}-v^{2})}
\bigg\{
\theta(x-x'){\rm e}^{ik_{e_{2}}^{S}(x-x')}+\theta(x'-x){\rm e}^{-ik_{e_{1}}^{S}(x-x')}
+
{\rm e}^{i(k_{e_{2}}^{S}x+k_{e_{1}}^{S}x')}[\tilde{b}_{62}\theta(x-x')+b_{51}\theta(x'-x)]
\bigg\}\\
&+
\frac{\eta {\rm e}^{\phi}}{i(u^{2}-v^{2})}
\Big[\theta(x-x')u^{2}\tilde{a}_{61}{\rm e}^{i(k_{e_{2}}^{S}x-k_{h_{2}}^{S}x')}+
\theta(x'-x)v^{2}a_{52}{\rm e}^{i(k_{e_{1}}^{S}x'-k_{h_{1}}^{S}x)}\Big]\\
&+
\frac{\eta uv {\rm e}^{i\phi}}{i(k_{h_{1}}^{S}+k_{h_{2}}^{S})(u^{2}-v^{2})}
\bigg\{
\theta(x-x'){\rm e}^{-ik_{h_{1}}^{S}(x-x')}+\theta(x'-x){\rm e}^{ik_{h_{2}}^{S}(x-x')}
+
{\rm e}^{i(k_{h_{1}}^{S}x+k_{h_{2}}^{S}x')}[\tilde{b}_{71}\theta(x-x')+b_{82}\theta(x'-x)]
\bigg\}\\
&+
\frac{\eta {\rm e}^{\phi}}{i(u^{2}-v^{2})}
\Big[\theta(x-x')v^{2}\tilde{a}_{72}{\rm e}^{i(k_{e_{1}}^{S}x'-k_{h_{1}}^{S}x)}+
\theta(x'-x)u^{2}a_{81}{\rm e}^{i(k_{e_{2}}^{S}x-k_{h_{2}}^{S}x')}\Big]
\,,\\
 [G_{eh}^{r}]_{\downarrow\uparrow}&=
 -
\frac{\eta uv {\rm e}^{i\phi}}{i(k_{e_{1}}^{S}+k_{e_{2}}^{S})(u^{2}-v^{2})}
\bigg\{
\theta(x-x'){\rm e}^{ik_{e_{1}}^{S}(x-x')}+\theta(x'-x){\rm e}^{-ik_{e_{2}}^{S}(x-x')}
+
{\rm e}^{i(k_{e_{1}}^{S}x+k_{e_{2}}^{S}x')}[\tilde{b}_{51}\theta(x-x')+b_{62}\theta(x'-x)]
\bigg\}\\
&-
\frac{\eta {\rm e}^{\phi}}{i(u^{2}-v^{2})}
\Big[\theta(x-x')u^{2}\tilde{a}_{52}{\rm e}^{i(k_{e_{1}}^{S}x-k_{h_{1}}^{S}x')}+
\theta(x'-x)v^{2}a_{61}{\rm e}^{i(k_{e_{2}}^{S}x'-k_{h_{2}}^{S}x)}\Big]\\
&-
\frac{\eta uv {\rm e}^{i\phi}}{i(k_{h_{1}}^{S}+k_{h_{2}}^{S})(u^{2}-v^{2})}
\bigg\{
\theta(x-x'){\rm e}^{-ik_{h_{2}}^{S}(x-x')}+\theta(x'-x){\rm e}^{ik_{h_{1}}^{S}(x-x')}
+
{\rm e}^{-i(k_{h_{1}}^{S}x'+k_{h_{2}}^{S}x)}[\tilde{b}_{82}\theta(x-x')+b_{71}\theta(x'-x)]
\bigg\}\\
&-
\frac{\eta {\rm e}^{\phi}}{i(u^{2}-v^{2})}
\Big[\theta(x-x')v^{2}\tilde{a}_{81}{\rm e}^{i(k_{e_{2}}^{S}x'-k_{h_{2}}^{S}x)}+
\theta(x'-x)u^{2}a_{72}{\rm e}^{i(k_{e_{1}}^{S}x-k_{h_{1}}^{S}x')}\Big]
 \,,\\
\end{split}
\end{equation}
where the coefficients $a_{ij}$ and $b_{ij}$ become phase-dependent  due to the finite phase difference across the junction and they are thus different than the coefficients found in NS junctions. 

We have found that in general $\tilde{b}_{ij}(\phi)=b_{ij}(\phi)$, $\tilde{a}_{ij}(\phi)=a_{ij}(-\phi)$, and
\begin{equation}
\begin{split}
b_{62}(\phi)&=b_{51}(\phi)\,,\quad b_{82}(\phi)=b_{71}(\phi)\,, \\
a_{72}(\phi)&=\tilde{a}_{61}(\phi)\,, \quad a_{81}(\phi)=\tilde{a}_{52}(\phi)\,,\quad 
\tilde{a}_{72}(\phi)=a_{61}(\phi)\,,\quad \tilde{a}_{81}(\phi)=a_{52}(\phi)\,,\\
b_{51}(\phi)&=\frac{4{\rm sin}^{2}(\phi/2)(k_{e_{1}}^{S}-k_{h_{1}}^{S})(k_{e_{1}}^{S}+k_{h_{2}}^{S})u^{2}v^{2}-iZ(k_{h_{1}}^{S}+k_{h_{2}}^{S}-iZ)}{K}\,,\\
b_{71}(\phi)&=-\frac{4{\rm sin}^{2}(\phi/2)(k_{e_{1}}^{S}-k_{h_{1}}^{S})(k_{e_{2}}^{S}+k_{h_{1}}^{S})u^{2}v^{2}-iZ(k_{e_{1}}^{S}+k_{e_{2}}^{S}+iZ)}{K}\,,\\
a_{52}(\phi)&=\frac{(1-{\rm e}^{-i\phi})(k_{e_{1}}^{S}+k_{h_{2}}^{S})uv ({\rm e}^{i\phi}u^{2}-v^{2})}{K}\,,\\
a_{61}(\phi)&=\frac{(1-{\rm e}^{-i\phi})(k_{e_{2}}^{S}+k_{h_{1}}^{S})uv ({\rm e}^{i\phi}u^{2}-v^{2})}{K}\,,
\end{split}
\end{equation}
where $K=(u^{4}+v^{4})(k_{h_{1}}^{S}+k_{h_{2}}^{S}-iZ)(k_{e_{1}}^{S}+k_{e_{2}}^{S}+iZ)-2u^{2}v^{2}[(k_{h_{2}}^{S}-k_{e_{2}}^{S}-iZ)(k_{e_{1}}^{S}-k_{h_{1}}^{S}+iZ)+(k_{e_{2}}^{S}+k_{h_{1}}^{S})(k_{e_{1}}^{S}+k_{h_{2}}^{S}){\rm cos}(\phi)]$.

The pairing amplitudes are finally found by employing Eqs.\,(\ref{decomposeSPIN}) in a similar way as for NS junctions. We therefore do not repeat the process here, but rather only write  the even- and odd-frequency components, which read as
\begin{equation}
\begin{split}
f_{0,{\rm B}}^{r,{\rm E}}(x,x',\omega)&=2B(\omega)\bigg[\frac{{\rm e}^{i\bar{k}|x-x'|}}{k_{e_{1}}^{S}+k_{e_{2}}^{S}}
+\frac{{\rm e}^{-i\bar{k}|x-x'|}}{k_{h_{1}}^{S}+k_{h_{2}}^{S}}\bigg]{\rm cos}[k_{\rm so}|x-x'|]{\rm e}^{-\kappa|x-x'|}{\rm e}^{i\phi_{\rm R}}\,,\\
f_{0,{\rm I}}^{r,{\rm E}}(x,x',\omega)&=B(\omega)
\bigg\{
\frac{{\rm e}^{i\bar{k}(x+x')}}{k_{e_{1}}^{S}+k_{e_{2}}^{S}}
\Big[ b_{51}{\rm e}^{ik_{\rm so}|x-x'|}+b_{62}{\rm e}^{-ik_{\rm so}|x-x'|}\Big]
+
\frac{{\rm e}^{-i\bar{k}(x+x')}}{k_{h_{1}}^{S}+k_{h_{2}}^{S}}
\Big[ b_{82}{\rm e}^{ik_{\rm so}|x-x'|}+b_{71}{\rm e}^{-ik_{\rm so}|x-x'|}\Big]
\,,\\
&+{\rm cos}[\bar{k}(x-x')] \bigg[
\Big(\tilde{a}_{61}\frac{u}{v}+a_{61}\frac{v}{u}\Big){\rm e}^{-ik_{\rm so}|x-x'|}
+
\Big(\tilde{a}_{52}\frac{u}{v}+a_{52}\frac{v}{u}\Big){\rm e}^{ik_{\rm so}|x-x'|}
\bigg]\bigg\}{\rm e}^{-\kappa(x+x')}{\rm e}^{i\phi_{\rm R}}\,,\\
f_{0,{\rm B}}^{r,{\rm O}}(x,x',\omega)&=0\,,\\
f_{0,{\rm I}}^{r,{\rm O}}(x,x',\omega)&=iB(\omega)\bigg\{
\Big[\tilde{a}_{61}\frac{u}{v}-a_{61}\frac{v}{u}\Big]{\rm e}^{-ik_{\rm so}|x-x'|}
+
\Big[\tilde{a}_{52}\frac{u}{v}-a_{52}\frac{v}{u}\Big]{\rm e}^{ik_{\rm so}|x-x'|}
\bigg\}{\rm sin}[\bar{k}(x-x')]{\rm e}^{-\kappa(x+x')}{\rm e}^{i\phi_{\rm R}}\,,\\
f_{3,{\rm B}}^{r,{\rm E}}(x,x',\omega)&=-2iB(\omega)\bigg[\frac{{\rm e}^{i\bar{k}|x-x'|}}{k_{e_{1}}^{S}+k_{e_{2}}^{S}}
+\frac{{\rm e}^{-i\bar{k}|x-x'|}}{k_{h_{1}}^{S}+k_{h_{2}}^{S}}\bigg]{\rm sgn}(x-x'){\rm sin}[k_{\rm so}|x-x'|]{\rm e}^{-\kappa|x-x'|}{\rm e}^{i\phi_{\rm R}}\,,\\
f_{3,{\rm I}}^{r,{\rm O}}(x,x',\omega)&=B(\omega)
\bigg\{
\frac{{\rm e}^{i\bar{k}(x+x')}}{k_{e_{1}}^{S}+k_{e_{2}}^{S}}
\Big[b_{62}{\rm e}^{-ik_{\rm so}|x-x'|}-b_{51}{\rm e}^{ik_{\rm so}|x-x'|}\Big]
+
\frac{{\rm e}^{-i\bar{k}(x+x')}}{k_{h_{1}}^{S}+k_{h_{2}}^{S}}
\Big[b_{71}{\rm e}^{-ik_{\rm so}|x-x'|}-b_{82}{\rm e}^{ik_{\rm so}|x-x'|}\Big]
\,,\\
&+{\rm cos}[\bar{k}(x-x')] \bigg[
\Big(\tilde{a}_{61}\frac{u}{v}+a_{61}\frac{v}{u}\Big){\rm e}^{-ik_{\rm so}|x-x'|}
-
\Big(\tilde{a}_{52}\frac{u}{v}+a_{52}\frac{v}{u}\Big){\rm e}^{ik_{\rm so}|x-x'|}
\bigg]\bigg\}{\rm sgn}(x-x'){\rm e}^{-\kappa(x+x')}{\rm e}^{i\phi_{\rm R}}\,,\\
f_{3,{\rm B}}^{r,{\rm O}}(x,x',\omega)&=0\,,\\
f_{3,{\rm I}}^{r,{\rm O}}(x,x',\omega)&=
iB(\omega)
 \bigg[
\Big(\tilde{a}_{61}\frac{u}{v}-a_{61}\frac{v}{u}\Big){\rm e}^{-ik_{\rm so}|x-x'|}
-
\Big(\tilde{a}_{52}\frac{u}{v}-a_{52}\frac{v}{u}\Big){\rm e}^{ik_{\rm so}|x-x'|}
\bigg]\bigg\}{\rm sin}[\bar{k}(x-x')]{\rm sgn}(x-x'){\rm e}^{-\kappa(x+x')}{\rm e}^{i\phi_{\rm R}}
\,.
\end{split}
\end{equation}
which correspond to ESE, OSO, ETO and OTE symmetries, respectively, and are reported in Eqs.\,(\ref{SNSpairing}) in the main text, where the results are also extensively analyzed.
%An important feature in previous expressions is that the odd-frequency components do not have a bulk (B) term and are solely proportional to Andreev coefficients $a_{ij}$, while the even-frequency ones include bulk terms, normal and Andreev reflection contributions. These results indeed support the idea that  there is a strong relation between odd-frequency pairing and Andreev reflection. From previous expressions we also can see that locally ($x=x'$) only ESE exists, and OSO, ETO and OTE amplitudes vanish. 
\end{widetext}
\end{document}